\documentclass[]{aa}
\nolinenumbers

\usepackage{graphicx}
\usepackage{txfonts}
\usepackage{placeins}

\usepackage[]{acronym}
\usepackage[
    colorlinks=true,
    citecolor=blue,
    linkcolor=blue,
    urlcolor=cyan]{hyperref}

\acrodef{nfw}[NFW]{\citet{Navarro1997}}
\acrodef{los}[LOS]{line-of-sight}
\acrodef{icm}[ICM]{intracluster medium}
\acrodef{psf}[PSF]{point spread function}
\acrodef{tm}[TM]{telescope module}
\acrodef{erass}[eRASS]{eROSITA All Sky Survey}
\acrodef{fof}[FoF]{friend-of-friend}
\acrodef{healpix}[HEALPix]{hierarchical equal area isolatitude pixelization}
\acrodef{whim}[WHIM]{warm-hot intergalactic medium}
\acrodef{igm}[IGM]{intergalactic medium}
\acrodef{sz}[SZ]{Sunyaev-Zeldovich}
\acrodef{nfw}[NFW]{Navarro-Frenk-White}
\acrodef{gnfw}[gNFW]{generalized NFW}
\acrodef{spt}[SPT]{South Pole Telescope}
\acrodef{act}[ACT]{Atacama Cosmology Telescope}
\acrodef{hmf}[HMF]{halo mass function}
\acrodef{the300}[THE300]{The Three Hundred}
\acrodef{agn}[AGN]{active galactic nuclei}
\acrodef{snr}[S/N]{signal-to-noise ratio}

\defcitealias{Lyskova2023}{L23}
\defcitealias{Vikhlinin2006}{V06}

\begin{document}

\title{The SRG/eROSITA All-Sky Survey}
\subtitle{Detection of shock-heated gas beyond the halo boundary into\\ the accretion region}

\titlerunning{Shock heated gas beyond halo boundary}
\authorrunning{X. Zhang et al.}
\author{
    X.~Zhang \inst{\ref{inst:mpe}},
    E.~Bulbul \inst{\ref{inst:mpe}},
    B.~Diemer \inst{\ref{inst:umd}},
    Y.~E.~Bahar \inst{\ref{inst:mpe}},
    J.~Comparat \inst{\ref{inst:mpe}},
    V.~Ghirardini \inst{\ref{inst:inaf-oas}},
    A.~Liu \inst{\ref{inst:bnu},\ref{inst:mpe}},
    N.~Malavasi \inst{\ref{inst:mpe}},
    T.~Mistele \inst{\ref{inst:mpe}},
    M.~Ramos-Ceja \inst{\ref{inst:mpe}},
    J.~S.~Sanders \inst{\ref{inst:mpe}},
    Y.~Zhang \inst{\ref{inst:mpe}},
    E.~Artis \inst{\ref{inst:mpe}},
    Z.~Ding \inst{\ref{inst:mpe}},
    L.~Fiorino \inst{\ref{inst:mpe}},
    M.~Kluge \inst{\ref{inst:mpe}}, 
    A.~Merloni \inst{\ref{inst:mpe}},
    K.~Nandra \inst{\ref{inst:mpe}},
    \and 
    S.~Zelmer \inst{\ref{inst:mpe}}  }

\institute{
    Max Planck Institute for Extraterrestrial Physics, Giessenbachstrasse 1, 85748 Garching, Germany\label{inst:mpe}\\
          \email{xzhang@mpe.mpg.de}
    \and
    Department of Astronomy, University of Maryland, College Park, MD 20742, USA\label{inst:umd}
     \and 
    INAF, Osservatorio di Astrofisica e Scienza dello Spazio, via Piero Gobetti 93/3, I-40129 Bologna, Italy\label{inst:inaf-oas} 
     \and
    Institute for Frontiers in Astronomy and Astrophysics, Beijing Normal University, Beijing 102206, China\label{inst:bnu}
         }

\date{Received ---; accepted ---}

\abstract{
The hot gas in the outskirts of galaxy cluster-sized halos, extending around and beyond the virial radius into nearby accretion regions, remains among one of the least explored baryon components of the large-scale cosmic structure. We present a stacking analysis of 680 galaxy clusters located in the western Galactic hemisphere, using data from the first two years of the \emph{Spectrum-Roentgen-Gamma}/eROSITA All-Sky Survey. The stacked X-ray surface brightness profile reveals a statistically significant signal extending out to $2\times r_\mathrm{200m}$ ($\sim4.5$~Mpc). The best-fit surface brightness profile is well described by a combination of terms describing orbiting and infalling gas, with a transition occurring around $r_\mathrm{200m}$. At this radius, the best-fit gas number density is $2.5 \times 10^{-5}$ cm$^{-3}$, corresponding to a baryon overdensity of 30. By integrating the gas density profile out to $r_\mathrm{200m}$, we inferred a gas fraction higher than the universal baryon fraction with the assumption of a typical halo concentration. However, correcting for possible clumping effects reduces the baryon fraction by more than 20\%. Additionally, we examined the distribution of hot gas in massive clusters in the IllustrisTNG simulations, from the halo center to the accretion region. This analysis reveals differences in radial gas profiles depending on whether the direction points toward voids or toward nearby cosmic filaments. 
Beyond $r_\mathrm{200m}$, the density profile along the filament direction exceeds that along the void direction. This pattern aligns with the observed transition radius between the one-halo and two-halo terms, suggesting that $r_\mathrm{200m}$ is the approximate radius marking the location at which cosmic filaments connect to galaxy clusters. Meanwhile, comparisons of the gas density and gas fraction profiles between the observation and the IllustrisTNG simulation suggest that the feedback processes in the stacking sample are more efficient at distributing gas to large radii than the IllustrisTNG model. 
}


\maketitle

\nolinenumbers

\section{Introduction}

Galaxy clusters, the most massive collapsed dark matter halos in the Universe, are positioned at the nodes of the cosmic web. Galaxy clusters comprise $T>10^7$~K hot baryonic gas as one sixth of their total mass and the remaining dark matter. While the gravitational potential is governed by dark matter, baryonic physics determines the properties of the hot gas. The cluster outskirts, here defined as the regions beyond $r_\mathrm{500c}$\footnote{Throughout this paper, we use spherical overdensity to define halos' mass and radius. For example, $r_\mathrm{500c}$ is the radius where the enclosed density is 500 times the critical density $\rho_\mathrm{c}$; and $M_\mathrm{200m}$ is the total mass enclosed by $r_\mathrm{200m}$, where the enclosed density is 200 times the mean matter density $\rho_\mathrm{m}$.}, are of particular interest as they hold critical information about both the distribution of dark matter and the thermodynamic state of the gas. 

\begin{figure}
    \centering
    \includegraphics[width=.9\linewidth]{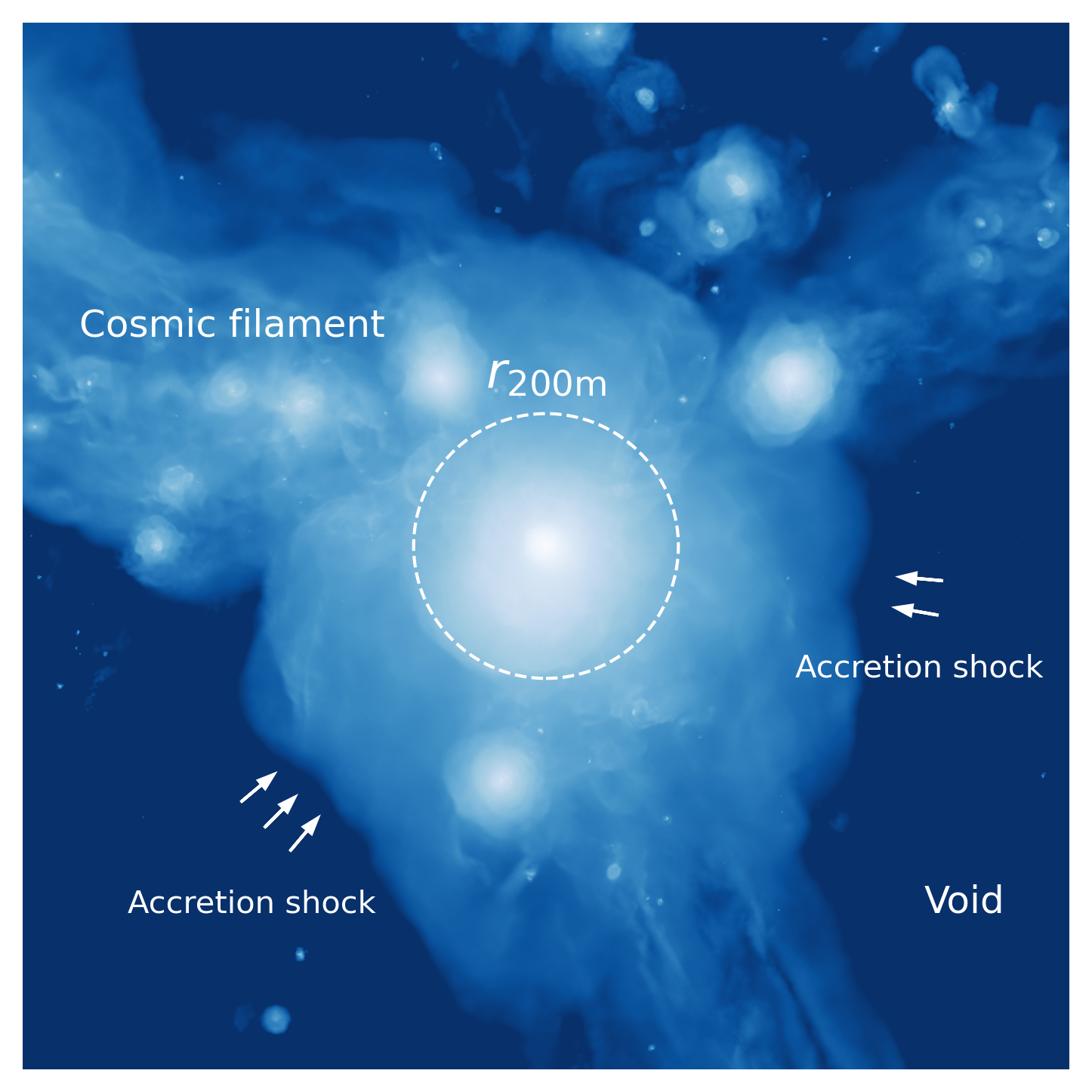}
    \caption{Spatial distribution of the X-ray emission from the hot gas around a massive dark matter halo. The dashed circle indicates the size of $r_\mathrm{200m}$. At large radii, the halo is connected to and accreting smaller nearby halos from cosmic filaments. This map is produced using gas particles from a $20\times20\times20$~Mpc$^{3}$ box around the \texttt{id=32} halo in the $z=0$ snapshot of the \mbox{TNG300-1} simulation (see Sect.~\ref{sect:simulation} for the details of map creation). The central halo is in a mass $M_\mathrm{500c}=2.8\times10^{14}M_\sun$ and an $r_\mathrm{200m}$ of 2.5~Mpc. Short arrows mark the accretion shock, i.e., the boundary between shock-heated gas and the cool intergalactic medium.}
    \label{fig:map_example}
\end{figure}

The radial density profile in the outskirts is characterized by halo mass accretion, for which the behavior of dark matter differs from that of baryons due to their collisional and collisionless nature \citep{Bertschinger1985}. In the case of collisionless dark matter, the infalling matter accumulates near the first apocenter of its orbit \citep[e.g.,][]{Fillmore1984,Bertschinger1985,Adhikari2014}. This leads to a phenomenon known as splashback, where the accreted matter causes a sharp decline in the outer density profile at radii around or beyond $r_\mathrm{200m}$. The exact location of this splashback radius is closely linked to the matter accretion rate \citep[e.g.,][]{Diemer2014,Diemer2017-2}. On the other hand, infalling collisional gas forms accretion shocks\footnote{We follow the convention and use the term ``accretion shock'' to refer to the ``external shock'' in \citet{Ryu2003}. In fact, the accretion shock is around all overdense environments, including cosmic filaments and sheets. In this work, we focus on the accretion shock around galaxy clusters.} \citep[e.g.,][]{Bertschinger1985,Ostriker1988,Shi2016}, heating the cool \ac{igm} to $T\gtrsim10^6$~K and creating a turbulent atmosphere outside the splashback radius \citep{Aung2021}. In reality, the distribution and thermodynamic properties of shock-heated gas\footnote{The terms of the \ac{icm} and \ac{whim} are usually used for the hot gas in galaxy clusters and cosmic filaments, respectively. Both gases are heated by cosmic accretion shocks as they accrete onto the cosmic web. At the radii where halos are connected with cosmic filaments, there is no clear boundary to distinguish the \ac{icm} and \ac{whim}. Therefore, in this work, we use the term ``shock-heated gas'' to refer to the hot gas in both halos and cosmic filaments.} are more complex than suggested by the self-similar spherical collapse scenario, for example, because they are subject to nonthermal pressure support, infalling gas from cosmic filaments, merging, and kinetic feedback from the halo center. Fig.~\ref{fig:map_example} demonstrates the X-ray emission from the hot shock-heated gas in and around a massive halo from a numerical simulation. It shows that the shock-heated gas fills the vast space in the outskirts and beyond. At radii beyond $r_\mathrm{200m}$, the hot gas distribution is complicated by the presence of small halos falling onto the central halo. At even larger radii, the accretion shock confines the hot gas, and cosmic filaments connect the halo to the large-scale structure. 

X-rays and the \ac{sz} effect are the two key observational techniques for exploring the properties of hot gas in the outskirts \citep[see][for reviews]{Reiprich2013, Walker2019}. Due to the rapidly declining X-ray surface brightness and \ac{sz} Compton-$y$ signal at large radii, most hot gas studies of the outskirts are limited to radii within $r_\mathrm{200c}$ \citep[e.g.,][]{Simionescu2011,walker2013,Eckert2013, Planck2013,Bulbul2016, Ghirardini2019,Mirakhor2020,McCall2024}. Beyond that radius, the gas clumping \citep{Nagai2011,Zhuravleva2013,Eckert2015,Angelinelli2021,Zhu2023}, the connection between clusters and cosmic filaments \citep{Rost2021,Gouin2022,Malavasi2020,Malavasi2023}, and the location of the accretion shock \citep{Lau2015,Baxter2021} are poorly constrained by individual pointing observations.

Recently, several stacking analyses of X-ray and \ac{sz} survey data demonstrate high \ac{snr} in the stacked profiles beyond $r_\mathrm{200c}$. In particular, \citet{Anbajagane2022,Anbajagane2024} stacked Atacama Cosmology Telescope and South Pole Telescope \ac{sz} survey data and discovered a $6\sigma$ pressure deficit with respect to the best-fit model at $\sim r_\mathrm{200m}$. \citet[][hereafter \citetalias{Lyskova2023}]{Lyskova2023} stacked 38 \emph{Planck} \ac{sz} selected clusters \citep{Planck2016-psz2,CHEX-MATE2021} using the eastern Galactic hemisphere \ac{erass} X-ray data, and obtained a gas density profile out to $3\times r_\mathrm{500c}$. 

For this work, we stacked the western Galactic hemisphere \ac{erass} data for more than 500 low-redshift clusters from a well-defined X-ray-selected cluster catalog detected in the first All-Sky Survey \citep{Bulbul2024,Kluge2024}. The larger sample allows us to investigate the circumcluster hot gas properties out to a larger radius. This article is organized as follows: in Sect.~\ref{sect:obs} we present the sample and the stacking analysis; we explain how we modeled the stacked profile in Sect.~\ref{sect:fitting}; in Sect.~\ref{sect:simulation} we explain how we use the numerical simulations to validate the stacking and modeling results; the discussion and conclusion are presented in Sects. \ref{sect:discussion} and \ref{sect:conclusion}, respectively. We adopted a flat $\Lambda$-cold-dark-matter cosmology with parameters $H_0=70$~km~s$^{-1}$~Mpc$^{-1}$, $\Omega_\mathrm{m}=0.3$, and $\Omega_\mathrm{\Lambda}=0.7$. The cosmic baryon fraction was adopted from \citet{Planck2020}, where $\Omega_\mathrm{b}/\Omega_\mathrm{m}=0.158$.

\section{Observation sample and data reduction}\label{sect:obs}

\subsection{Sample selection}

\begin{figure}
    \centering
    \includegraphics[width=0.45\textwidth]{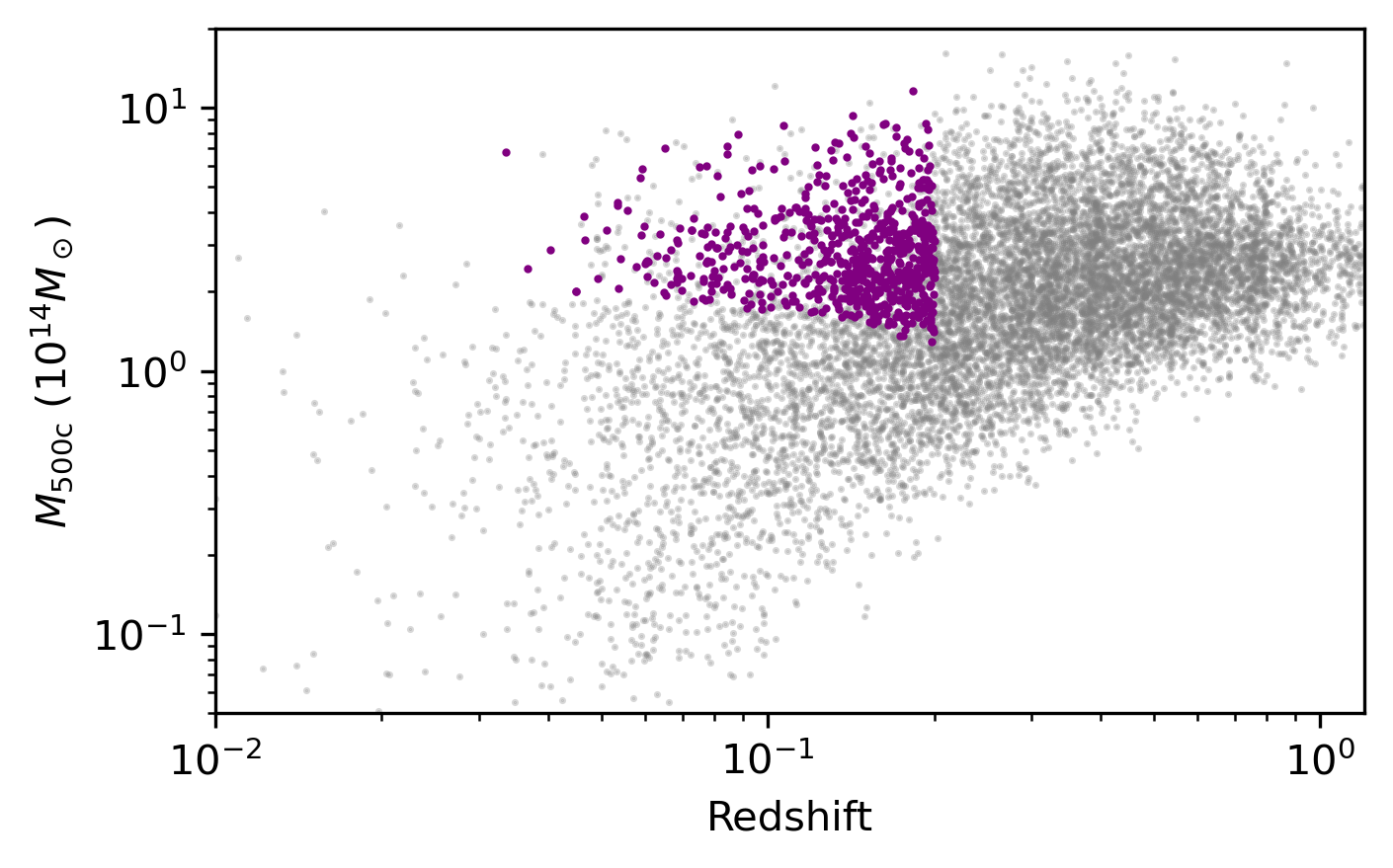}\\
    \includegraphics[width=0.45\textwidth]{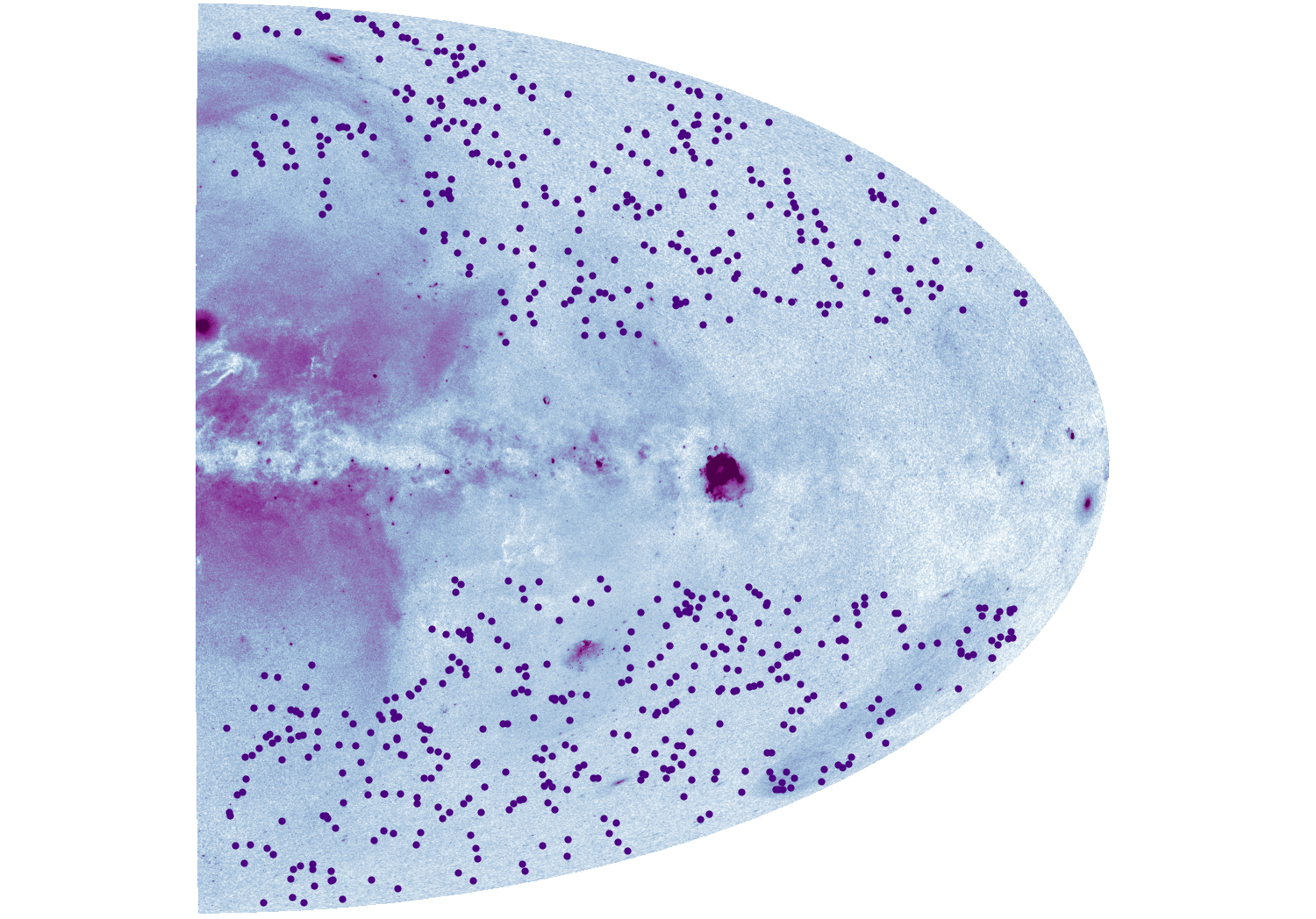}
    \caption{\emph{Top}: Mass-redshift distribution of the full eRASS1 galaxy cluster and group sample (gray) from \citet{Bulbul2024} and the clusters used in this work (purple).
    \emph{Bottom}: Locations of the selected sample on the west Galactic hemisphere eRASS1 X-ray sky.}
    \label{fig:sample}
\end{figure}

We selected our analysis sample from the first half-year survey of \ac{erass} (hereafter eRASS1) primary galaxy clusters and groups catalog \citep{Bulbul2024}, which is based on extended sources in the eRASS1 primary catalog \citep{Merloni2024} with further optical confirmation \citep{Kluge2024}. The overdensity masses $M_\mathrm{500c}$ of the clusters were estimated using the 0.2--2.3~keV count rate to the weak lensing calibrated mass scaling relation from the eRASS1 cluster abundance cosmology analysis \citep{Ghirardini2024,Grandis2024,Kleinebreil2025,Okabe2025}. We selected a luminosity-limited sample in the low redshift Universe based on the following criteria: 
\begin{enumerate}
    \item Luminosity $L_\mathrm{0.5-2keV}>2\times10^{43}$~erg~s$^{-1}$, which corresponds to a mass threshold of $M_\mathrm{500c}\approx2\times10^{14}M_\sun$ based on the scaling relations from the eRASS1 cosmology;
    \item Redshift $0.03<z<0.2$;
    \item Optical richness $\lambda>20$ to eliminate left over contamination in the sample;
    \item Galactic latitude $|b|>20\degr$ to avoid high galactic absorption;
    \item The median value of the 0.6--1~keV band count rate \citep{Zheng2024} in a 0.5$\degr$--3$\degr$ annulus $<6$~cts~s$^{-1}$~deg$^{-2}$ to avoid high Galactic foreground emission;
    \item At least a $3.5\degr$ angular distance to the eROSITA-DE footprint boundary for proper stray light estimation (see Sect.~\ref{sect:stray_light}).
\end{enumerate}
Using the criteria above, we selected 694 galaxy clusters. We visually checked their locations on the sky map. Two objects are affected by the Virgo Cluster emission and were therefore removed. Twelve objects were removed because they are the less massive clusters in cluster pairs. After this process, a sample of 680 clusters remained in a redshift range of 0.034 to 0.2 and a $M_\mathrm{500c}$ range of $1.3-11.6\times10^{14}M_\sun$. The median values of the sample redshift and $M_\mathrm{500c}$ are 0.15 and $2.6\times10^{14}M_\sun$, respectively. The mass-redshift distribution, as well as the sky position distribution of the sample, are plotted in Fig.~\ref{fig:sample}. It shows that the 0.6--1~keV count rate threshold we applied successfully selects objects in low-foreground emission regions. 

\subsection{Data reduction and surface brightness profile stacking}

We analyzed the first four scans of the eROSITA All-Sky Survey (hereafter eRASS:4), which were collected from 12th December 2019 to 19th December 2021. The data were processed with the eSASS \citep{Brunner2022} pipeline version 020, which is similar to the version 010 used for eRASS1 data release with improvements on boresight correction, detector noise suppression, and subpixel position computation \citep[see][for the details]{Merloni2024}. We only selected the events from \acp{tm} 1,2,3,4,6 (hereafter TM8) to avoid the systematic uncertainties caused by the optical light leak in \acp{tm} 5 and 7 \citep{Predehl2021}. We used the tools in the eSASS software package, version \texttt{eSASSusers\_211214\_0\_4}, to generate \ac{erass} data products.

We adopted the full soft band 0.2--2.3 keV to maximize the signal-to-noise ratio. For each cluster, we created the count image using \texttt{evtool} and generated the corresponding vignetting corrected exposure map using \texttt{expmap}. The count map and exposure map are centered at the cluster X-ray centroid and are extended to an angular distance of $10\times r_\mathrm{500c}+3.5\degr$, where the $10\times r_\mathrm{500c}$ aperture is used for analysis, and the additional $3.5\degr$ annulus is for calculating the stray light from the sources outside the analysis region. 

\subsubsection{Source masks}\label{sect:mask_source}
We masked out different source types for each cluster field. These sources include 

\begin{enumerate}
    \item X-ray point sources in eRASS:4 catalog, where the source detection method and analysis are described in \citet{Merloni2024}. 
    We note that in a cluster field, especially in the central bright region, the source-detection configuration used for the master catalog yields spurious point source entries \citep{Merloni2024}. We removed these spurious sources by running an additional wavelet detection process in the $r<r_{500}$ region. We first ran the software \textsc{wvdecomp}\footnote{\url{https://github.com/avikhlinin/wvdecomp}} with both detection and filtering thresholds $(3, 3, 3, 4, 4)$ at scales $(8, 16, 32, 64, 128)\arcsec$. Then we ran the software \textsc{sextractor} on the wavelet-filtered image to identify sources, with a detection threshold setting of 10. We cross-matched the wavelet-detected sources with those in the eRASS:4 master catalog. In the $r<0.8\times r_{500c}$ region, we only masked out sources identified by both methods. We visually inspected the resulting mask maps and verified the robustness of spurious source cleaning using the parameters mentioned above.
    
    \item Galaxy clusters and groups. We followed \citet{Zhang2024} to mask out 1) X-ray selected galaxy clusters and groups in the eRASS1 cluster catalog \citep{Bulbul2024} with masking radii $1.5r_\mathrm{500c}$; and 2) optically selected richness $\lambda>20$ clusters\footnote{It is a proprietary catalog of redMaPPer \citep{Rykoff2014} run on the DESI Legacy Imaging Surveys \citep{Dey2019} DR10 data. See \citet{Kluge2024} for the description.} with masking radii $1.5r_\lambda$ (see eq. 4 in \citealt{Rykoff2014} for definition).
\end{enumerate}

\subsubsection{Stray light}\label{sect:stray_light}

The cluster emission in the regions we are interested in is below the sky background. Therefore, the stray light from eROSITA, the $3.5\degr$-radius halo around any source produced by single-reflected photons, could affect the stacked profile and needs to be removed.
We followed the recipe described in appendix A of \citet{Churazov2023} to correct for stray light contamination. In short, for each object, we first estimated the sky background level as the average count rate in the source-free region. Then we convolved the sky background-subtracted and vignetting-corrected count rate image with a kernel of the stray light profile \citep[eq. A.2 in][]{Churazov2023} to obtain a first-order approximation of the stray light count rate. The normalization of the stray light kernel was computed as the fraction of the 2D stray light profile volume with respect to the 2D volume of the total 
\ac{psf} \citep[eq. A.1 + eq. A.2 in][]{Churazov2023}. 

\subsubsection{Profile stacking}

The image products we created have pixel widths of $8\arcsec$, resulting in a total number of $6.5\times10^8$ pixels for stacking. Meanwhile, this pixel size is smaller than the $30\arcsec$ \ac{psf} half-energy width of the \ac{erass}. To boost the calculation speed and save the memory space, we binned the original pixels to a $N_\mathrm{side}=4096$ \ac{healpix} scheme following the method described in \citet{Zhang2024}. 

For each object, we extracted a count profile and an exposure profile from the count and exposure maps, respectively. The radii of the profiles are scaled to $r_\mathrm{200m}$ of the object, which was converted from $r_\mathrm{500c}$ by assuming an \ac{nfw} profile with the halo concentration parameter $c_\mathrm{200c}=4$. This value approximately represents the concentration of halos in the cluster mass range of the sample \citep[e.g.,][]{Child2018,Diemer2019,Ishiyama2021,Okabe2025}. The conversion factor has a redshift dependence, ranging from 2.48 at $z=0.034$ to 2.24 at $z=0.2$ within our sample. 

We stacked the surface brightness profiles using a weight inversely proportional to the projected sky solid angle. The weight of each object $w=D_\mathrm{A}^2r_{500c}^{-2}$ corrects for the potential bias from nearby objects and massive objects that are of a large angular size, where $D_\mathrm{A}$ is the angular diameter distance as a function of redshift. The stacked surface brightness profile is
\begin{equation}\label{eq:weight_average}
    S_\mathrm{X}(r) = \frac{\sum_i w_iN_i(r)}{\sum_i w_it_i(r)\Omega_i(r)},
\end{equation}
where $N_i(r)$ the $i$th count profile, $t_i(r)\Omega_i(r)$ the $i$th exposure profile in unit of s$^{-1}$~deg$^{-2}$. We used the \ac{healpix} oriented bootstrap resampling method described in \citet{Zhang2024} to estimate the uncertainty of the stacked profile. We generated 500 bootstrapping samples of the \ac{healpix} pixels from the full pixel list, and for each bootstrapping sample, we calculated a stacked profile using Eq.~\ref{eq:weight_average}. Throughout this paper, we use the mean and covariance matrix of the 500 bootstrapping sample profiles to represent the stacked profile and its uncertainty. 

\begin{figure}
    \centering
    \includegraphics[width=.95\linewidth]{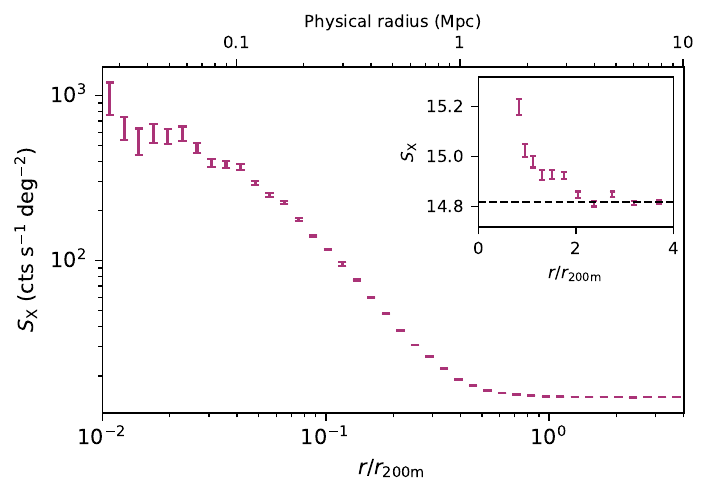}
    \caption{Stacked eROSITA surface brightness profile in the 0.2--2.3~keV band after the stray light component was removed. The radial distance to the cluster center is scaled to the overdensity radius $r_\mathrm{200m}$. The corresponding physical radius given the sample median mass and redshift is labeled at the top of the figure. The top-right inset provides a zoomed-in view of the profile within a zoomed surface brightness range around the background level, with the dashed horizontal line indicating the average surface brightness between 3 and 4~$r_\mathrm{200m}$. The profile shows significant X-ray emission extended to approximately $2\times r_\mathrm{200m}$. }
    \label{fig:profile_main}
\end{figure}

\section{Stacked eROSITA surface brightness profile and modeling}\label{sect:fitting}

The stacked profile of the 680 galaxy clusters from 0.01 to 4 $r_\mathrm{200m}$ is shown in Fig.~\ref{fig:profile_main}. As expected, the background level is consistent with the values in the 0.2-2.3~keV reported by \citet{Zheng2024}, which is $\sim15$~cts~s$^{-1}$~deg$^{-2}$ in the Galactic anti-center region with the instrumental background included. The high \ac{snr} of the stacked profile allows us to probe the extremely weak signal at the background-dominant radii. The inset in Fig.~\ref{fig:profile_main} shows that there is still a significant positive signal above the local background level at  $r>r_\mathrm{200m}$. Based on bootstrap sampling, the averaged surface brightness in the 1-2~$r_\mathrm{200m}$ range is higher than in the 2-4~$r_\mathrm{200m}$ range with a significance of $12\sigma$. We conclude that this constitutes a significant detection of the signal up to $2\times r_\mathrm{200m}$, which is approximately 4.5~Mpc given the median mass and redshift ranges of our sample. At $\sim2\times r_\mathrm{200m}$, there is a plausible bump. We investigated it by separating each cluster into four $90\degr$ sectors and stacking each of the sectors (see Appendix \ref{app:four_sector}). We find that this feature is contributed by fluctuations in the first sector and is not universal across all sectors. Meanwhile, in Appendix \ref{app:null_test}, we performed a test of stacking signals with cluster positions randomly distributed on the sky. This experiment shows that the uncertainty from uncorrelated components is less than 0.3\% of the background level. The test indicates that the detected signal does not arise from a background fluctuation in this background-dominated regime, but instead originates from X-ray emission in the circumcluster accretion region. In the following subsections, we model the stacked surface brightness profile in the full radial range out to $4r_\mathrm{200m}$.

\subsection{Model formalism}\label{sect:model}

Numerical simulations suggest that both the dark matter and gas density profiles from halo center to a few $r_\mathrm{200m}$ can be separated as inner and outer parts \citep[e.g.,][]{Diemand2008, ONeil2021, Diemer2022, Garcia2023}. The inner profile is contributed by the matter orbiting within the halo, and the outer profile is contributed by infalling matter in the ambient accretion region, for example, nearby cosmic filaments and halos therein. The transition between the two regions is characterized by a sharp change in slope. This feature has been observed in \ac{sz} stacking \citep[e.g.][]{Anbajagane2022,Anbajagane2024} and weak lensing shear stacking \citep[e.g.][]{Chang2018}. 

Numerous models for the inner and outer profiles have been proposed, some of which we later test in Section~\ref{sect:sim_2d}. To fit our observed data, we adopted a relatively simple prescription based on the halo model, which is commonly used to describe large-scale clustering \citep[see][for a review]{Cooray2002}. Specifically, for the one-halo term, we adopted a \ac{gnfw} profile,
\begin{equation}
    n_\mathrm{H}^\mathrm{gNFW}(r)=n_\mathrm{H,0}\times\frac{1}{\left(r/r_\mathrm{s}\right)^{\gamma}\left[1 + \left(r/r_\mathrm{s}\right)^\alpha\right]^{(\beta-\gamma)/\alpha}},
\end{equation}
where $\alpha$, $\beta$, $\gamma$, and $r_\mathrm{s}$ are the four shape parameters. The \ac{gnfw} model has been widely used to describe the pressure profile of cluster gas \citep[e.g.,][]{Nagai2007, Arnaud2010, Bulbul2010}. Because the X-ray-emitting gas in the off-filament directions terminates at the accretion shock, we also included this feature in the model. We assumed that the hot gas density in the downstream region after shock compression is the mean baryon density $\left<\rho_\mathrm{b}\right>$. The modified hydrogen density profile is then
\begin{equation}\label{eq:prof_boundary}
    n_\mathrm{H}(r) = \begin{cases}
        n_\mathrm{H}^\mathrm{gNFW}(r) & n_\mathrm{H}(r)\ge\left< n_\mathrm{H}\right>\\
        0 & \mathrm{elsewhere},
    \end{cases}
\end{equation}
where $\left<n_\mathrm{H}\right>$ is the mean hydrogen number density converted from $\left<\rho_\mathrm{b}\right>$. For our observed sample, $\left<n_\mathrm{H}\right>=3\times10^{-7}$~cm$^{-3}$ at the sample's median redshift of $0.15$. The gas density can be converted to an X-ray emission
\begin{equation}\label{eq:n_to_epsilon}
    \epsilon_\mathrm{1h}(r) = n_\mathrm{e}n_\mathrm{H}\Lambda_\mathrm{cf}(T,Z),
\end{equation}
where we adopt $n_\mathrm{e}=1.2n_\mathrm{H}$ for the hot gas and $\Lambda_\mathrm{cf}(T,Z)$ is the temperature and metallicity-dependent cooling function. We projected the 3D emission onto a 2D plane. The projected surface X-ray photon rate is
\begin{equation}\label{eq:projection}
    \Sigma_\mathrm{X}(r)=2\times\int_0^{l_\mathrm{max}}\epsilon_\mathrm{X}\left(\sqrt{l^2+r^2}\right) \mathrm{d}l,
\end{equation}
where $l$ is the projection \ac{los}, $l_\mathrm{max}$ is the one side projection depth. With a flat sky approximation, the observed surface brightness
\begin{align}\label{eq:sb_conversion}
    S_\mathrm{X}\left[\mathrm{cts\ s^{-1}\ deg^{-2}}\right]
    =&\Sigma_\mathrm{X}\left[\mathrm{cts\ s^{-1}\ kpc^{-2}}\right]\times
    A_\mathrm{eff}\left[\mathrm{cm}^2\right]\\ \nonumber
    &\times\frac{2.55\times10^{-48}}{\left(1+z\right)^3},
\end{align}
where $A_\mathrm{eff}$ is the averaged effective area of the observation band. We calculated $\Lambda_\mathrm{cf}A_\mathrm{eff}$ in the 0.2--2.3~keV band 1) using the APEC model with ATOMDB v3.09 \citep{Smith2001, Foster2020}, 2) with an assumption of $Z=0.3Z_\sun$ and \citet{Lodders09} abundance table, 3) using the effective area curve of the five used eROSITA telescope modules, 4) applying a $K$-correction at $z=0.15$, and 5) taking an averaged foreground the average HI column density of $3\times10^{20}$~cm$^2$ of the sample into account. The value of $\Lambda_\mathrm{cf}A_\mathrm{eff}$ varies between 3 and 6~ph~s$^{-1}$~cm$^{5}$ in the temperature range $T>3\times10^6$~K. In this work, we adopted $\Lambda_\mathrm{cf}A_\mathrm{eff}=4.9\times10^{-12}$~ph~s$^{-1}$~cm$^{5}$, which is the averaged value in the $3\times10^6-5\times10^{7}$~K temperature range. In Appendix \ref{app:cooling}, we present a more detailed investigation of the impact of different metallicity, foreground Galactic absorption, and radial temperature variation on the cooling function value. We conclude that the possible systematic uncertainty on the $\Lambda_\mathrm{cf}A_\mathrm{eff}$ is $\sim10\%$. 

For the two-halo term, we started with theoretical calculations of the X-ray emission from nearby halos that are spatially correlated with the central halo. The distribution of neighbor halos around a halo can be described
\begin{equation}\label{eq:neighbor_halo_density}
    \frac{\mathrm{d}n}{\mathrm{d}V}(r,M_\mathrm{c})=\xi_\mathrm{mm}^\mathrm{lin}(r)b(M_\mathrm{c})\int \frac{\mathrm{d}n}{\mathrm{d}M\mathrm{d}V}b(M)\mathrm{d}M,
\end{equation}
where $M_\mathrm{c}$ is the mass of the central halo, $\xi_\mathrm{mm}^\mathrm{lin}$ is the matter-matter correlation function in the linear regime, $\mathrm{d}n/\mathrm{d}M\mathrm{d}V$ is the \ac{hmf}, $b$ is the mass dependent halo bias parameter, which quantifies the excess clustering of halos over the clustering of dark matter. We adopted \ac{hmf} from \citet{Tinker2008} and halo bias from \citet{Tinker2010}. The profile of $\xi_\mathrm{mm}^\mathrm{lin}$ and the models of halo bias and \ac{hmf} are numerically implemented in the package \textsc{colossus}\footnote{\url{https://bdiemer.bitbucket.io/colossus}} \citep{Diemer2018-colossus}. With a halo luminosity-mass scaling relation, we converted the halo distribution to the emission distribution, 
\begin{equation}\label{eq:2h-3d}
    \epsilon_\mathrm{2h}(r,M_\mathrm{c})=\xi_\mathrm{mm}^\mathrm{lin}(r)b(M_\mathrm{c})\int_{M_\mathrm{min}}^{M_\mathrm{max}} \frac{\mathrm{d}n}{\mathrm{d}M\mathrm{d}V}b(M)L_\mathrm{X}(M)\mathrm{d}M,
\end{equation}
where $\epsilon_\mathrm{2h}\equiv\mathrm{d}L_\mathrm{X,2h}/\mathrm{d}V$, $L_\mathrm{X}(M)$ is the luminosity-mass scaling relation, $M_\mathrm{min}$ and $M_\mathrm{max}$ are the integration limits. We used a survey selection function to describe the halo masking. Therefore, we rewrote Eq.~\ref{eq:2h-3d} to
\begin{align}\label{eq:2h-3d_with_selection}
    \epsilon_\mathrm{2h}(r,M_\mathrm{c},z)=
    &\xi_\mathrm{mm}^\mathrm{lin}(r,z)b(M_\mathrm{c},z)\nonumber\\ \times\int_0^{M_\mathrm{max}}&\frac{\mathrm{d}n(M,z)}{\mathrm{d}M\mathrm{d}V}b(M,z)\left[1-P_\mathrm{det}(M,z)\right]L_\mathrm{X}(M,z) \mathrm{d}M,
\end{align}
where $P_\mathrm{det}(M,z)$ is the eRASS1 X-ray selection function \citep{Clerc2024}, $M_\mathrm{max}$ is the mass threshold of masking objects in observations, which is approximately $10^{14.3}M_\sun$ given our $\lambda=20$ richness threshold for source masking and the richness-mass scaling relation from \citet{Ghirardini2024}. For the $L_\mathrm{X}-M$ scaling relation, we adopted the one from \citet{Bulbul2019} and applied a factor of 1.4 to convert 0.5--2~keV luminosity to 0.2--2.3~keV luminosity. Similar to the one-halo component, we applied Eqs. \ref{eq:projection} and \ref{eq:sb_conversion} to $\epsilon_\mathrm{2h}$ to obtain the projected two-halo term model surface brightness, where we used an integration limit $l_\mathrm{max}=100$~Mpc. We created the individual object two-halo model, and then stacked them using Eq.~\ref{eq:weight_average} to obtain a synthesis model $S_\mathrm{X,halo}^\mathrm{2h}(r)$ for fitting. To account for the emission from unvirialized filament gas, an additional free normalization parameter $A_\mathrm{2h}$ was also included in the fitting. The final projected two-halo model can be expressed as 
\begin{equation}\label{eq:2h_final}
    S_\mathrm{X}^\mathrm{2h}(r)=A_\mathrm{2h}S_\mathrm{X,halo}^\mathrm{2h}(r).
\end{equation}

We used a constant profile to account for background components not associated with our stacking objects. The uncorrelated components include the instrumental background, Galactic foreground, and the cosmic X-ray background. 

\subsection{Surface brightness profile fitting}
We modeled the stacked eRASS:4 profiles using the Bayesian inference package \textsc{pocoMC} \citep{Karamanis2022_method,Karamanis2022_pocomc}. The likelihood is defined as $\mathcal{L}=-\chi^2/2$, where $\chi^2=(S_\mathrm{x}^\mathrm{data}-S_\mathrm{x}^\mathrm{model})^{\mathrm{T}}\mathcal{C}^{-1}(S_\mathrm{x}^\mathrm{data}-S_\mathrm{x}^\mathrm{model})$ is a generalized form that takes the profile covariance into account. We applied flat priors for all parameters. We adopted the median and the 16th to 84th percentiles of the posterior as the best-fit parameter and the uncertainty range. 

\begin{figure*}
    \centering
    \includegraphics[width=0.45\linewidth]{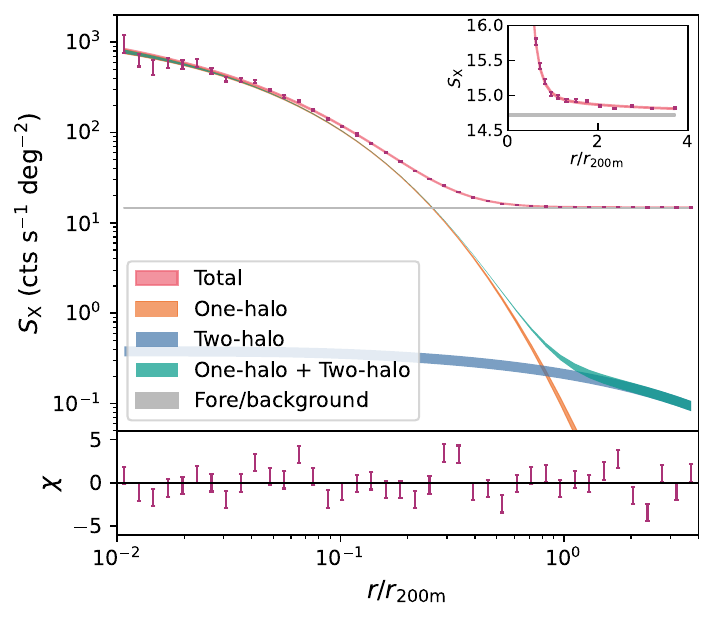} 
    \includegraphics[width=0.45\linewidth]{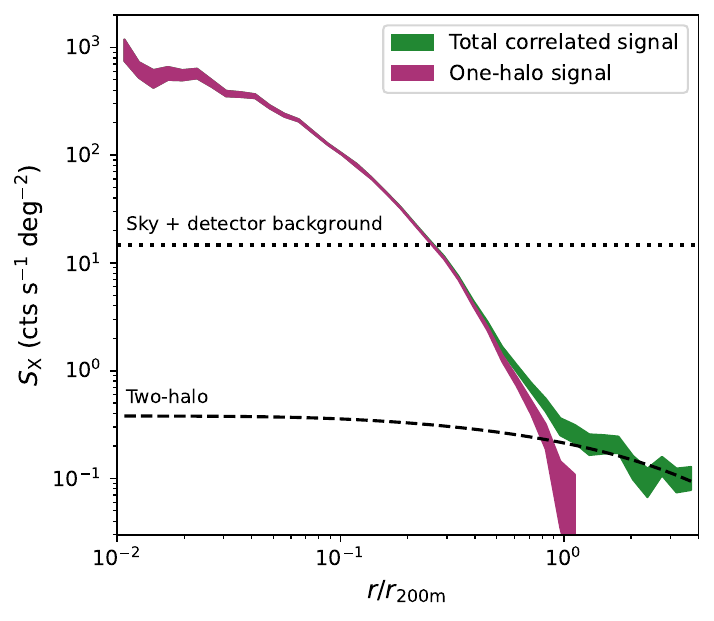} 
    \caption{\emph{Left}: Best-fit results of the stacked 0.2--2.3~keV eROSITA surface brightness profile, where we adopted the \ac{gnfw} model for the one-halo component. The width of each component denotes the $1\sigma$ scatter of the posterior samples. The inset shows a zoomed-in view around the background level. 
    \emph{Right}: Surface brightness profiles of the background-subtracted correlated signal (green) and the one-halo term signal. The systematic uncertainties of the uncorrelated background and the fitting uncertainties of the constant model and two-halo term normalization were propagated when calculating the profile uncertainty. 
    }
    \label{fig:obs_prof_fitting}
\end{figure*}

We fit the model combination of the \ac{gnfw}, two-halo, and a constant to the 0.2--2.3~keV stacked profile. The best-fit parameters are listed in Table \ref{tab:parameter}. In the left panel of Fig.~\ref{fig:obs_prof_fitting}, we plot each best-fit component of the stacked profile as well as the fit residual. The fitting illustrates that the $r_\mathrm{200m}$ is approximately the boundary between the one-halo dominant and two-halo dominant regions. To check the necessity of including the two-halo term in the fitting, we also fit the profile using \texttt{\ac{gnfw} + constant} model combination. The best-fit components are plotted in Fig.~\ref{fig:profile_fit_no2h}. Without the two-halo term, there are more residuals in the range of  $1-2\times r_\mathrm{200m}$. Based on the Bayesian model comparison framework, we calculated the Bayes factor of the two fits, which is $\exp(11.5)$. It suggests that the fitting by including the two-halo term is decisively favored according to the Jeffreys' scale \citep{Jeffreys1961} or a similar scale later suggested by \citet{Kass1995}. The normalization of the two-halo term is $0.22\pm0.03$~cts~s$^{-1}$~deg$^{-2}$ at $r_\mathrm{200m}$, corresponding to $\sim1\times10^{35}$ erg~s$^{-1}$~kpc$^{-2}$ at the sample median redshift, which is at the same order of the magnitude of that from the recent galaxy-X-ray cross-correlation study \citep{Comparat2025}. 

The fitting decomposes the total stacked signal into the uncorrelated signal, the two-halo term, and the one-halo term, allowing us to obtain the background-subtracted surface brightness profiles. In the right panel of Fig. \ref{fig:obs_prof_fitting}, we plot the net correlated surface brightness profile and the net one-halo term surface brightness profile. The constant background and two-halo term normalization uncertainties from the fitting, as well as the systematic uncertainties of the uncorrelated background estimated by random position stacking in Appendix \ref{app:null_test}, were propagated to compute the net profile uncertainties. The net surface brightness profile of the one-halo term decreases dramatically and enters the noise-dominant regime at $r_\mathrm{200m}$. 

\begin{table}[]
\caption{Best-fit parameters of the \ac{gnfw} gas density profile and the two-halo term normalization.}\label{tab:parameter}
    \centering
    \begin{tabular}{lcc}
    \hline\hline
    Parameter & Unit & Value \\
    \hline
    $\log n_\mathrm{H,0}$& cm$^{-3}$ & $-2.92\pm0.18$\\
    $r_\mathrm{s}$ & $r_\mathrm{200m}$ & $1.18\pm0.30$\\
    $\alpha$ & - & $0.739\pm0.054$\\
    $\beta$ & - & $5.69\pm0.61$\\
    $\gamma$ & - & $0.389\pm0.085$\\
    $S_\mathrm{X}^\mathrm{2h}(r_\mathrm{200m})$ & cts s$^{-1}$ deg$^{-2}$& $0.22\pm0.03$\\
    \hline
    \end{tabular}

\end{table}

\subsection{Density profile and comparison with the literature}

\begin{figure*}
    \centering
    \includegraphics[width=0.45\textwidth]{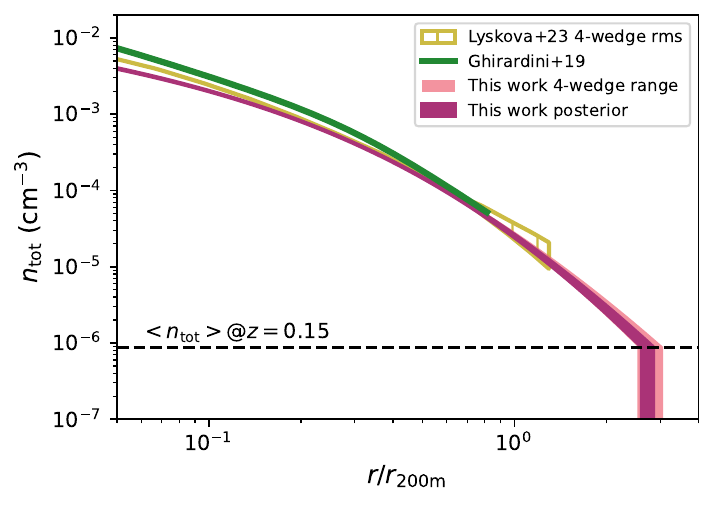}
    \includegraphics[width=0.45\textwidth]{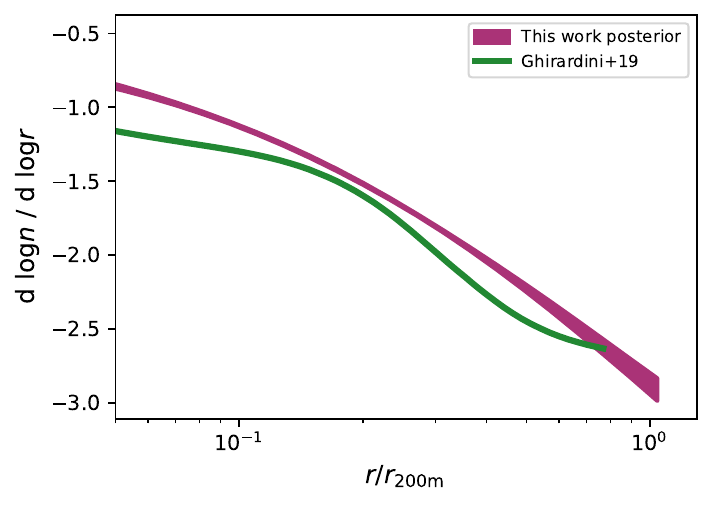}
    \caption{\emph{Left}: Best-fit \ac{gnfw} gas number density profile $1\sigma$ posterior range (purple) and the range of individual fittings of the four wedges (red) inferred by the eROSITA observations. The result from \citetalias{Lyskova2023} (yellow) up to $3\times r_\mathrm{500c}$, and the result from \citet{Ghirardini2019} up to $2\times r_\mathrm{500c}$ (green). 
    \emph{Right}: The logarithmic radial density gradient from this work in purple and from \citet{Ghirardini2019} in green. } 
    \label{fig:ntot_vs_lit}
\end{figure*}

We present the best-fit one-halo term gas number density profile in the left panel of Fig.~\ref{fig:ntot_vs_lit}. We note that the profile we show is the total gas number density $n_\mathrm{tot}$, where $n_\mathrm{tot}=2.3\times n_\mathrm{H}$ for typical halo gas. We present the profile uncertainty in two ways: the purple band is the $1\sigma$ scatter of the posterior sample from model inference; the red band is the range of the fitting results from four individual sectors. Due to the large pixel size we used for stacking, which is $0.86\arcmin$, or 140 kpc physical scale at sample median redshift 0.15, we only show the density profile beyond $0.05\times r_\mathrm{200m}$. Our best-fit density profile illustrates that the gas density at a large radius of $r_\mathrm{200m}$ is still about $2.5\times10^{-5}$~cm$^{-3}$, corresponding to a baryon density contrast of 30. The cutoff of the profile at the cosmic averaged baryon density is introduced by the condition we applied in Eq.~\ref{eq:prof_boundary}. We discuss it later in Sect.~\ref{sect:boundary}. In the right panel of Fig.~\ref{fig:ntot_vs_lit}, we present the best-fit density slope profile. The slope gradually steepens from -1 in central regions to -3 at $r_\mathrm{200m}$. 

Our best-fit density profile is overall consistent with the density profiles derived from the \textit{XMM-Newton} and eROSITA observations of the \textit{Planck}-selected samples out to two and three $r_\mathrm{500c}$ (\citealt{Ghirardini2019}; \citetalias{Lyskova2023}), as shown in the left panel of Fig.~\ref{fig:ntot_vs_lit}. The normalization and slopes derived from the eROSITA observations of \textit{Planck}-selected clusters in \citetalias{Lyskova2023} are consistent with our fits in the radial range of $0.4-1\times r_\mathrm{200m}$. Nonetheless, within the central region $r<0.4\times r_\mathrm{200m}$, we find a slightly shallower slope and lower normalization. By contrast, the density profiles from the \emph{XMM-Newton} observations in \citet{Ghirardini2019} show the highest normalization and the steepest slope compared to both our results and those of \citetalias{Lyskova2023}. A likely explanation for the offset observed in the central regions is the different mass ranges of the utilized samples. \citetalias{Lyskova2023} stacked eROSITA observations of the \textit{Planck} \ac{sz} sample \citep{Planck2016-psz2} with a mass threshold of $M_\mathrm{500c}> 2\times10^{14}$~$M_\sun$ and a mean mass of $4.1\times10^{14}$~$M_\sun$, while \citet{Ghirardini2019} used the X-COP sample \citep{Eckert2017}, which consists of 12 \textit{Planck}-selected clusters with masses significantly higher than both our sample and that of \citetalias{Lyskova2023}. Thus, both comparison samples probe higher-mass clusters than those studied in this work, which could account for the observed offset. 

\section{Stacked profiles in IllustrisTNG simulation}\label{sect:simulation}

In this section, we explore the galaxy cluster halos in the IllustrisTNG simulation to 1) validate our stacking and modeling processes; 2) compare the observations to the simulations to understand underlying physical processes in this unexplored region with previous observations; 3) study the 3D structure of the shock-heated gas from the galaxy cluster center to the accretion region. The IllustrisTNG is a suite of cosmological magnetohydrodynamical simulations \citep{Naiman2018, Springel2018, Nelson2018, Marinacci2018, Pillepich2018} of galaxy formation in a fixed cosmology \citep{Planck2016}. Halo and galaxy catalogs are based on the \ac{fof} algorithm and subhalo/galaxy identification with Subfind \citep{Springel2001}. 

For this work, we used the $z=0$ snapshot of the highest-resolution version on the largest box, \mbox{TNG300-1}, to study the X-ray properties in cluster outskirts. Specifically, we selected 159 halos with $M_\mathrm{500c} > 10^{14}$~$M_\sun$ from the \ac{fof} catalog, whose mass range is comparable to the observation sample. We used the code \textsc{hydrotools} \citep{Diemer2017-SFH, Diemer2018-HIHII, Tacchella2019} to extract all gas cells out to $10$~Mpc and calculate thermodynamic properties:
(1) the total gas number density $n_\mathrm{tot}=n_\mathrm{H}+n_\mathrm{e}+n_\mathrm{He}$, the sum of hydrogen, electron, and helium number density. The $n_\mathrm{H}$ and $n_\mathrm{He}$ were converted from the mass density by assuming mass fractions of 0.76 and 0.24, respectively. The $n_\mathrm{e}$ were calculated using the electron abundance of the cell. We ignored the contribution from metal elements in the total number density. 
(2) the gas temperature $T$, which was converted from the internal energy of the gas cell. 
(3) the electron entropy $K_\mathrm{e}=k_\mathrm{B}T_\mathrm{e}n_\mathrm{e}^{-2/3}$, where $k_\mathrm{B}$ is the Boltzmann constant, $T_\mathrm{e}=T$ by assuming the thermal equilibrium. 
(4) the gas pressure $P_\mathrm{gas}=n_\mathrm{tot}k_\mathrm{B}T$.

In addition, we calculated the 0.2--2.3~keV X-ray emission properties under the assumption of collisional ionization equilibrium and using the ATOMDB~v3.09 APEC code \citep{Smith2001, Foster2020}. For each gas cell, 
\begin{align}
    L_\mathrm{X}=&V_\mathrm{cell}n_\mathrm{H}n_\mathrm{e}\Lambda_{Z=0}(T)\nonumber \\
    &+Z_\mathrm{cell}V_\mathrm{cell}n_\mathrm{H}n_\mathrm{e}\left[\Lambda_{Z=Z_\sun}(T) - \Lambda_{Z=0}(T)\right],
\end{align}
where $V_\mathrm{cell}$ is the cell volume, $Z_\mathrm{cell}$ is the cell metallicity in solar unit, $\Lambda_{Z=0}$ and $\Lambda_{Z=Z_\sun}$ are the cooling function with zero metallicity and solar metallicity \citep{Lodders09}, respectively.

\subsection{Projected emission profile and fitting method validation}\label{sect:sim_2d}

In this subsection, we validated the profile fitting framework using numerical simulation data. We aim to investigate whether the fitting results from the profile modeling framework we applied in Sect.~\ref{sect:fitting} could match the true 3D density profiles up to the accretion shock. 

In each $20\times20\times20$~Mpc$^3$ box centered at each halo, we projected the 0.2--2.3~keV X-ray emission on the XY, YZ, and XZ planes, respectively. Then we created nearby (sub)halo masks for each object and each projection angle, using $M_\mathrm{500c}$ thresholds of $10^{13}$, $10^{13.5}$, and $10^{14}$~$M_\sun$. We note that some of the bright X-ray halos in binary or multi-object systems are not identified as the main halo. We therefore excluded Subfind halos instead of \ac{fof} halos with a masking radius of $1.5r_\mathrm{500c}$. Though the overdensity mass is not in the Subfind catalog, it can be converted from $M_\mathrm{subfind}$. For most central halos, the ratio between $M_\mathrm{500c}$ and $M_\mathrm{subfind}$ is about $2/3$. For other subhalos, though we do not exactly know this ratio, we continued using this mask routine, which still successfully removes particles in the central regions of galaxies. For each projection direction and halo mask threshold, we stacked the radially averaged projected emission profile
\begin{equation}\label{eq:weight_average_sim}
    \Sigma_\mathrm{x}(r)=\frac{\sum_i w_i \Sigma_{\mathrm{x},i}(r)}{\sum_i w_i},
\end{equation}
where $r$ is the radius in a unit of $r_\mathrm{200m}$, $w_i$ is the weight, and here we adopted $r_\mathrm{500c}^{-2}$ to reduce the bias from the large physical size objects. We calculated the median profile from the three projection angles as the final stacked profile.

\begin{figure*}
    \centering
    \includegraphics[width=0.99\linewidth]{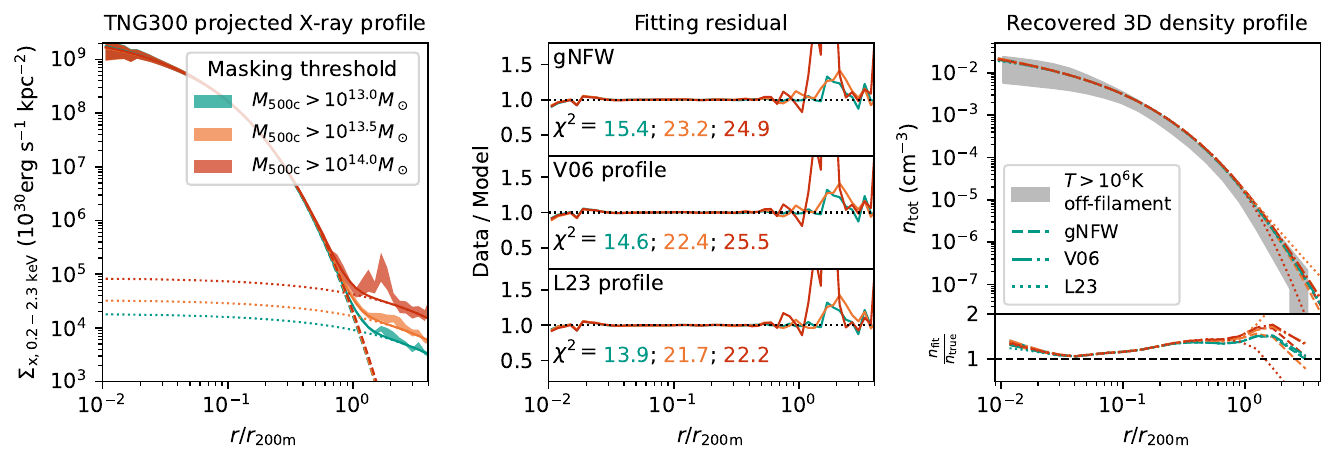}
    \caption{\emph{Left}: Stacked surface X-ray emission profiles of the \mbox{TNG300-1} galaxy cluster halos. The teal, orange, and red colors denote the nearby-halos masking thresholds of $10^{13}$, $10^{13.5}$, and $10^{14}M_\sun$, respectively. The shaded regions represent the bootstrapped uncertainties, reflecting the sample scatter. The best-fit gNFW profiles, the two-halo terms, and the total profiles are plotted as dashed, dotted, and solid lines, respectively. The sharp peaks in the outer profile are contributed to by bright nearby halos that are below the masking threshold. 
    \emph{Middle}: The residuals of the fittings of the three different one-halo models. The $\chi^2$ values for each fit are labeled. 
    \emph{Right}: The best-fit total number density profiles and their comparison to the $T>10^6$~K gas density profile in the off-filament directions in 3D (see Sect.~\ref{sect:profiles_3d}). Different colors denote the different nearby halo masking thresholds of the stacked profiles, as in the \emph{left} and \emph{middle} panels. The dashed, dash-dotted, and dotted lines denote the results of the gNFW, \citetalias{Vikhlinin2006}, and \citetalias{Lyskova2023} models, respectively.}
    \label{fig:2d_em}
\end{figure*}

The stacked 0.2--2.3~keV $\Sigma_\mathrm{x}$ profile with the three nearby-halo masking thresholds is plotted in the left panel of Fig.~\ref{fig:2d_em} as shaded regions. The three stacked profiles are identical in the $r<0.7\times r_\mathrm{200m}$ radial range and turn flat at $r\gtrsim r_\mathrm{200m}$. The outer parts of the profiles are contributed by gaseous (sub)halos with masses below the masking threshold from the far outskirts to cosmic filaments, together with the unvirialized gas in filaments. Different halo masking thresholds result in different $\Sigma_\mathrm{x}$ normalizations of the outer profile. Meanwhile, the outer parts of the three stacked profiles show several enhancements, where the one in the $10^{14}M_\sun$ masking threshold profile is the most significant. These enhancements are from bright X-ray halos below the masking threshold, and as the threshold decreases, their significance to the overall emission becomes negligible. 

We followed Sect.~\ref{sect:fitting} to fit the one-halo and two-halo models to the projected 2D simulated $\Sigma_\mathrm{X}$ profiles. For the one-halo term, in addition to the gNFW model we used in Sect.~\ref{sect:fitting}, we included two more models for testing:
\begin{itemize}
    \item The \citet[][hereafter \citetalias{Vikhlinin2006}]{Vikhlinin2006} profile, which is 
    \begin{equation}
    n_\mathrm{H} = n_\mathrm{H,0}\times\frac{\left( r/r_\mathrm{c}\right)^{-\alpha}}{\left[1 + (r/r_\mathrm{c})^2\right]^{3\beta-\alpha/2}}\times\frac{1}{\left[1+(r/r_\mathrm{s})^{\gamma}\right]^{\epsilon/\gamma}}\ ,
    \end{equation}
    where $n_\mathrm{H,0}$ is the hydrogen number density normalization, $\alpha$, $\beta$, $\epsilon$, $\gamma$, $r_\mathrm{c}$, and $r_\mathrm{s}$ are the six parameters that control the profile shape.
    \item The \citetalias{Lyskova2023} profile, which is a modification of the \citetalias{Vikhlinin2006} profile with an additional slope change at the large radii, 
    \begin{align}\label{eq:lyskova23}
         n_\mathrm{H} = n_\mathrm{H,0}&\times\frac{\left( r/r_\mathrm{c}\right)^{-\alpha}}{\left[1 + (r/r_\mathrm{c})^2\right]^{3\beta-\alpha/2}}\times\frac{1}{\left[1+(r/r_\mathrm{s})^{\gamma}\right]^{\epsilon/\gamma}}\nonumber\\ &\times\frac{\left[1+(r/r_\mathrm{s2})^\gamma\right]^{\epsilon/\gamma}}{\left[1+(r/r_\mathrm{s2})^\xi\right]^{\lambda/\xi}} \ ,
    \end{align}
    where $\lambda$, $\xi$, $r_\mathrm{s2}$ are three additional shape parameters. 
\end{itemize}
We adopted a constant value $\Lambda_\mathrm{cf}=8\times10^{-24}$~erg~s$^{-1}$~cm$^3$ by assuming a constant temperature and abundance profile for the density to emissivity conversion. This is approximately the radiative cooling in the 0.2--2.3~keV band with $T=2\times10^7$~K and $Z=0.3Z_\sun$. As discussed in Sect.~\ref{sect:fitting}, a choice of a lower metallicity of $0.2Z_\sun$ could result in a slightly lower $\Lambda_\mathrm{cf}$. For the two-halo term formalism, there are several settings different from those we used in Sect.~\ref{sect:model}. First, we adopted the \mbox{TNG300-1} $L_\mathrm{X}-M$ scaling relation from \citet{Pop2022}. Second, because the simulation data we extracted for each cluster are in $20\times20\times20$~Mpc$^3$ boxes, we used $l_\mathrm{max}=10$~Mpc as the one side projection depth when calculating the model $\Sigma_\mathrm{X}$ using Eq.~\ref{eq:projection}. Third, because we masked nearby halos in simulations only based on the mass, we did not further apply additional selection functions such as the one in Eq.~\ref{eq:2h-3d_with_selection}. 

We fit the three stacked profiles with the different mass exclusion thresholds, with the combined one-halo and two-halo terms. For each stacked profile, the three one-halo models were used separately for validation. We used the package \textsc{iminuit} \citep{iminuit} for minimizing the $\chi^2$ values. The left panel of Fig.~\ref{fig:2d_em} shows the results of the fits with the gNFW model as the one-halo component. The middle panel of Fig.~\ref{fig:2d_em} presents the residuals and best-fit $\chi^2$ values of the fits using the gNFW, \citetalias{Vikhlinin2006}, and \citetalias{Lyskova2023} models on the three stacked profiles with different mass exclusion thresholds. The residuals show that all three one-halo models, together with the two-halo term, can well fit the projected emission profile from halo center to $4\times r_\mathrm{200m}$. The three one-halo profile models return similar residuals and $\chi^2$ values. The large residual in fittings with a nearby halo masking threshold of $10^{14}$~$M_\sun$ is due to several unmasked bright objects that are below the threshold. The right panel of Fig.~\ref{fig:2d_em} shows the best-fit one-halo density profiles of the gNFW, \citetalias{Vikhlinin2006}, and \citetalias{Lyskova2023} models in dashed, dashed-dotted, and dotted lines, respectively, where different colors denote the different nearby halo masking thresholds. We overplot the scatter of the 3D density profiles in the off-filament directions as the gray shaded region (see later Sect.~\ref{sect:profiles_3d} for details). All the best-fit 3D density profiles using the three models are close to the true profiles. Among the three models we tested, the \citetalias{Lyskova2023} model exhibits the most scatter, due to its additional flexibility in changing the slope at large radii, which allows it to overfit the features of the outer profile dominated by the two-halo term. Meanwhile, the fits using the \ac{gnfw} model are in good agreement with those using the \citetalias{Vikhlinin2006} model, despite having two fewer shape parameters. This comparison also highlights that the \ac{gnfw} model can well describe the gas density profile out to large radii. Complex models with more shape parameters do not significantly improve the fitting residual. 

This analysis shows that by adding a two-halo term to account for the nearby infalling (sub)halos and circumcluster gas in the cosmic filament directions, all three one-halo density models can well recover the true density profile in the off-filament directions out to the accretion shock in numerical simulations. In other words, the two-halo term accounts for the major gas-clumping effects, i.e., the presence of nearby (sub)halos and isotropic gas distribution due to cosmic filaments at large radii. The mild offset between the best-fit profile normalization and the median of the true 3D density profile can be explained by additional minor clumping effects, e.g., halo triaxiality, turbulence-induced density fluctuations, and the presence of inner shocks and cold fronts. Moreover, even if we assumed a constant cooling function and ignored variations in temperature and metallicity, the fits successfully recover the true number density profile, suggesting a minor impact from these two effects. We therefore conclude that our models, validated against the IllustrisTNG simulations, reliably recover the density profiles from the cluster cores out to the circumcluster region beyond $r_{200m}$.

\subsection{Comparison between simulations and observations}\label{sect:obs_vs_sim}

\begin{figure*}
    \centering
    \includegraphics[width=0.4\textwidth]{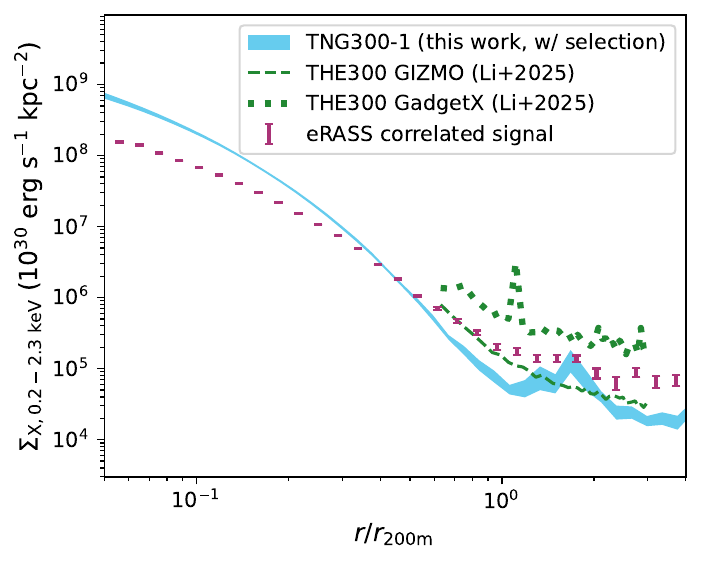}
    \includegraphics[width=0.4\textwidth]{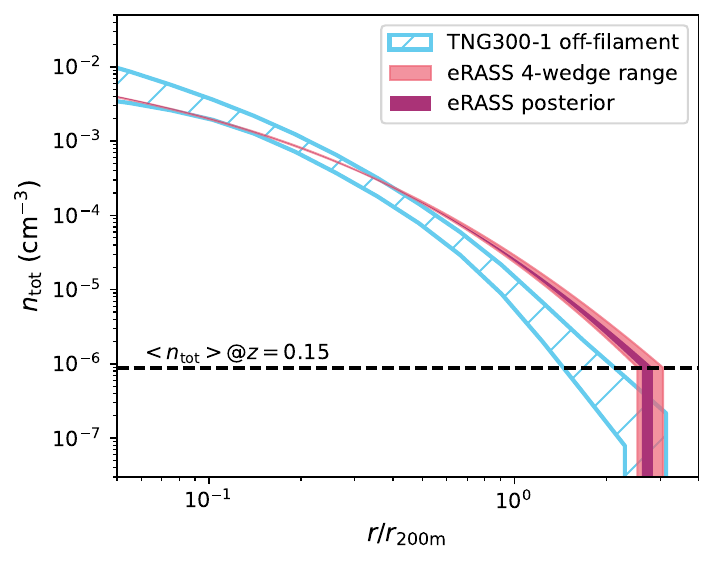}
    \caption{
    \emph{Left}: Comparison of the projected emission profiles from observations and simulations. The observed stacked eROSITA profile from this work is shown in purple error bars. The profiles extracted from the \mbox{TNG300-1} simulation are shown as a cyan band. The \mbox{TNG300-1} profile is more centrally peaked than the observed profile. For comparison, profiles of The Three Hundred Gizmo-Simba and Gadget-X simulations from \citet{Li2025} are shown as dashed and dotted green lines, respectively. 
    \emph{Right}: Best-fit \ac{gnfw} gas number density profile $1\sigma$ posterior range (purple) and the range of individual fittings of the four wedges (red). We overplot the \mbox{TNG300-1} gas number density profile in the off-filament directions (cyan-hatched region). }
    \label{fig:sim_vs_obs}
\end{figure*}

Different baryon-physics models, especially feedback models in hydrodynamic simulations, lead to large discrepancies in the gas distribution \citep[e.g.,][]{Moser2022,Schaller2025}. The observed \ac{erass} profiles allow us to test the baryon-physics models of the \mbox{TNG300-1} simulations. The normalization and slope of the inner profile characterize the spatial distribution of hot gas within the halo, reflecting its thermodynamic structure and underlying thermal and nonthermal effects. The outer profile reflects both the amounts of hot gas and gas-rich halos in the circumcluster or filament regions. In this section, we compare the observed profiles with the simulated profiles in the \mbox{TNG300-1}. Before the comparison, we applied the eRASS:1 cluster selection to the simulation sample. We adopted the mass and redshift dependent selection function $P_\mathrm{det}(M,z)$ from \citet{Clerc2024}. For each halo in the simulation sample, the detection probability as a function of mass is 
\begin{equation}
    P_\mathrm{det}(M)=\int_{0.03}^{0.2} P_\mathrm{det}(M,z)P(z)\mathrm{d}z,
\end{equation}
where $P(z)\propto \mathrm{d}V_\mathrm{cov}/\mathrm{d}z$ is the p.d.f. of the halo at different redshifts, which is scaled with the differential comoving volume. When calculating a selection-applied simulation profile, we used $P_\mathrm{det}(M)$ as the averaging weight. We note again that the simulation data are from the $z=0$ snapshot. Here we ignored the redshift evolution of halo properties and the \ac{hmf} difference from 0.2 to 0. 

We first compare the projected X-ray emission profile $\Sigma_\mathrm{X}$ between observations and simulations in the left panel of Fig.~\ref{fig:sim_vs_obs}. The \ac{erass} observed $\Sigma_\mathrm{X}$ profile (purple error bars) is the total $S_\mathrm{X}$ profile with the best-fit uncorrelated component (see Sect.~\ref{sect:fitting}) subtracted and converted using Eq.~\ref{eq:sb_conversion}. It includes both the one-halo and the two-halo components. In addition to the \mbox{TNG300-1} simulations analyzed here, we include a comparison with two \ac{the300} profiles reported by \citet{Li2025}, from the Gizmo-Simba and Gadget-X runs (shown as dashed and dotted green lines, respectively). The same panel also displays the \mbox{TNG300-1} simulation $\Sigma_\mathrm{X}$ profile, with a nearby halo masking threshold of $10^{14}M_\sun$ indicated by the cyan-shaded region. In the inner region $r\lesssim r_\mathrm{200m}$, the \mbox{TNG300-1} $\Sigma_\mathrm{X}$ profile is more centrally peaked compared to the stacked observations. 
In the radius range of $1-4r_\mathrm{200m}$, the normalization of the \mbox{TNG300-1} profile is a factor of 3 lower than the observations. This difference can be partly attributable to the limited projection depth of the \mbox{TNG300-1} analysis (20~Mpc), which substantially reduces the projected two-halo signal. 

Although the three simulation profiles are all calculated at $z=0$ and with a nearby halo masking threshold of $M_\mathrm{500c}=10^{14}M_\sun$, the emissivity profiles of the two \ac{the300} simulations have normalizations higher than those in the \mbox{TNG300-1} in cluster outskirts to the circumcluster region, as shown in the left panel of Fig.~\ref{fig:sim_vs_obs}. This could be partially attributed to the mass difference of the sample and the mass exclusion limit of the surrounding infalling halos. \ac{the300} simulations only include massive halos with $M_\mathrm{200c}>10^{14.8}M_\sun$, whose halo bias is higher than the sample used in the \mbox{\mbox{TNG300-1}}. However, the one order-of-magnitude difference between the two profiles, Gizmo-Simba and Gadget-X, is due solely to the different simulation settings. As reported by \citet{Li2025}, the Gadget-X run contains a higher fraction of dense gas at large radii. The comparison of the three profiles at large radii suggests that differences in the adoption of simulation codes, physical models, and model parameters could lead to a significant difference in the intensity of the circumcluster X-ray emission. The new stacked observations of the eROSITA clusters allow us to test physical models in these unexplored regions, from the cluster's far outskirts to cosmic filaments.

As the next step, in the right panel of Fig. \ref{fig:sim_vs_obs}, we compare the one-halo term density profile from the eROSITA observations with the \mbox{TNG300-1} $T>10^6$~K hot gas density profiles in the off-filament direction (see the next subsection for details), where we use a blue hatched band to denote the $1\sigma$ scatter of the simulation profiles. As also shown in the $\Sigma_\mathrm{X}$ profile comparison, the observed gas density profile is less centrally peaked than that predicted by the \mbox{TNG300-1} simulation, suggesting that feedback processes in cluster-scale halos may be stronger and more efficient than those implemented in the IllustrisTNG model, displacing part of the gas toward the cluster outskirts. Similar discrepancies have been reported at lower halo masses in the galaxy group regime. The stacked kinematic \ac{sz} (kSZ) signal of luminous red galaxies measured by ACT appears more extended than the IllustrisTNG predictions \citep{Hadzhiyska2025}. Similarly, the eROSITA-selected galaxy groups tend to exhibit higher entropy from $r_\mathrm{2500c}$ to $r_\mathrm{500c}$, corresponding to a lower central gas density compared to the MillenniumTNG simulations \citep{Bahar2024}, whose feedback model is similar to that of the \mbox{TNG300-1} simulations used in this work. These results are also consistent with the systematically elevated normalization of the halo $L_\mathrm{X}-M$ relation \citep{Pop2022} and the higher gas mass fractions within $r_\mathrm{500c}$ of the optically selected groups \citep{Popesso2024}. A detailed study of gas distribution with feedback models across a large mass range of eROSITA-selected galaxy clusters and groups will be presented in our upcoming studies (Ding et al. in prep., Clerc et al. in prep.).

\subsection{Thermodynamic profiles in 3D}
\label{sect:profiles_3d}

In this section, we investigate the thermodynamic properties of the gas from the halo center to the accretion regions in the \mbox{TNG300-1} simulations. We extracted gas particles out to $5\ r_\mathrm{200m}$, and created masks for particles belonging to nearby halos and subhalos. 

The spherically averaged gas density profiles from the cluster outskirts out to several $r_\mathrm{200m}$ from hydrodynamic simulations consistently find that the profiles steepen from the cluster center to around $r_\mathrm{200m}$, before flattening again due to the influence of cosmic filaments \citep{ONeil2021, Angelinelli2022, Towler2024}.
However, in the following, we shall argue that at the radii with the presence of cosmic filaments, the gas properties toward the into-filament and off-filament (or into-void) environments could be significantly different and cannot be reflected by the spherically averaged profiles. 
Therefore, we aim to explore these quantities in both the filament and off-filament directions. Following \citet{Mansfield2017} and \citet{Aung2021}, we adopted an $N_\mathrm{side}=8$ \ac{healpix} scheme to group particles into 768 directions with respect to the halo center. For each \ac{los} direction, we binned the volume-averaged radial profiles of $T$, $K_\mathrm{e}$, $P_\mathrm{gas}$, and $n_\mathrm{tot}$ using 20 logarithmically spaced bins from 0.01 to 5 $r_\mathrm{200m}$. 
The temperature range of $10^5-10^{5.5}$~K is usually the boundary between cool-to-warm and warm-hot gas \citep[e.g.,][]{Cen1999,vandeVoort2011}. 
Here we adopted a temperature threshold of $10^5$~K to classify them into the in-filament and off-filament \ac{los} directions. The into-filament/off-filament \ac{los} profiles are those with the last radial bin temperature higher/lower than $10^5$~K. 

\begin{figure*}
    \includegraphics[width=\linewidth]{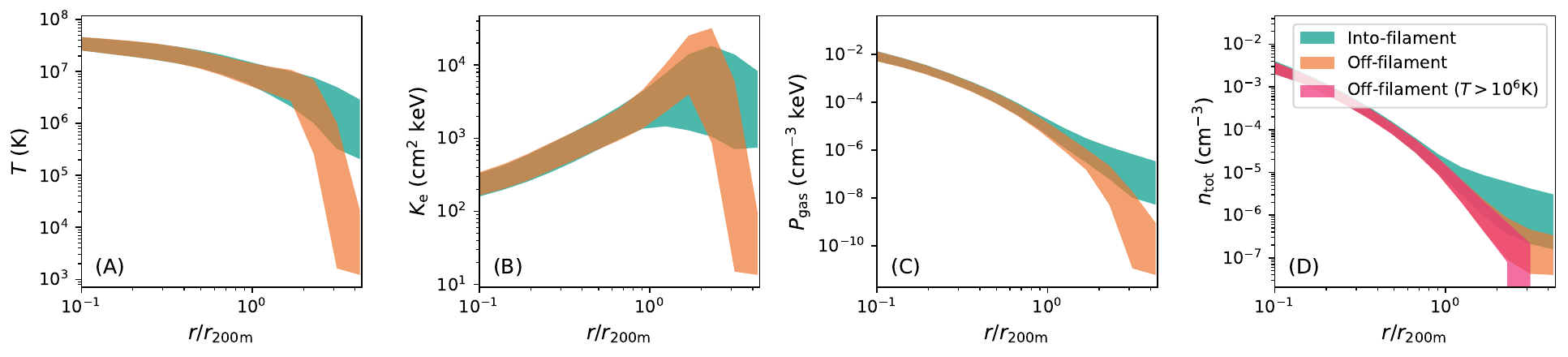}
    \caption{One-$\sigma$ scatter of the temperature (A), electron entropy (B), gas pressure (C), and gas density (D) line-of-sight profiles grouped into the filament (teal color) and off-filament directions (orange color) in the \mbox{TNG300-1} simulations. All four profile types show a strong discrepancy in the two directions. In addition, we plot the $T>10^6$~K gas density profile in the off-filament direction in magenta in the bottom right panel. The hot gas is confined by the accretion shock within $3\times r_\mathrm{200m}$. }
    \label{fig:los_3d_tkpn}
\end{figure*}

Fig.~\ref{fig:los_3d_tkpn} shows the 68\% scatter of the $768\times159$ \ac{los} profiles of $T$, $K_\mathrm{e}$, $P_\mathrm{gas}$, and $n_\mathrm{tot}$ in the two different directions, where nearby $M_\mathrm{500c}>10^{13}M_\sun$ (sub)halos were excluded. For all the thermodynamic quantities of $T$, $K_\mathrm{e}$, and $P_\mathrm{gas}$, the into-filament and off-filament \ac{los} profiles agree with each other within $r_\mathrm{200m}$ and show a strong discrepancy beyond that radius. The into-filament temperature profiles keep $T\sim10^6$~K out to the maximum radius we extracted, while the off-filament-direction temperature profiles have more than two orders of magnitude drops at the radii of their accretion shocks, $\sim2-3$~$r_\mathrm{200m}$. Similarly, the discrepancy between the two directions and the feature of the accretion shock is also presented in the $K_\mathrm{e}$ and $P_\mathrm{gas}$ profiles. The $n_\mathrm{tot}$ profile in the two directions deviates at $r_\mathrm{200m}$, but neither shows a jump at the shock radii. The absence of a strong jump in the total gas number density but a sharp drop in temperature is compatible with the Rankine-Hugoniot shock conditions, which impose limits of
$\lim_{\mathcal{M}\to\infty}\rho_2/\rho_1=4$ on the density jump and $\lim_{\mathcal{M}\to\infty}T_2/T_1=5/16\times\mathcal{M}^2$ on the temperature jump. For a high-$\mathcal{M}$ accretion shock, the temperature can be enhanced by more than one order of magnitude, but the density enhancement is limited to 4. This change is smaller than the scatter of \ac{los} $n_\mathrm{tot}$ profiles among the cluster sample and could be easily smeared out due to the triaxiality of the shock surface. We also plot the off-filament direction gas number density profiles using gas cells with $T>10^6$~K, which corresponds to the temperature range of X-ray-emitting gas. The gas number density profiles of the $T>10^6$~K phase follow the total gas number density profile up to the termination at the accretion shock, which reflects that the hot gas in the off-filament direction is confined by the accretion shock. Outside the accretion shock, all gas is at temperatures below $10^6$~K. This behavior matches the order-of-magnitude drop in the temperature profile.

\section{Discussion}\label{sect:discussion}

\subsection{Clumping effects}

When measuring the halo gas density profile by deprojecting the 2D X-ray emission profile under the spherical symmetry assumption, gas clumping can overestimate the measured gas density, especially in the cluster outskirts. The clumping effects are quantified by a clumping factor 
\begin{equation}
    C(r)\equiv\left<\rho_\mathrm{gas}(r)^2\right>/\left<\rho_\mathrm{gas}(r)\right>^2.
\end{equation}
In a broad sense, the clumping factor $C(r)$ describes not only the presence of clump structures in and around the central halos e.g., gas rich subhalos and nearby cosmic filaments, but also other effects that violate the spherical symmetry assumption, e.g., large scale inhomogeneity, halo ellipticity, and gas density fluctuations due to turbulence, which are also termed as ``residual clumpiness'' \citep{Roncarelli2013}. 
Numerical simulations have quantified that $C(r)$ is mild at intermediate radii due to the residual clumpiness and increases dramatically in the far outskirts due to the anisotropic gas distribution in the presence of connected cosmic filaments \citep[e.g.,][]{Nagai2011,Angelinelli2021}. Pointing observations using \emph{XMM-Newton} and \emph{Chandra} also confirmed the mild gas clumping up to $\sim r_\mathrm{200c}$ \citep[e.g.,][]{Eckert2015,Zhu2023,Kovacs2023}. 

In this work, we used the two-halo term to account for the emission from nearby halos and cosmic filaments, which is validated by the analysis of the \mbox{TNG300-1} simulation in Sect. \ref{sect:sim_2d}. With this method, we reduced the impact of the main clumpiness that leads to an overestimation of the gas density at $r_\mathrm{200m}$ by a factor of a few \citep{Nagai2011,Angelinelli2021}. However, residual clumping effects remain. Additionally, because we fit a single model to the stacked emission profile, the variance in the density profile of the stacking sample also introduces a similar ``clumping'' effect, i.e., the measured averaged density profile is weighted by individual source surface brightnesses scaled by $\rho_\mathrm{gas}(r)^2$. Although gas clumpiness varies across numerical simulation suites, our analysis of the \mbox{TNG300-1} simulation sheds light on the possible residual clumpiness strength. The ratios between the best-fit density profiles and the sample median density profile (right panel of Fig. \ref{fig:2d_em}) indicate an overestimate of the gas density of 1.5 in the radial range of $r_\mathrm{500c}<r<r_\mathrm{200m}$. This possible clumping factor also agrees with the residual clumping factor value at $\sim1.5\times r_\mathrm{200c}$ reported by \citet{Roncarelli2013}.

\subsection{Characteristic radii at the halo boundary}\label{sect:boundary}

One important question in studying halos and structure formation is where a halo's boundary lies. The infall, orbit, and turnaround of collisionless dark matter give rise to the splashback feature, which is now widely considered a novel definition of the halo boundary. However, the infall of collisional gas does not exhibit the splashback feature because it gets shocked before it can enter the halo. In this section, we discuss the possible characteristic radii of the gas content that can be constrained by our observations, and that could be used as halo boundaries, and the implications of these results. 

The most prominent feature in both theoretical calculations and numerical simulations that can serve as the halo gas boundary is the accretion shock. In fact, the accretion shock in simulations is a gigantic structure enclosing the nodes, filaments, and sheets \citep[e.g.][]{Ryu2003, Schaal2016}. For halos of galaxy clusters, in particular, our analysis of the \ac{los} dependent thermodynamic profiles using \mbox{TNG300-1} shows that the accretion shock is the boundary of hot gas in the off-filament direction. 

In contrast to numerical simulations, it is extremely difficult to measure the accretion shock radius in observations. For example, \citet{Anbajagane2022} presented an \ac{sz} staking study but found only a marginal feature attributed to an average accretion shock at $4.6\times r_\mathrm{200m}$. 
The projected surface brightness profile of the stacked eROSITA observations itself (see the left panel of Fig.~\ref{fig:sim_vs_obs}) shows no clear surface brightness features corresponding to the accretion shock beyond $r_{200m}$. Nevertheless, the profile fitting in this work provides an indirect clue to the accretion shock radius $r_\mathrm{shock}$, where the gas density reaches the post-shock density condition, $r_\mathrm{shock}=r|_{\rho(r)=\rho_\mathrm{post}}$. The first assumption adopted is that $\rho_\mathrm{post}=\left<\rho_\mathrm{b}\right>$ in Eq. \ref{eq:prof_boundary}, which is supported by the \mbox{TNG300-1} simulation. Our analysis of the simulation shows that the density profile in the off-filament direction (see the panel D of Fig.~\ref{fig:los_3d_tkpn}) reaches the accretion shock with a gas density $2\times10^{-7}$~cm$^{-3}$, slightly lower than the cosmic mean baryon density $\left<\rho_\mathrm{b}\right>$. In addition to our analysis of the \mbox{TNG300-1} simulation, \citet{Vurm2023} analyzed the gas structure around the most massive halo in the C-EAGLE simulation, and their results also show that the post-shock gas density is slightly below $\left<\rho_\mathrm{b}\right>$. The second assumption is that the best-fit gNFW density profile within $r_\mathrm{200m}$ can be extrapolated to larger radii, as supported by our \mbox{TNG300-1} analysis. 
From our fitting results in Sect. \ref{sect:fitting}, the gas density profile of the one-halo term reaches $\left<\rho_\mathrm{b}\right>$ at $\sim3\times r_\mathrm{200m}$, indicating a possible $r_\mathrm{shock}$ based on our assumption. We clarify that this is a rough, indirect estimate based on the two assumptions mentioned above. A more generalized estimate can be expressed as 
\begin{equation}
    \log r_\mathrm{shock}/r_\mathrm{200m} \sim \frac{\log \rho_\mathrm{post}/\rho_\mathrm{gas}(r_\mathrm{200m})}{\mathrm{d}\log \rho_\mathrm{gas}(r)/\mathrm{d}\log r}.
\end{equation}
Given the gas density contrast $\Delta_\mathrm{b}\sim30$ and the density slope $\mathrm{d}\log \rho_\mathrm{gas}(r)/\mathrm{d}\log r\sim-3$ at $r_\mathrm{200m}$, unless there is a sudden steepening of the gas density outside $r_\mathrm{200m}$ or the post-shock density is much higher than $\left<\rho_\mathrm{b}\right>$, the $r_\mathrm{shock}/r_\mathrm{200m}$ ratio is a factor of a few and not too close to 1. 

Though the accretion shock is a prominent feature separating the void and the overdense cosmic web, mass accretion of galaxy clusters is mostly from the connected cosmic filaments. The high velocity infalling flows from cosmic filaments penetrate the outer atmosphere of the halo \citep[e.g.][]{Malavasi2023, Vurm2023, Rost2024}. The model fitting of our observed profile directly reflects the radius of the halo-filament connection. 
In Fig.~\ref{fig:obs_prof_fitting}, the one-halo and two-halo terms intersect at $\sim0.9\times r_\mathrm{200m}$, shifting to $\sim1.1\times r_\mathrm{200m}$ if we assume that 50\% of the two-halo emission originates from nearby gas-rich halos. We could therefore still conclude that $r_\mathrm{200m}$ approximately marks the halo–filament connection radius. Our analysis of the \mbox{TNG300-1} simulation also suggests that the pressure and density \ac{los} profiles in the filament direction deviate from those in the off-filament direction at roughly $r_\mathrm{200m}$.
Our results of halo-filament connection radius are close to the ``gas splashback radius'' reported by \citet{ONeil2021,Towler2024}, who used the criteria of the steepest slope of radial gas density profile in the logarithmic space to characterize the radius. We argue that the presence of the steepest slope is necessary but not sufficient for claiming a splashback feature. The splashback is a feature of collisionless dark matter particles; the reported steepest slope is due to nearby filaments that flatten the spherically averaged radial density profile. Meanwhile, the halo-filament connection radius at $r_\mathrm{200m}$ is approximately the inner shock location reported by \citet{Anbajagane2022, Anbajagane2024} and the polarized stacked radio emission reported in \citet{Vernstrom2023}, suggesting the discovered shock signal is due to the gas inflows from cosmic filaments.

\subsection{Two-halo term normalization}

When we fit the two-halo term, the model (Eq. \ref{eq:2h_final}) includes a free normalization parameter $A_\mathrm{2h}$ and a theoretical prediction of the profile. For the component of theoretical prediction, its profile shape is from $\xi_\mathrm{mm}^\mathrm{lin}$, and the normalization depends on the halo bias, \ac{hmf}, $L_\mathrm{X}-M$ relation, and the selection function of masked sources. In principle, if the theoretical prediction of the two-halo term is accurate, the best-fit value of $A_\mathrm{2h}$ indicates that, in addition to the X-ray emission from correlated nearby halos, the amount of X-ray emission from unvirialized gas in nearby cosmic filaments is required to match the observation. In our case, the predicted normalization is $3.7\times10^{34}$~erg~s$^{-1}$~kpc$^{-2}$ and the best-fit $\log A_\mathrm{2h}=0.37\pm0.06$, which means that only $\sim40\%$ of the two-halo term emission is from the model prediction. We note that our theoretical prediction uses models of the \citet{Tinker2010} halo bias $b$, the \citet{Tinker2008} \ac{hmf}, and the \citet{Bulbul2019} $L_\mathrm{X}-M$ relation. To understand the impact of model adoption on the theoretical prediction of the two-halo normalization, we tested additional combinations of \ac{hmf} models from \citet{Watson2013}, \citet{Bocquet2016}, \citet{Despali2016}; halo bias models from \citet{Bhattacharya2011}, \citet{Comparat2017}, \citet{Pillepich2010}; scaling relation model from \citet{Lovisari2020} and \citet{Chiu2022}. The predicted two-halo emission ranges from $1.7\times10^{34}$ to $7\times10^{34}$~erg~s$^{-1}$~kpc$^{-2}$. The large range of the predicted two-halo normalization indicates that, though the fraction of unvirialized gas emission is positive, it is difficult to have a precise constraint.

We note that the present two-halo model formalism focuses on the effects relevant to our analysis. In this context, we did not explicitly incorporate the scatter in the $L_\mathrm{X}-M$ and $\Lambda-M$ scaling relations. Consequently, the mass-dependent selection function and mass cut in Eq. 9 provide an approximate representation of the source masking scheme. Second, the two-halo term is modeled without including a contribution from unresolved \ac{agn}. In addition, the X-ray properties of circumcluster halos may reflect aspects of their assembly history or ongoing ram-pressure stripping. Effects not captured by the current framework could influence theoretical predictions. Nevertheless, the robust detection of the two-halo signal at a background level of $\lesssim2\%$ highlights the importance of using next-generation X-ray missions, such as NewAthena (Cruise et al. 2025), to probe physical processes associated with cluster formation and ongoing mass accretion in these regions. This work provides a foundational investigation in the X-ray regime.

\subsection{Gas fraction out to $r_\mathrm{200m}$}

The reported gas mass fraction of halos within $r_\mathrm{500c}$ is lower than the cosmic baryon fraction, and the depletion of the gas content is a function of halo mass \citep[see the review of][and references therein]{Eckert2021}. With the best-fit gas density profile in Sect.~\ref{sect:fitting}, we can measure the baryon density and therefore mass distribution beyond $r_\mathrm{500c}$ by calculating a gas mass fraction profile. 

The gas fraction is defined as 
\begin{equation}
    f_\mathrm{gas}(r) \equiv\frac{M_\mathrm{gas}(r)}{M_\mathrm{tot}(r)}, 
\end{equation}
~where $M_\mathrm{gas}(r)$ and $M_\mathrm{tot}(r)$ are the gas mass and total mass within the radius $r$. We estimated $M_\mathrm{tot}(r)$ by assuming that the total mass of the halo follows an \ac{nfw} profile. Due to the self-similar properties of the \ac{nfw} model, the matter density profile in units of spherical overdensity radii only depends on the concentration parameter. 

\begin{figure}
    \centering
    \includegraphics[width=\linewidth]{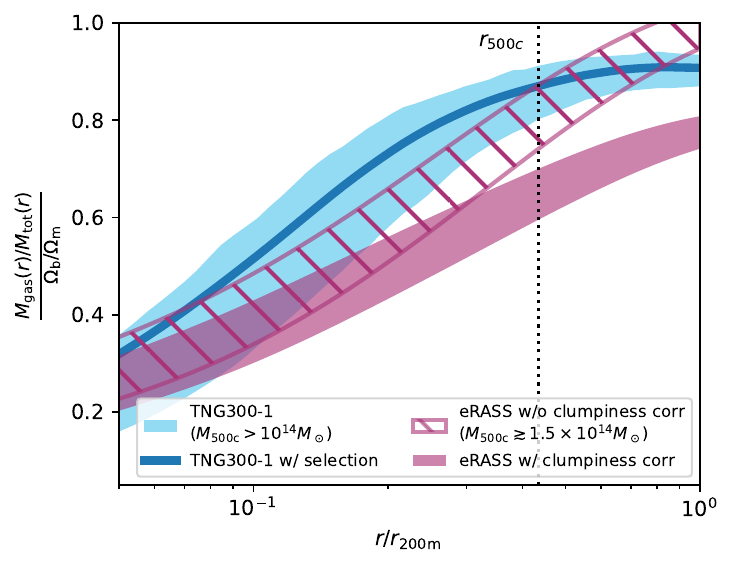}
    \caption{Measured eROSITA gas fraction profiles with respect to the cosmic baryon fraction up to $r_\mathrm{200m}$ and the comparison with simulations. The observed profile from observation with and without residual clumpiness correction is plotted in the hatched and filled magenta bands, respectively. The bandwidth represents the uncertainty from halo concentration $c_\mathrm{200c}$ assumptions of from 3 to 5. The $1\sigma$ scatter of gas fraction profiles of $M_\mathrm{500c}>10^{14}M_\sun$ halos in the \mbox{TNG300-1} simulations is plotted in a blue band, and the averaged gas fraction of halos with the selection function applied is plotted in the thick blue line. }
    \label{fig:gas_fraction}
\end{figure}

We used the hydrogen mass fraction of 0.71 to calculate $f_\mathrm{gas}$. This value is calculated with the assumptions of \citet{Lodders09} abundance for $Z=0.3Z_\sun$. Fig.~\ref{fig:gas_fraction} shows the $f_\mathrm{gas}$ profile with respect to the cosmic baryon fraction with the adoption of $c_\mathrm{200c}$ in [3,5] in the hatched magenta band. The $f_\mathrm{gas}$ value continuously increases beyond $r_\mathrm{500c}$ and reaches $\gtrsim100\%$ of the cosmic baryon fraction at $r_\mathrm{200m}$. Since the stellar mass fraction additionally amounts to a few to ten percent of the cosmic baryon fraction \citep[e.g.][]{Gonzalez2013}, the resulting total baryon fraction within $r_\mathrm{200m}$, including both the X-ray emitting gas and stellar components, is higher than the cosmic value, indicating possible gas clumping effects that overestimate the gas density. 
In an alternative calculation, we corrected for the gas density overestimation due to residual clumpiness. We applied the ratio between the best-fit and the sample median from the simulation analysis in Sect. \ref{sect:sim_2d} to the best-fit observed gas density profile. The resulting $f_\mathrm{gas}$ profile is plotted as the filled magenta band in Fig.~\ref{fig:gas_fraction}. The comparison shows that correcting for clumping significantly reduces the measured $f_\mathrm{gas}$ in the outskirts. At $r_\mathrm{200m}$, the $f_\mathrm{gas}$ profile with clumpiness correction is $\sim80\%$ of the cosmic mean baryon fraction. 
The profile also shows that the $f_\mathrm{gas,r500c}$ with clumpiness correction is $\sim65\%$ of the cosmic mean baryon fraction, and is in line with the $f_\mathrm{gas}-M$ relation from group size halos \citep{Sun2009,Lovisari2015, Bahar2022, Bulbul2024, Siegel2025} to massive halos \citep{Bulbul2012, Eckert2019, Bulbul2019,Liu2022,Bulbul2024} given the median mass $M_\mathrm{500c}=2.6\times10^{14}M_\sun$ of our sample, though there is a $30\%$ range of the measurements in literature (see fig. 7 in the review of \citealt{Eckert2021}). 

To compare the observed results with the \mbox{TNG300-1} simulations. We calculated $f_\mathrm{gas}$ for 159 galaxy clusters and show the $1\sigma$ scatter as the blue filled band in Fig.~\ref{fig:gas_fraction}. Meanwhile, the averaged gas mass fraction with the selection function applied is plotted as the thick blue line. The $f_\mathrm{gas}$ profiles of the \mbox{TNG300-1} simulation reach an asymptotic value of 90\% within $r_\mathrm{200m}$, and show a strong discrepancy with the observed $f_\mathrm{gas}$ profile with clumpiness correction. The discrepancy is consistent with that observed in the density profile comparison in Sect. \ref{sect:obs_vs_sim}, reflecting that the IllustrisTNG model produces more concentrated gas density profiles. The observed $80\%$ gas fraction with respect to the cosmic mean baryon density at $r_\mathrm{200m}$ agrees more with simulations with stronger feedback models. For example, the Magneticum simulation shows similar values of $f_\mathrm{gas}$ beyond $r_\mathrm{200c}$ in bins of cluster masses \citep{Angelinelli2022}. 

It is important to bear in mind that the residual clumpiness correction used in this section is based on our analysis of the \mbox{TNG300-1} simulation. The strength of the correction reflects the large-scale gas inhomogeneity and the variance in density profiles among individual cluster halos in IllustrisTNG and may differ in other simulation suites. Moreover, we note an important caveat: beyond the systematic uncertainties in the total mass profile arising from the assumed concentration parameter ($c_\mathrm{200c}=[3,5]$), the gas mass profile is subject to additional systematics not included in the total error budget. In particular, in this work, the conversion from gas density to X-ray emissivity relies on the assumption of a constant cooling function. If the true gas metallicity in the outskirts is lower than the assumed value, which is supported by numerical simulations, this would lead to an underestimation of both the gas mass profile and the $f_\mathrm{gas}$. Our investigations of the metallicity and temperature-dependent cooling function in Appendix \ref{app:cooling} suggest a possible $10\%$ underestimation if the metallicity is $0.2Z_\sun$.

\section{Conclusions}\label{sect:conclusion}

This study investigates the distribution of shock-heated gas from the outskirts of galaxy clusters into the large-scale accretion regions. We performed a stacking analysis of the two-year eROSITA All-Sky survey observations of 680 galaxy clusters selected in X-rays from the eRASS1 survey, spanning a mass range of $M_\mathrm{500c}\gtrsim10^{14}M_\sun$ and a redshift interval of $0.03<z<0.2$ \citep{Bulbul2024}. At the $1-2\times r_\mathrm{200m}$ ($\sim2.3-4.5$~Mpc) radii, we detected a statistically significant $12\sigma$ excess X-ray signal above the background. This represents the first detection of X-ray emission associated with galaxy clusters into the accretion region at large radii.

We modeled the stacked surface brightness profile as a superposition of one-halo, two-halo, and constant components, corresponding respectively to X-ray emission from the primary cluster halo, correlated structures such as infalling halos and filamentary gas, and the uncorrelated foreground/background. The best-fit results show that at $r_\mathrm{200m}$, the gas density is $\sim2.5\times10^{-5}$~cm$^{-3}$, corresponding to a baryon density contrast of $\sim30$. Comparing the deprojected gas density profile derived from the one-halo term of the virialized gas to previous results in the literature, in the inner regions $r<0.1\times r_\mathrm{200m}$, the profile shows slightly lower gas densities than those reported by \citet{Ghirardini2019} and \citetalias{Lyskova2023}, likely due to differences in the sample selection. In the intermediate radii, between $0.4\times r_\mathrm{200m}$ and $0.6\times r_\mathrm{200m}$ (approximately $r_\mathrm{500c}$ to $r_\mathrm{200c}$), our results are consistent with those reported in the literature.  The transition between the one-halo and two-halo regimes occurs around $\sim r_\mathrm{200m}$, suggesting that beyond this radius, the observed signal primarily is from unvirialized gas in connected filaments and nearby halos, under the detection limit of eROSITA.

We analyzed the \mbox{TNG300-1} simulation to investigate the anisotropic distribution of the intracluster and circumcluster gas and to compare it with the observations. The thermodynamic profiles of the shock-heated gas exhibit clear directional dependence relative to the cosmic filaments. Along off-filament directions, the high-entropy, hot gas is sharply truncated and confined by the accretion shock at $r>r_\mathrm{200m}$. In contrast, profiles along filament directions remain elevated out to the maximum extraction radius, consistent with continuous gas inflow from the cosmic filament. Furthermore, we used the \mbox{TNG300-1} data to validate our modeling framework for stacked eROSITA observations. Incorporating a two-halo term that accounts for shock-heated gas and X-ray halos in filaments results in a best-fit one-halo profile that accurately recovers the gas density profile in off-filament directions.

The observed gas density profile is more extended, while the profiles predicted by the \mbox{TNG300-1} numerical simulations are more concentrated. This trend is also reflected in the gas fraction profile of the stacked cluster of galaxies. The gas fraction profile of the \mbox{TNG300-1} cluster sample exhibits a steep rise beyond $>0.1r_\mathrm{200m}$ compared to the observations. If we apply a correction for residual clumping in the density profiles, the observed gas fractions at all radii are lower than in the \mbox{TNG300-1} simulation. The differences in observed density and gas fraction profiles suggest that the feedback processes in $M_\mathrm{500c}>10^{14}M_\sun$ halos may be stronger than modeled in IllustrisTNG. This behavior is consistent with the recently observed trends at lower mass halos in galaxy group regimes in ACT kSZ measurements \citep{Hadzhiyska2025, Siegel2025} and the stacked eROSITA observation of X-ray and optically selected galaxy groups \citep{Bahar2024, Popesso2024}.

Comparing observations with simulations, we identified two characteristic radii in the accretion region of the cluster outskirts that mark transitions in the thermodynamic state of the intracluster and intergalactic gas, potentially defining physical halo boundaries in this region. The first is the halo–filament intersection radius, located at approximately $r_\mathrm{200m}$, where the two-halo term for the unvirialized gas and infalling halos in the filaments begins to dominate the stacked X-ray surface brightness profile over the one-halo term of the virialized intracluster gas, corresponding to the scale at which the gas density and pressure profiles along cosmic filament directions exceed those in the off-filament directions in the \mbox{TNG300-1} simulations. The second is the accretion shock radius, estimated to lie at $\sim3\times r_\mathrm{200m}$, inferred by extrapolating the one-halo gas density to the cosmic mean baryon density, consistent with the simulation predictions from the IllustrisTNG. However, the current depth of the eROSITA All-Sky Survey and the background levels limit our study to regions beyond $r_\mathrm{200m}$ and to the direct detection of accretion-shock features.

This work presents the \ac{erass} stacking analysis of the shock-heated gas in cluster far outskirts and surrounding circumcluster accretion regions. The results represent the average properties of the sample studied here. Detailed studies on individual systems require high sensitivity, low instrumental background, an understanding of the foreground/background large-scale structure, and deep exposure. The upcoming NewAthena mission \citep{Nandra2013,Cruise2025} in the late 2030s will be particularly well-suited for such targeted studies of individual galaxy clusters, providing new insights into the cluster–filament connection, the anisotropic distribution of gas at cluster boundaries, and the emission contribution from unvirialized gas within filaments.

\begin{acknowledgements}
The authors thank the referee for their insightful comments, which helped improve the manuscript. 
The authors acknowledge Lars Hernquist and Volker Springel for helpful discussions.
This work is based on data from eROSITA, the soft X-ray instrument aboard SRG, a joint Russian-German science mission supported by the Russian Space Agency (Roskosmos), in the interests of the Russian Academy of Sciences represented by its Space Research Institute (IKI), and the Deutsches Zentrum für Luft- und Raumfahrt (DLR). The SRG spacecraft was built by Lavochkin Association (NPOL) and its subcontractors, and is operated by NPOL with support from the Max Planck Institute for Extraterrestrial Physics (MPE).

The development and construction of the eROSITA X-ray instrument was led by MPE, with contributions from the Dr. Karl Remeis Observatory Bamberg \& ECAP (FAU Erlangen-Nuernberg), the University of Hamburg Observatory, the Leibniz Institute for Astrophysics Potsdam (AIP), and the Institute for Astronomy and Astrophysics of the University of Tübingen, with the support of DLR and the Max Planck Society. The Argelander Institute for Astronomy of the University of Bonn and the Ludwig-Maximilians-Universität München also participated in the science preparation for eROSITA. The eROSITA data shown here were processed using the eSASS/NRTA software system developed by the German eROSITA consortium.
X. Zhang, E. Bulbul, E. Artis, and S. Zelmer acknowledge financial support from the European Research Council (ERC) Consolidator Grant under the European Union’s Horizon 2020 research and innovation program (grant agreement CoG DarkQuest No 101002585). A.L. acknowledges the support from the National Natural Science Foundation of China (Grant No. 12588202). A.L. is supported by the China Manned Space Program with grant no. CMS-CSST-2025-A04.
\end{acknowledgements}

\bibliographystyle{aa} 
\bibliography{aa57423-25}

@ARTICLE{Adhikari2014,
       author = {{Adhikari}, Susmita and {Dalal}, Neal and {Chamberlain}, Robert T.},
        title = "{Splashback in accreting dark matter halos}",
      journal = {\jcap},
         year = 2014,
        month = nov,
       volume = {2014},
       number = {11},
        pages = {019-019},
          doi = {10.1088/1475-7516/2014/11/019},
archivePrefix = {arXiv},
       eprint = {1409.4482},
 primaryClass = {astro-ph.CO},
       adsurl = {https://ui.adsabs.harvard.edu/abs/2014JCAP...11..019A}
}

@ARTICLE{Anbajagane2022,
       author = {{Anbajagane}, D. and {Chang}, C. and {Jain}, B. and {Adhikari}, S. and {Baxter}, E.~J. and {Benson}, B.~A. and {Bleem}, L.~E. and {Bocquet}, S. and {Calzadilla}, M.~S. and {Carlstrom}, J.~E. and {Chang}, C.~L. and {Chown}, R. and {Crawford}, T.~M. and {Crites}, A.~T. and {Cui}, W. and {de Haan}, T. and {Di Mascolo}, L. and {Dobbs}, M.~A. and {Everett}, W.~B. and {George}, E.~M. and {Grandis}, S. and {Halverson}, N.~W. and {Holder}, G.~P. and {Holzapfel}, W.~L. and {Hrubes}, J.~D. and {Lee}, A.~T. and {Luong-Van}, D. and {McDonald}, M.~A. and {McMahon}, J.~J. and {Meyer}, S.~S. and {Millea}, M. and {Mocanu}, L.~M. and {Mohr}, J.~J. and {Natoli}, T. and {Omori}, Y. and {Padin}, S. and {Pryke}, C. and {Reichardt}, C.~L. and {Ruhl}, J.~E. and {Saro}, A. and {Schaffer}, K.~K. and {Shirokoff}, E. and {Staniszewski}, Z. and {Stark}, A.~A. and {Vieira}, J.~D. and {Williamson}, R.},
        title = "{Shocks in the stacked Sunyaev-Zel'dovich profiles of clusters II: Measurements from SPT-SZ + Planck Compton-y map}",
      journal = {\mnras},
         year = 2022,
        month = aug,
       volume = {514},
       number = {2},
        pages = {1645-1663},
          doi = {10.1093/mnras/stac1376},
archivePrefix = {arXiv},
       eprint = {2111.04778},
 primaryClass = {astro-ph.CO},
       adsurl = {https://ui.adsabs.harvard.edu/abs/2022MNRAS.514.1645A}
}

@ARTICLE{Anbajagane2024,
       author = {{Anbajagane}, D. and {Chang}, C. and {Baxter}, E.~J. and {Charney}, S. and {Lokken}, M. and {Aguena}, M. and {Allam}, S. and {Alves}, O. and {Amon}, A. and {An}, R. and {Andrade-Oliveira}, F. and {Bacon}, D. and {Battaglia}, N. and {Bechtol}, K. and {Becker}, M.~R. and {Benson}, B.~A. and {Bernstein}, G.~M. and {Bleem}, L. and {Bocquet}, S. and {Bond}, J.~R. and {Brooks}, D. and {Carnero Rosell}, A. and {Carrasco Kind}, M. and {Chen}, R. and {Choi}, A. and {Costanzi}, M. and {Crawford}, T.~M. and {Crocce}, M. and {da Costa}, L.~N. and {Pereira}, M.~E.~S. and {Davis}, T.~M. and {De Vicente}, J. and {Desai}, S. and {Devlin}, M.~J. and {Diehl}, H.~T. and {Doel}, P. and {Doux}, C. and {Drlica-Wagner}, A. and {Elvin-Poole}, J. and {Ferrero}, I. and {Fert{\'e}}, A. and {Flaugher}, B. and {Fosalba}, P. and {Friedel}, D. and {Frieman}, J. and {Garc{\'\i}a-Bellido}, J. and {Gatti}, M. and {Giannini}, G. and {Grandis}, S. and {Gruen}, D. and {Gruendl}, R.~A. and {Gutierrez}, G. and {Harrison}, I. and {Hill}, J.~C. and {Hilton}, M. and {Hinton}, S.~R. and {Hollowood}, D.~L. and {Honscheid}, K. and {Jain}, B. and {James}, D.~J. and {Jarvis}, M. and {Kuehn}, K. and {Lin}, M. and {MacCrann}, N. and {Marshall}, J.~L. and {McCullough}, J. and {McMahon}, J.~J. and {Mena-Fern{\'a}ndez}, J. and {Menanteau}, F. and {Miquel}, R. and {Moodley}, K. and {Mroczkowski}, T. and {Myles}, J. and {Naess}, S. and {Navarro-Alsina}, A. and {Ogando}, R.~L.~C. and {Page}, L.~A. and {Palmese}, A. and {Pandey}, S. and {Patridge}, B. and {Pieres}, A. and {Plazas Malag{\'o}n}, A.~A. and {Porredon}, A. and {Prat}, J. and {Reichardt}, C. and {Reil}, K. and {Rodriguez-Monroy}, M. and {Rollins}, R.~P. and {Romer}, A.~K. and {Rykoff}, E.~S. and {Sanchez}, E. and {S{\'a}nchez}, C. and {Sanchez Cid}, D. and {Schaan}, E. and {Schubnell}, M. and {Secco}, L.~F. and {Sevilla-Noarbe}, I. and {Sheldon}, E. and {Shin}, T. and {Sif{\'o}n}, C. and {Smith}, M. and {Staggs}, S.~T. and {Suchyta}, E. and {Swanson}, M.~E.~C. and {Tarle}, G. and {To}, C. and {Troxel}, M.~A. and {Tutusaus}, I. and {Vavagiakis}, E.~M. and {Weaverdyck}, N. and {Weller}, J. and {Wiseman}, P. and {Wollack}, E.~J. and {Yanny}, B.},
        title = "{Cosmological shocks around galaxy clusters: a coherent investigation with DES, SPT, and ACT}",
      journal = {\mnras},
         year = 2024,
        month = jan,
       volume = {527},
       number = {3},
        pages = {9378-9404},
          doi = {10.1093/mnras/stad3726},
archivePrefix = {arXiv},
       eprint = {2310.00059},
 primaryClass = {astro-ph.GA},
       adsurl = {https://ui.adsabs.harvard.edu/abs/2024MNRAS.527.9378A}
}

@ARTICLE{Angelinelli2021,
       author = {{Angelinelli}, M. and {Ettori}, S. and {Vazza}, F. and {Jones}, T.~W.},
        title = "{Proprieties of clumps and filaments around galaxy clusters}",
      journal = {\aap},
         year = 2021,
        month = sep,
       volume = {653},
          eid = {A171},
        pages = {A171},
          doi = {10.1051/0004-6361/202140471},
archivePrefix = {arXiv},
       eprint = {2102.01096},
 primaryClass = {astro-ph.CO},
       adsurl = {https://ui.adsabs.harvard.edu/abs/2021A&A...653A.171A}
}

@ARTICLE{Angelinelli2022,
       author = {{Angelinelli}, M. and {Ettori}, S. and {Dolag}, K. and {Vazza}, F. and {Ragagnin}, A.},
        title = "{Mapping `out-of-the-box' the properties of the baryons in massive halos}",
      journal = {\aap},
         year = 2022,
        month = jul,
       volume = {663},
          eid = {L6},
        pages = {L6},
          doi = {10.1051/0004-6361/202244068},
archivePrefix = {arXiv},
       eprint = {2206.08382},
 primaryClass = {astro-ph.GA},
       adsurl = {https://ui.adsabs.harvard.edu/abs/2022A&A...663L...6A}
}

@ARTICLE{Arnaud2010,
       author = {{Arnaud}, M. and {Pratt}, G.~W. and {Piffaretti}, R. and {B{\"o}hringer}, H. and {Croston}, J.~H. and {Pointecouteau}, E.},
        title = "{The universal galaxy cluster pressure profile from a representative sample of nearby systems (REXCESS) and the Y$_{SZ}$ - M$_{500}$ relation}",
      journal = {\aap},
         year = 2010,
        month = jul,
       volume = {517},
          eid = {A92},
        pages = {A92},
          doi = {10.1051/0004-6361/200913416},
archivePrefix = {arXiv},
       eprint = {0910.1234},
 primaryClass = {astro-ph.CO},
       adsurl = {https://ui.adsabs.harvard.edu/abs/2010A&A...517A..92A}
}

@ARTICLE{Aung2021,
       author = {{Aung}, Han and {Nagai}, Daisuke and {Lau}, Erwin T.},
        title = "{Shock and splash: gas and dark matter halo boundaries around {\ensuremath{\Lambda}}CDM galaxy clusters}",
      journal = {\mnras},
         year = 2021,
        month = dec,
       volume = {508},
       number = {2},
        pages = {2071-2078},
          doi = {10.1093/mnras/stab2598},
archivePrefix = {arXiv},
       eprint = {2012.00977},
 primaryClass = {astro-ph.CO},
       adsurl = {https://ui.adsabs.harvard.edu/abs/2021MNRAS.508.2071A}
}

@ARTICLE{Bahar2022,
       author = {{Bahar}, Y. Emre and {Bulbul}, Esra and {Clerc}, Nicolas and {Ghirardini}, Vittorio and {Liu}, Ang and {Nandra}, Kirpal and {Pacaud}, Florian and {Chiu}, I. -Non and {Comparat}, Johan and {Ider-Chitham}, Jacob and {Klein}, Mathias and {Liu}, Teng and {Merloni}, Andrea and {Migkas}, Konstantinos and {Okabe}, Nobuhiro and {Ramos-Ceja}, Miriam E. and {Reiprich}, Thomas H. and {Sanders}, Jeremy S. and {Schrabback}, Tim},
        title = "{The eROSITA Final Equatorial-Depth Survey (eFEDS). X-ray properties and scaling relations of galaxy clusters and groups}",
      journal = {\aap},
         year = 2022,
        month = may,
       volume = {661},
          eid = {A7},
        pages = {A7},
          doi = {10.1051/0004-6361/202142462},
archivePrefix = {arXiv},
       eprint = {2110.09534},
 primaryClass = {astro-ph.CO},
       adsurl = {https://ui.adsabs.harvard.edu/abs/2022A&A...661A...7B}
}

@ARTICLE{Bahar2024,
       author = {{Bahar}, Y.~E. and {Bulbul}, E. and {Ghirardini}, V. and {Sanders}, J.~S. and {Zhang}, X. and {Liu}, A. and {Clerc}, N. and {Artis}, E. and {Balzer}, F. and {Biffi}, V. and {Bose}, S. and {Comparat}, J. and {Dolag}, K. and {Garrel}, C. and {Hadzhiyska}, B. and {Hern{\'a}ndez-Aguayo}, C. and {Hernquist}, L. and {Kluge}, M. and {Krippendorf}, S. and {Merloni}, A. and {Nandra}, K. and {Pakmor}, R. and {Popesso}, P. and {Ramos-Ceja}, M. and {Seppi}, R. and {Springel}, V. and {Weller}, J. and {Zelmer}, S.},
        title = "{The SRG/eROSITA All-Sky Survey: Constraints on AGN feedback in galaxy groups}",
      journal = {\aap},
         year = 2024,
        month = nov,
       volume = {691},
          eid = {A188},
        pages = {A188},
          doi = {10.1051/0004-6361/202449399},
archivePrefix = {arXiv},
       eprint = {2401.17276},
 primaryClass = {astro-ph.CO},
       adsurl = {https://ui.adsabs.harvard.edu/abs/2024A&A...691A.188B}
}

@ARTICLE{Baxter2021,
       author = {{Baxter}, Eric J. and {Adhikari}, Susmita and {Vega-Ferrero}, Jes{\'u}s and {Cui}, Weiguang and {Chang}, Chihway and {Jain}, Bhuvnesh and {Knebe}, Alexander},
        title = "{Shocks in the stacked Sunyaev-Zel'dovich profiles of clusters - I. Analysis with the Three Hundred simulations}",
      journal = {\mnras},
         year = 2021,
        month = dec,
       volume = {508},
       number = {2},
        pages = {1777-1787},
          doi = {10.1093/mnras/stab2720},
archivePrefix = {arXiv},
       eprint = {2101.04179},
 primaryClass = {astro-ph.CO},
       adsurl = {https://ui.adsabs.harvard.edu/abs/2021MNRAS.508.1777B}
}

@ARTICLE{Bertschinger1985,
       author = {{Bertschinger}, E.},
        title = "{Self-similar secondary infall and accretion in an Einstein-de Sitter universe}",
      journal = {\apjs},
         year = 1985,
        month = may,
       volume = {58},
        pages = {39-65},
          doi = {10.1086/191028},
       adsurl = {https://ui.adsabs.harvard.edu/abs/1985ApJS...58...39B}
}

@ARTICLE{Bhattacharya2011,
       author = {{Bhattacharya}, Suman and {Heitmann}, Katrin and {White}, Martin and {Luki{\'c}}, Zarija and {Wagner}, Christian and {Habib}, Salman},
        title = "{Mass Function Predictions Beyond {\ensuremath{\Lambda}}CDM}",
      journal = {\apj},
         year = 2011,
        month = may,
       volume = {732},
       number = {2},
          eid = {122},
        pages = {122},
          doi = {10.1088/0004-637X/732/2/122},
archivePrefix = {arXiv},
       eprint = {1005.2239},
 primaryClass = {astro-ph.CO},
       adsurl = {https://ui.adsabs.harvard.edu/abs/2011ApJ...732..122B}
}

@ARTICLE{Bocquet2016,
       author = {{Bocquet}, Sebastian and {Saro}, Alex and {Dolag}, Klaus and {Mohr}, Joseph J.},
        title = "{Halo mass function: baryon impact, fitting formulae, and implications for cluster cosmology}",
      journal = {\mnras},
         year = 2016,
        month = mar,
       volume = {456},
       number = {3},
        pages = {2361-2373},
          doi = {10.1093/mnras/stv2657},
archivePrefix = {arXiv},
       eprint = {1502.07357},
 primaryClass = {astro-ph.CO},
       adsurl = {https://ui.adsabs.harvard.edu/abs/2016MNRAS.456.2361B}
}

@article{Brunner2022,
	adsurl = {https://ui.adsabs.harvard.edu/abs/2022A&A...661A...1B},
	archiveprefix = {arXiv},
	author = {{Brunner}, H. and {Liu}, T. and {Lamer}, G. and {Georgakakis}, A. and {Merloni}, A. and {Brusa}, M. and {Bulbul}, E. and {Dennerl}, K. and {Friedrich}, S. and {Liu}, A. and {Maitra}, C. and {Nandra}, K. and {Ramos-Ceja}, M.~E. and {Sanders}, J.~S. and {Stewart}, I.~M. and {Boller}, T. and {Buchner}, J. and {Clerc}, N. and {Comparat}, J. and {Dwelly}, T. and {Eckert}, D. and {Finoguenov}, A. and {Freyberg}, M. and {Ghirardini}, V. and {Gueguen}, A. and {Haberl}, F. and {Kreykenbohm}, I. and {Krumpe}, M. and {Osterhage}, S. and {Pacaud}, F. and {Predehl}, P. and {Reiprich}, T.~H. and {Robrade}, J. and {Salvato}, M. and {Santangelo}, A. and {Schrabback}, T. and {Schwope}, A. and {Wilms}, J.},
	doi = {10.1051/0004-6361/202141266},
	eid = {A1},
	eprint = {2106.14517},
	journal = {\aap},
	month = may,
	pages = {A1},
	primaryclass = {astro-ph.HE},
	title = {{The eROSITA Final Equatorial Depth Survey (eFEDS). X-ray catalogue}},
	volume = {661},
	year = 2022}

@ARTICLE{Bulbul2010,
       author = {{Bulbul}, G. Esra and {Hasler}, Nicole and {Bonamente}, Massimiliano and {Joy}, Marshall},
        title = "{An Analytic Model of the Physical Properties of Galaxy Clusters}",
      journal = {\apj},
         year = 2010,
        month = sep,
       volume = {720},
       number = {2},
        pages = {1038-1044},
          doi = {10.1088/0004-637X/720/2/1038},
archivePrefix = {arXiv},
       eprint = {0911.2827},
 primaryClass = {astro-ph.CO},
       adsurl = {https://ui.adsabs.harvard.edu/abs/2010ApJ...720.1038B}
}

@ARTICLE{Bulbul2012,
       author = {{Bulbul}, G. Esra and {Smith}, Randall K. and {Foster}, Adam and {Cottam}, Jean and {Loewenstein}, Michael and {Mushotzky}, Richard and {Shafer}, Richard},
        title = "{High-resolution XMM-Newton Spectroscopy of the Cooling Flow Cluster A3112}",
      journal = {\apj},
         year = 2012,
        month = mar,
       volume = {747},
       number = {1},
          eid = {32},
        pages = {32},
          doi = {10.1088/0004-637X/747/1/32},
archivePrefix = {arXiv},
       eprint = {1110.4422},
 primaryClass = {astro-ph.CO},
       adsurl = {https://ui.adsabs.harvard.edu/abs/2012ApJ...747...32B}
}

@ARTICLE{Bulbul2016,
       author = {{Bulbul}, Esra and {Randall}, Scott W. and {Bayliss}, Matthew and {Miller}, Eric and {Andrade-Santos}, Felipe and {Johnson}, Ryan and {Bautz}, Mark and {Blanton}, Elizabeth L. and {Forman}, William R. and {Jones}, Christine and {Paterno-Mahler}, Rachel and {Murray}, Stephen S. and {Sarazin}, Craig L. and {Smith}, Randall K. and {Ezer}, Cemile},
        title = "{Probing the Outskirts of the Early-Stage Galaxy Cluster Merger A1750}",
      journal = {\apj},
         year = 2016,
        month = feb,
       volume = {818},
       number = {2},
          eid = {131},
        pages = {131},
          doi = {10.3847/0004-637X/818/2/131},
archivePrefix = {arXiv},
       eprint = {1510.00017},
 primaryClass = {astro-ph.CO},
       adsurl = {https://ui.adsabs.harvard.edu/abs/2016ApJ...818..131B}
}

@ARTICLE{Bulbul2019,
       author = {{Bulbul}, Esra and {Chiu}, I. -Non and {Mohr}, Joseph J. and {McDonald}, Michael and {Benson}, Bradford and {Bautz}, Mark W. and {Bayliss}, Matthew and {Bleem}, Lindsey and {Brodwin}, Mark and {Bocquet}, Sebastian and {Capasso}, Raffaella and {Dietrich}, J{\"o}rg P. and {Forman}, Bill and {Hlavacek-Larrondo}, Julie and {Holzapfel}, W.~L. and {Khullar}, Gourav and {Klein}, Matthias and {Kraft}, Ralph and {Miller}, Eric D. and {Reichardt}, Christian and {Saro}, Alex and {Sharon}, Keren and {Stalder}, Brian and {Schrabback}, Tim and {Stanford}, Adam},
        title = "{X-Ray Properties of SPT-selected Galaxy Clusters at 0.2 < z < 1.5 Observed with XMM-Newton}",
      journal = {\apj},
         year = 2019,
        month = jan,
       volume = {871},
       number = {1},
          eid = {50},
        pages = {50},
          doi = {10.3847/1538-4357/aaf230},
archivePrefix = {arXiv},
       eprint = {1807.02556},
 primaryClass = {astro-ph.CO},
       adsurl = {https://ui.adsabs.harvard.edu/abs/2019ApJ...871...50B}
}

@ARTICLE{Bulbul2024,
       author = {{Bulbul}, E. and {Liu}, A. and {Kluge}, M. and {Zhang}, X. and {Sanders}, J.~S. and {Bahar}, Y.~E. and {Ghirardini}, V. and {Artis}, E. and {Seppi}, R. and {Garrel}, C. and {Ramos-Ceja}, M.~E. and {Comparat}, J. and {Balzer}, F. and {B{\"o}ckmann}, K. and {Br{\"u}ggen}, M. and {Clerc}, N. and {Dennerl}, K. and {Dolag}, K. and {Freyberg}, M. and {Grandis}, S. and {Gruen}, D. and {Kleinebreil}, F. and {Krippendorf}, S. and {Lamer}, G. and {Merloni}, A. and {Migkas}, K. and {Nandra}, K. and {Pacaud}, F. and {Predehl}, P. and {Reiprich}, T.~H. and {Schrabback}, T. and {Veronica}, A. and {Weller}, J. and {Zelmer}, S.},
        title = "{The SRG/eROSITA All-Sky Survey. The first catalog of galaxy clusters and groups in the Western Galactic Hemisphere}",
      journal = {\aap},
         year = 2024,
        month = may,
       volume = {685},
          eid = {A106},
        pages = {A106},
          doi = {10.1051/0004-6361/202348264},
archivePrefix = {arXiv},
       eprint = {2402.08452},
 primaryClass = {astro-ph.CO},
       adsurl = {https://ui.adsabs.harvard.edu/abs/2024A&A...685A.106B}
}

@ARTICLE{Cen1999,
       author = {{Cen}, Renyue and {Ostriker}, Jeremiah P.},
        title = "{Where Are the Baryons?}",
      journal = {\apj},
         year = 1999,
        month = mar,
       volume = {514},
       number = {1},
        pages = {1-6},
          doi = {10.1086/306949},
archivePrefix = {arXiv},
       eprint = {astro-ph/9806281},
 primaryClass = {astro-ph},
       adsurl = {https://ui.adsabs.harvard.edu/abs/1999ApJ...514....1C}
}

@ARTICLE{Chang2018,
       author = {{Chang}, C. and {Baxter}, E. and {Jain}, B. and {S{\'a}nchez}, C. and {Adhikari}, S. and {Varga}, T.~N. and {Fang}, Y. and {Rozo}, E. and {Rykoff}, E.~S. and {Kravtsov}, A. and {Gruen}, D. and {Hartley}, W. and {Huff}, E.~M. and {Jarvis}, M. and {Kim}, A.~G. and {Prat}, J. and {MacCrann}, N. and {McClintock}, T. and {Palmese}, A. and {Rapetti}, D. and {Rollins}, R.~P. and {Samuroff}, S. and {Sheldon}, E. and {Troxel}, M.~A. and {Wechsler}, R.~H. and {Zhang}, Y. and {Zuntz}, J. and {Abbott}, T.~M.~C. and {Abdalla}, F.~B. and {Allam}, S. and {Annis}, J. and {Bechtol}, K. and {Benoit-L{\'e}vy}, A. and {Bernstein}, G.~M. and {Brooks}, D. and {Buckley-Geer}, E. and {Carnero Rosell}, A. and {Carrasco Kind}, M. and {Carretero}, J. and {D'Andrea}, C.~B. and {da Costa}, L.~N. and {Davis}, C. and {Desai}, S. and {Diehl}, H.~T. and {Dietrich}, J.~P. and {Drlica-Wagner}, A. and {Eifler}, T.~F. and {Flaugher}, B. and {Fosalba}, P. and {Frieman}, J. and {Garc{\'\i}a-Bellido}, J. and {Gaztanaga}, E. and {Gerdes}, D.~W. and {Gruendl}, R.~A. and {Gschwend}, J. and {Gutierrez}, G. and {Honscheid}, K. and {James}, D.~J. and {Jeltema}, T. and {Krause}, E. and {Kuehn}, K. and {Lahav}, O. and {Lima}, M. and {March}, M. and {Marshall}, J.~L. and {Martini}, P. and {Melchior}, P. and {Menanteau}, F. and {Miquel}, R. and {Mohr}, J.~J. and {Nord}, B. and {Ogando}, R.~L.~C. and {Plazas}, A.~A. and {Sanchez}, E. and {Scarpine}, V. and {Schindler}, R. and {Schubnell}, M. and {Sevilla-Noarbe}, I. and {Smith}, M. and {Smith}, R.~C. and {Soares-Santos}, M. and {Sobreira}, F. and {Suchyta}, E. and {Swanson}, M.~E.~C. and {Tarle}, G. and {Weller}, J. and {DES Collaboration}},
        title = "{The Splashback Feature around DES Galaxy Clusters: Galaxy Density and Weak Lensing Profiles}",
      journal = {\apj},
         year = 2018,
        month = sep,
       volume = {864},
       number = {1},
          eid = {83},
        pages = {83},
          doi = {10.3847/1538-4357/aad5e7},
archivePrefix = {arXiv},
       eprint = {1710.06808},
 primaryClass = {astro-ph.CO},
       adsurl = {https://ui.adsabs.harvard.edu/abs/2018ApJ...864...83C}
}

@ARTICLE{CHEX-MATE2021,
       author = {{CHEX-MATE Collaboration} and {Arnaud}, M. and {Ettori}, S. and {Pratt}, G.~W. and {Rossetti}, M. and {Eckert}, D. and {Gastaldello}, F. and {Gavazzi}, R. and {Kay}, S.~T. and {Lovisari}, L. and {Maughan}, B.~J. and {Pointecouteau}, E. and {Sereno}, M. and {Bartalucci}, I. and {Bonafede}, A. and {Bourdin}, H. and {Cassano}, R. and {Duffy}, R.~T. and {Iqbal}, A. and {Maurogordato}, S. and {Rasia}, E. and {Sayers}, J. and {Andrade-Santos}, F. and {Aussel}, H. and {Barnes}, D.~J. and {Barrena}, R. and {Borgani}, S. and {Burkutean}, S. and {Clerc}, N. and {Corasaniti}, P. -S. and {Cuillandre}, J. -C. and {De Grandi}, S. and {De Petris}, M. and {Dolag}, K. and {Donahue}, M. and {Ferragamo}, A. and {Gaspari}, M. and {Ghizzardi}, S. and {Gitti}, M. and {Haines}, C.~P. and {Jauzac}, M. and {Johnston-Hollitt}, M. and {Jones}, C. and {K{\'e}ruzor{\'e}}, F. and {Le Brun}, A.~M.~C. and {Mayet}, F. and {Mazzotta}, P. and {Melin}, J. -B. and {Molendi}, S. and {Nonino}, M. and {Okabe}, N. and {Paltani}, S. and {Perotto}, L. and {Pires}, S. and {Radovich}, M. and {Rubino-Martin}, J. -A. and {Salvati}, L. and {Saro}, A. and {Sartoris}, B. and {Schellenberger}, G. and {Streblyanska}, A. and {Tarr{\'\i}o}, P. and {Tozzi}, P. and {Umetsu}, K. and {van der Burg}, R.~F.~J. and {Vazza}, F. and {Venturi}, T. and {Yepes}, G. and {Zarattini}, S.},
        title = "{The Cluster HEritage project with XMM-Newton: Mass Assembly and Thermodynamics at the Endpoint of structure formation. I. Programme overview}",
      journal = {\aap},
         year = 2021,
        month = jun,
       volume = {650},
          eid = {A104},
        pages = {A104},
          doi = {10.1051/0004-6361/202039632},
archivePrefix = {arXiv},
       eprint = {2010.11972},
 primaryClass = {astro-ph.CO},
       adsurl = {https://ui.adsabs.harvard.edu/abs/2021A&A...650A.104C}
}

@ARTICLE{Child2018,
       author = {{Child}, Hillary L. and {Habib}, Salman and {Heitmann}, Katrin and {Frontiere}, Nicholas and {Finkel}, Hal and {Pope}, Adrian and {Morozov}, Vitali},
        title = "{Halo Profiles and the Concentration-Mass Relation for a {\ensuremath{\Lambda}}CDM Universe}",
      journal = {\apj},
         year = 2018,
        month = may,
       volume = {859},
       number = {1},
          eid = {55},
        pages = {55},
          doi = {10.3847/1538-4357/aabf95},
archivePrefix = {arXiv},
       eprint = {1804.10199},
 primaryClass = {astro-ph.CO},
       adsurl = {https://ui.adsabs.harvard.edu/abs/2018ApJ...859...55C}
}

@ARTICLE{Chiu2022,
       author = {{Chiu}, I. -Non and {Ghirardini}, Vittorio and {Liu}, Ang and {Grandis}, Sebastian and {Bulbul}, Esra and {Bahar}, Y. Emre and {Comparat}, Johan and {Bocquet}, Sebastian and {Clerc}, Nicolas and {Klein}, Matthias and {Liu}, Teng and {Li}, Xiangchong and {Miyatake}, Hironao and {Mohr}, Joseph and {More}, Surhud and {Oguri}, Masamune and {Okabe}, Nobuhiro and {Pacaud}, Florian and {Ramos-Ceja}, Miriam E. and {Reiprich}, Thomas H. and {Schrabback}, Tim and {Umetsu}, Keiichi},
        title = "{The eROSITA Final Equatorial-Depth Survey (eFEDS). X-ray observable-to-mass-and-redshift relations of galaxy clusters and groups with weak-lensing mass calibration from the Hyper Suprime-Cam Subaru Strategic Program survey}",
      journal = {\aap},
         year = 2022,
        month = may,
       volume = {661},
          eid = {A11},
        pages = {A11},
          doi = {10.1051/0004-6361/202141755},
archivePrefix = {arXiv},
       eprint = {2107.05652},
 primaryClass = {astro-ph.CO},
       adsurl = {https://ui.adsabs.harvard.edu/abs/2022A&A...661A..11C}
}

@ARTICLE{Churazov2023,
       author = {{Churazov}, E. and {Khabibullin}, I. and {Bykov}, A.~M. and {Lyskova}, N. and {Sunyaev}, R.},
        title = "{Tempestuous life beyond R$_{500}$: X-ray view on the Coma cluster with SRG/eROSITA. II. Shock and relic}",
      journal = {\aap},
         year = 2023,
        month = feb,
       volume = {670},
          eid = {A156},
        pages = {A156},
          doi = {10.1051/0004-6361/202244021},
archivePrefix = {arXiv},
       eprint = {2205.07511},
 primaryClass = {astro-ph.CO},
       adsurl = {https://ui.adsabs.harvard.edu/abs/2023A&A...670A.156C}
}

@ARTICLE{Clerc2024,
       author = {{Clerc}, N. and {Comparat}, J. and {Seppi}, R. and {Artis}, E. and {Bahar}, Y.~E. and {Balzer}, F. and {Bulbul}, E. and {Dauser}, T. and {Garrel}, C. and {Ghirardini}, V. and {Grandis}, S. and {Kirsch}, C. and {Kluge}, M. and {Liu}, A. and {Pacaud}, F. and {Ramos-Ceja}, M.~E. and {Reiprich}, T.~H. and {Sanders}, J. and {Wilms}, J. and {Zhang}, X.},
        title = "{The SRG/eROSITA All-Sky Survey. X-ray selection function models for the eRASS1 galaxy cluster cosmology}",
      journal = {\aap},
         year = 2024,
        month = jul,
       volume = {687},
          eid = {A238},
        pages = {A238},
          doi = {10.1051/0004-6361/202449447},
archivePrefix = {arXiv},
       eprint = {2402.08457},
 primaryClass = {astro-ph.CO},
       adsurl = {https://ui.adsabs.harvard.edu/abs/2024A&A...687A.238C}
}

@ARTICLE{Comparat2017,
       author = {{Comparat}, Johan and {Prada}, Francisco and {Yepes}, Gustavo and {Klypin}, Anatoly},
        title = "{Accurate mass and velocity functions of dark matter haloes}",
      journal = {\mnras},
         year = 2017,
        month = aug,
       volume = {469},
       number = {4},
        pages = {4157-4174},
          doi = {10.1093/mnras/stx1183},
archivePrefix = {arXiv},
       eprint = {1702.01628},
 primaryClass = {astro-ph.CO},
       adsurl = {https://ui.adsabs.harvard.edu/abs/2017MNRAS.469.4157C}
}

@ARTICLE{Comparat2025,
       author = {{Comparat}, Johan and {Merloni}, Andrea and {Ponti}, Gabriele and {Shreeram}, Soumya and {Zhang}, Yi and {Reiprich}, Thomas H. and {Liu}, Ang and {Seppi}, Riccardo and {Zhang}, Xiaoyuan and {Clerc}, Nicolas and {Nicola}, Andrina and {Nandra}, Kirpal and {Salvato}, Mara and {Malavasi}, Nicola},
        title = "{Cross-correlation between soft X-rays and galaxies: A new benchmark for galaxy evolution models}",
      journal = {\aap},
         year = 2025,
        month = may,
       volume = {697},
          eid = {A173},
        pages = {A173},
          doi = {10.1051/0004-6361/202554208},
archivePrefix = {arXiv},
       eprint = {2503.19796},
 primaryClass = {astro-ph.GA},
       adsurl = {https://ui.adsabs.harvard.edu/abs/2025A&A...697A.173C}
}

@ARTICLE{Cooray2002,
       author = {{Cooray}, Asantha and {Sheth}, Ravi},
        title = "{Halo models of large scale structure}",
      journal = {\physrep},
         year = 2002,
        month = dec,
       volume = {372},
       number = {1},
        pages = {1-129},
          doi = {10.1016/S0370-1573(02)00276-4},
archivePrefix = {arXiv},
       eprint = {astro-ph/0206508},
 primaryClass = {astro-ph},
       adsurl = {https://ui.adsabs.harvard.edu/abs/2002PhR...372....1C}
}

@ARTICLE{Cruise2025,
       author = {{Cruise}, Mike and {Guainazzi}, Matteo and {Aird}, James and {Carrera}, Francisco J. and {Costantini}, Elisa and {Corrales}, Lia and {Dauser}, Thomas and {Eckert}, Dominique and {Gastaldello}, Fabio and {Matsumoto}, Hironori and {Osten}, Rachel and {Petrucci}, Pierre-Olivier and {Porquet}, Delphine and {Pratt}, Gabriel W. and {Rea}, Nanda and {Reiprich}, Thomas H. and {Simionescu}, Aurora and {Spiga}, Daniele and {Troja}, Eleonora},
        title = "{The NewAthena mission concept in the context of the next decade of X-ray astronomy}",
      journal = {Nature Astronomy},
         year = 2025,
        month = jan,
       volume = {9},
        pages = {36-44},
          doi = {10.1038/s41550-024-02416-3},
archivePrefix = {arXiv},
       eprint = {2501.03100},
 primaryClass = {astro-ph.IM},
       adsurl = {https://ui.adsabs.harvard.edu/abs/2025NatAs...9...36C}
}

@ARTICLE{Despali2016,
       author = {{Despali}, Giulia and {Giocoli}, Carlo and {Angulo}, Raul E. and {Tormen}, Giuseppe and {Sheth}, Ravi K. and {Baso}, Giacomo and {Moscardini}, Lauro},
        title = "{The universality of the virial halo mass function and models for non-universality of other halo definitions}",
      journal = {\mnras},
         year = 2016,
        month = mar,
       volume = {456},
       number = {3},
        pages = {2486-2504},
          doi = {10.1093/mnras/stv2842},
archivePrefix = {arXiv},
       eprint = {1507.05627},
 primaryClass = {astro-ph.CO},
       adsurl = {https://ui.adsabs.harvard.edu/abs/2016MNRAS.456.2486D}
}

@ARTICLE{Dey2019,
       author = {{Dey}, Arjun and {Schlegel}, David J. and {Lang}, Dustin and {Blum}, Robert and {Burleigh}, Kaylan and {Fan}, Xiaohui and {Findlay}, Joseph R. and {Finkbeiner}, Doug and {Herrera}, David and {Juneau}, St{\'e}phanie and {Landriau}, Martin and {Levi}, Michael and {McGreer}, Ian and {Meisner}, Aaron and {Myers}, Adam D. and {Moustakas}, John and {Nugent}, Peter and {Patej}, Anna and {Schlafly}, Edward F. and {Walker}, Alistair R. and {Valdes}, Francisco and {Weaver}, Benjamin A. and {Y{\`e}che}, Christophe and {Zou}, Hu and {Zhou}, Xu and {Abareshi}, Behzad and {Abbott}, T.~M.~C. and {Abolfathi}, Bela and {Aguilera}, C. and {Alam}, Shadab and {Allen}, Lori and {Alvarez}, A. and {Annis}, James and {Ansarinejad}, Behzad and {Aubert}, Marie and {Beechert}, Jacqueline and {Bell}, Eric F. and {BenZvi}, Segev Y. and {Beutler}, Florian and {Bielby}, Richard M. and {Bolton}, Adam S. and {Brice{\~n}o}, C{\'e}sar and {Buckley-Geer}, Elizabeth J. and {Butler}, Karen and {Calamida}, Annalisa and {Carlberg}, Raymond G. and {Carter}, Paul and {Casas}, Ricard and {Castander}, Francisco J. and {Choi}, Yumi and {Comparat}, Johan and {Cukanovaite}, Elena and {Delubac}, Timoth{\'e}e and {DeVries}, Kaitlin and {Dey}, Sharmila and {Dhungana}, Govinda and {Dickinson}, Mark and {Ding}, Zhejie and {Donaldson}, John B. and {Duan}, Yutong and {Duckworth}, Christopher J. and {Eftekharzadeh}, Sarah and {Eisenstein}, Daniel J. and {Etourneau}, Thomas and {Fagrelius}, Parker A. and {Farihi}, Jay and {Fitzpatrick}, Mike and {Font-Ribera}, Andreu and {Fulmer}, Leah and {G{\"a}nsicke}, Boris T. and {Gaztanaga}, Enrique and {George}, Koshy and {Gerdes}, David W. and {Gontcho}, Satya Gontcho A. and {Gorgoni}, Claudio and {Green}, Gregory and {Guy}, Julien and {Harmer}, Diane and {Hernandez}, M. and {Honscheid}, Klaus and {Huang}, Lijuan Wendy and {James}, David J. and {Jannuzi}, Buell T. and {Jiang}, Linhua and {Joyce}, Richard and {Karcher}, Armin and {Karkar}, Sonia and {Kehoe}, Robert and {Kneib}, Jean-Paul and {Kueter-Young}, Andrea and {Lan}, Ting-Wen and {Lauer}, Tod R. and {Le Guillou}, Laurent and {Le Van Suu}, Auguste and {Lee}, Jae Hyeon and {Lesser}, Michael and {Perreault Levasseur}, Laurence and {Li}, Ting S. and {Mann}, Justin L. and {Marshall}, Robert and {Mart{\'\i}nez-V{\'a}zquez}, C.~E. and {Martini}, Paul and {du Mas des Bourboux}, H{\'e}lion and {McManus}, Sean and {Meier}, Tobias Gabriel and {M{\'e}nard}, Brice and {Metcalfe}, Nigel and {Mu{\~n}oz-Guti{\'e}rrez}, Andrea and {Najita}, Joan and {Napier}, Kevin and {Narayan}, Gautham and {Newman}, Jeffrey A. and {Nie}, Jundan and {Nord}, Brian and {Norman}, Dara J. and {Olsen}, Knut A.~G. and {Paat}, Anthony and {Palanque-Delabrouille}, Nathalie and {Peng}, Xiyan and {Poppett}, Claire L. and {Poremba}, Megan R. and {Prakash}, Abhishek and {Rabinowitz}, David and {Raichoor}, Anand and {Rezaie}, Mehdi and {Robertson}, A.~N. and {Roe}, Natalie A. and {Ross}, Ashley J. and {Ross}, Nicholas P. and {Rudnick}, Gregory and {Safonova}, Sasha and {Saha}, Abhijit and {S{\'a}nchez}, F. Javier and {Savary}, Elodie and {Schweiker}, Heidi and {Scott}, Adam and {Seo}, Hee-Jong and {Shan}, Huanyuan and {Silva}, David R. and {Slepian}, Zachary and {Soto}, Christian and {Sprayberry}, David and {Staten}, Ryan and {Stillman}, Coley M. and {Stupak}, Robert J. and {Summers}, David L. and {Sien Tie}, Suk and {Tirado}, H. and {Vargas-Maga{\~n}a}, Mariana and {Vivas}, A. Katherina and {Wechsler}, Risa H. and {Williams}, Doug and {Yang}, Jinyi and {Yang}, Qian and {Yapici}, Tolga and {Zaritsky}, Dennis and {Zenteno}, A. and {Zhang}, Kai and {Zhang}, Tianmeng and {Zhou}, Rongpu and {Zhou}, Zhimin},
        title = "{Overview of the DESI Legacy Imaging Surveys}",
      journal = {\aj},
         year = 2019,
        month = may,
       volume = {157},
       number = {5},
          eid = {168},
        pages = {168},
          doi = {10.3847/1538-3881/ab089d},
archivePrefix = {arXiv},
       eprint = {1804.08657},
 primaryClass = {astro-ph.IM},
       adsurl = {https://ui.adsabs.harvard.edu/abs/2019AJ....157..168D}
}

@ARTICLE{Diemand2008,
   author = {{Diemand}, J. and {Kuhlen}, M.},
    title = "{Infall Caustics in Dark Matter Halos?}",
  journal = {\apjl},
archivePrefix = "arXiv",
   eprint = {0804.4185},
     year = 2008,
    month = jun,
   volume = 680,
    pages = {L25-L28},
      doi = {10.1086/589688},
   adsurl = {http://adsabs.harvard.edu/abs/2008ApJ...680L..25D}
}

@ARTICLE{Diemer2014,
       author = {{Diemer}, Benedikt and {Kravtsov}, Andrey V.},
        title = "{Dependence of the Outer Density Profiles of Halos on Their Mass Accretion Rate}",
      journal = {\apj},
         year = 2014,
        month = jul,
       volume = {789},
       number = {1},
          eid = {1},
        pages = {1},
          doi = {10.1088/0004-637X/789/1/1},
archivePrefix = {arXiv},
       eprint = {1401.1216},
 primaryClass = {astro-ph.CO},
       adsurl = {https://ui.adsabs.harvard.edu/abs/2014ApJ...789....1D}
}

@ARTICLE{Diemer2017-2,
       author = {{Diemer}, Benedikt and {Mansfield}, Philip and {Kravtsov}, Andrey V. and {More}, Surhud},
        title = "{The Splashback Radius of Halos from Particle Dynamics. II. Dependence on Mass, Accretion Rate, Redshift, and Cosmology}",
      journal = {\apj},
         year = 2017,
        month = jul,
       volume = {843},
       number = {2},
          eid = {140},
        pages = {140},
          doi = {10.3847/1538-4357/aa79ab},
archivePrefix = {arXiv},
       eprint = {1703.09716},
 primaryClass = {astro-ph.CO},
       adsurl = {https://ui.adsabs.harvard.edu/abs/2017ApJ...843..140D}
}

@ARTICLE{Diemer2017-SFH,
   author = {{Diemer}, B. and {Sparre}, M. and {Abramson}, L.~E. and {Torrey}, P.
	},
    title = "{Log-normal Star Formation Histories in Simulated and Observed Galaxies}",
  journal = {\apj},
archivePrefix = "arXiv",
   eprint = {1701.02308},
     year = 2017,
    month = apr,
   volume = 839,
      eid = {26},
    pages = {26},
      doi = {10.3847/1538-4357/aa68e5},
   adsurl = {http://adsabs.harvard.edu/abs/2017ApJ...839...26D}
}

@ARTICLE{Diemer2018-colossus,
       author = {{Diemer}, Benedikt},
        title = "{COLOSSUS: A Python Toolkit for Cosmology, Large-scale Structure, and Dark Matter Halos}",
      journal = {\apjs},
         year = 2018,
        month = dec,
       volume = {239},
       number = {2},
          eid = {35},
        pages = {35},
          doi = {10.3847/1538-4365/aaee8c},
archivePrefix = {arXiv},
       eprint = {1712.04512},
 primaryClass = {astro-ph.CO},
       adsurl = {https://ui.adsabs.harvard.edu/abs/2018ApJS..239...35D}
}

@ARTICLE{Diemer2018-HIHII,
       author = {{Diemer}, Benedikt and {Stevens}, Adam R.~H. and {Forbes}, John C. and
        {Marinacci}, Federico and {Hernquist}, Lars and {Lagos}, Claudia
        del P. and {Sternberg}, Amiel and {Pillepich}, Annalisa and
        {Nelson}, Dylan and {Popping}, Gerg{\"o} and {Villaescusa-
        Navarro}, Francisco and {Torrey}, Paul and {Vogelsberger}, Mark},
        title = "{Modeling the Atomic-to-molecular Transition in Cosmological Simulations
        of Galaxy Formation}",
      journal = {The Astrophysical Journal Supplement Series},
         year = 2018,
        month = Oct,
       volume = {238},
          eid = {33},
        pages = {33},
          doi = {10.3847/1538-4365/aae387},
 primaryClass = {astro-ph.GA},
       adsurl = {https://ui.adsabs.harvard.edu/#abs/2018ApJS..238...33D}
}

@ARTICLE{Diemer2019,
       author = {{Diemer}, Benedikt and {Joyce}, Michael},
        title = "{An Accurate Physical Model for Halo Concentrations}",
      journal = {\apj},
         year = 2019,
        month = feb,
       volume = {871},
       number = {2},
          eid = {168},
        pages = {168},
          doi = {10.3847/1538-4357/aafad6},
archivePrefix = {arXiv},
       eprint = {1809.07326},
 primaryClass = {astro-ph.CO},
       adsurl = {https://ui.adsabs.harvard.edu/abs/2019ApJ...871..168D}
}

@ARTICLE{Diemer2022,
       author = {{Diemer}, Benedikt},
        title = "{A dynamics-based density profile for dark haloes - I. Algorithm and basic results}",
      journal = {\mnras},
         year = 2022,
        month = jun,
       volume = {513},
       number = {1},
        pages = {573-594},
          doi = {10.1093/mnras/stac878},
archivePrefix = {arXiv},
       eprint = {2112.03921},
 primaryClass = {astro-ph.CO},
       adsurl = {https://ui.adsabs.harvard.edu/abs/2022MNRAS.513..573D}
}

@ARTICLE{Eckert2013,
       author = {{Eckert}, D. and {Molendi}, S. and {Vazza}, F. and {Ettori}, S. and {Paltani}, S.},
        title = "{The X-ray/SZ view of the virial region. I. Thermodynamic properties}",
      journal = {\aap},
         year = 2013,
        month = mar,
       volume = {551},
          eid = {A22},
        pages = {A22},
          doi = {10.1051/0004-6361/201220402},
archivePrefix = {arXiv},
       eprint = {1301.0617},
 primaryClass = {astro-ph.CO},
       adsurl = {https://ui.adsabs.harvard.edu/abs/2013A&A...551A..22E}
}

@ARTICLE{Eckert2015,
       author = {{Eckert}, D. and {Roncarelli}, M. and {Ettori}, S. and {Molendi}, S. and {Vazza}, F. and {Gastaldello}, F. and {Rossetti}, M.},
        title = "{Gas clumping in galaxy clusters}",
      journal = {\mnras},
         year = 2015,
        month = mar,
       volume = {447},
       number = {3},
        pages = {2198-2208},
          doi = {10.1093/mnras/stu2590},
archivePrefix = {arXiv},
       eprint = {1310.8389},
 primaryClass = {astro-ph.CO},
       adsurl = {https://ui.adsabs.harvard.edu/abs/2015MNRAS.447.2198E}
}

@ARTICLE{Eckert2017,
       author = {{Eckert}, D. and {Ettori}, S. and {Pointecouteau}, E. and {Molendi}, S. and {Paltani}, S. and {Tchernin}, C.},
        title = "{The XMM cluster outskirts project (X‑ COP )}",
      journal = {Astronomische Nachrichten},
         year = 2017,
        month = mar,
       volume = {338},
       number = {293},
        pages = {293-298},
          doi = {10.1002/asna.201713345},
archivePrefix = {arXiv},
       eprint = {1611.05051},
 primaryClass = {astro-ph.CO},
       adsurl = {https://ui.adsabs.harvard.edu/abs/2017AN....338..293E}
}

@ARTICLE{Eckert2019,
       author = {{Eckert}, D. and {Ghirardini}, V. and {Ettori}, S. and {Rasia}, E. and {Biffi}, V. and {Pointecouteau}, E. and {Rossetti}, M. and {Molendi}, S. and {Vazza}, F. and {Gastaldello}, F. and {Gaspari}, M. and {De Grandi}, S. and {Ghizzardi}, S. and {Bourdin}, H. and {Tchernin}, C. and {Roncarelli}, M.},
        title = "{Non-thermal pressure support in X-COP galaxy clusters}",
      journal = {\aap},
         year = 2019,
        month = jan,
       volume = {621},
          eid = {A40},
        pages = {A40},
          doi = {10.1051/0004-6361/201833324},
archivePrefix = {arXiv},
       eprint = {1805.00034},
 primaryClass = {astro-ph.CO},
       adsurl = {https://ui.adsabs.harvard.edu/abs/2019A&A...621A..40E}
}

@ARTICLE{Eckert2021,
       author = {{Eckert}, Dominique and {Gaspari}, Massimo and {Gastaldello}, Fabio and {Le Brun}, Amandine M.~C. and {O'Sullivan}, Ewan},
        title = "{Feedback from Active Galactic Nuclei in Galaxy Groups}",
      journal = {Universe},
         year = 2021,
        month = may,
       volume = {7},
       number = {5},
          eid = {142},
        pages = {142},
          doi = {10.3390/universe7050142},
archivePrefix = {arXiv},
       eprint = {2106.13259},
 primaryClass = {astro-ph.GA},
       adsurl = {https://ui.adsabs.harvard.edu/abs/2021Univ....7..142E}
}

@ARTICLE{Fillmore1984,
       author = {{Fillmore}, J.~A. and {Goldreich}, P.},
        title = "{Self-similar gravitational collapse in an expanding universe}",
      journal = {\apj},
         year = 1984,
        month = jun,
       volume = {281},
        pages = {1-8},
          doi = {10.1086/162070},
       adsurl = {https://ui.adsabs.harvard.edu/abs/1984ApJ...281....1F}
}

@ARTICLE{Foster2020,
       author = {{Foster}, Adam R. and {Heuer}, Keri},
        title = "{PyAtomDB: Extending the AtomDB Atomic Database to Model New Plasma Processes and Uncertainties}",
      journal = {Atoms},
         year = 2020,
        month = aug,
       volume = {8},
       number = {3},
        pages = {49},
          doi = {10.3390/atoms8030049},
       adsurl = {https://ui.adsabs.harvard.edu/abs/2020Atoms...8...49F}
}

@ARTICLE{Garcia2023,
       author = {{Garc{\'\i}a}, Rafael and {Salazar}, Edgar and {Rozo}, Eduardo and {Adhikari}, Susmita and {Aung}, Han and {Diemer}, Benedikt and {Nagai}, Daisuke and {Wolfe}, Brandon},
        title = "{A better way to define dark matter haloes}",
      journal = {\mnras},
         year = 2023,
        month = may,
       volume = {521},
       number = {2},
        pages = {2464-2476},
          doi = {10.1093/mnras/stad660},
archivePrefix = {arXiv},
       eprint = {2207.11827},
 primaryClass = {astro-ph.CO},
       adsurl = {https://ui.adsabs.harvard.edu/abs/2023MNRAS.521.2464G}
}

@ARTICLE{Ghirardini2019,
       author = {{Ghirardini}, V. and {Eckert}, D. and {Ettori}, S. and {Pointecouteau}, E. and {Molendi}, S. and {Gaspari}, M. and {Rossetti}, M. and {De Grandi}, S. and {Roncarelli}, M. and {Bourdin}, H. and {Mazzotta}, P. and {Rasia}, E. and {Vazza}, F.},
        title = "{Universal thermodynamic properties of the intracluster medium over two decades in radius in the X-COP sample}",
      journal = {\aap},
         year = 2019,
        month = jan,
       volume = {621},
          eid = {A41},
        pages = {A41},
          doi = {10.1051/0004-6361/201833325},
archivePrefix = {arXiv},
       eprint = {1805.00042},
 primaryClass = {astro-ph.CO},
       adsurl = {https://ui.adsabs.harvard.edu/abs/2019A&A...621A..41G}
}

@ARTICLE{Ghirardini2024,
       author = {{Ghirardini}, V. and {Bulbul}, E. and {Artis}, E. and {Clerc}, N. and {Garrel}, C. and {Grandis}, S. and {Kluge}, M. and {Liu}, A. and {Bahar}, Y.~E. and {Balzer}, F. and {Chiu}, I. and {Comparat}, J. and {Gruen}, D. and {Kleinebreil}, F. and {Krippendorf}, S. and {Merloni}, A. and {Nandra}, K. and {Okabe}, N. and {Pacaud}, F. and {Predehl}, P. and {Ramos-Ceja}, M.~E. and {Reiprich}, T.~H. and {Sanders}, J.~S. and {Schrabback}, T. and {Seppi}, R. and {Zelmer}, S. and {Zhang}, X. and {Bornemann}, W. and {Brunner}, H. and {Burwitz}, V. and {Coutinho}, D. and {Dennerl}, K. and {Freyberg}, M. and {Friedrich}, S. and {Gaida}, R. and {Gueguen}, A. and {Haberl}, F. and {Kink}, W. and {Lamer}, G. and {Li}, X. and {Liu}, T. and {Maitra}, C. and {Meidinger}, N. and {Mueller}, S. and {Miyatake}, H. and {Miyazaki}, S. and {Robrade}, J. and {Schwope}, A. and {Stewart}, I.},
        title = "{The SRG/eROSITA all-sky survey: Cosmology constraints from cluster abundances in the western Galactic hemisphere}",
      journal = {\aap},
         year = 2024,
        month = sep,
       volume = {689},
          eid = {A298},
        pages = {A298},
          doi = {10.1051/0004-6361/202348852},
archivePrefix = {arXiv},
       eprint = {2402.08458},
 primaryClass = {astro-ph.CO},
       adsurl = {https://ui.adsabs.harvard.edu/abs/2024A&A...689A.298G}
}

@ARTICLE{Gonzalez2013,
       author = {{Gonzalez}, Anthony H. and {Sivanandam}, Suresh and {Zabludoff}, Ann I. and {Zaritsky}, Dennis},
        title = "{Galaxy Cluster Baryon Fractions Revisited}",
      journal = {\apj},
         year = 2013,
        month = nov,
       volume = {778},
       number = {1},
          eid = {14},
        pages = {14},
          doi = {10.1088/0004-637X/778/1/14},
archivePrefix = {arXiv},
       eprint = {1309.3565},
 primaryClass = {astro-ph.CO},
       adsurl = {https://ui.adsabs.harvard.edu/abs/2013ApJ...778...14G}
}

@ARTICLE{Gouin2022,
       author = {{Gouin}, C. and {Gallo}, S. and {Aghanim}, N.},
        title = "{Gas distribution from clusters to filaments in IllustrisTNG}",
      journal = {\aap},
         year = 2022,
        month = aug,
       volume = {664},
          eid = {A198},
        pages = {A198},
          doi = {10.1051/0004-6361/202243032},
archivePrefix = {arXiv},
       eprint = {2201.00593},
 primaryClass = {astro-ph.CO},
       adsurl = {https://ui.adsabs.harvard.edu/abs/2022A&A...664A.198G}
}

@ARTICLE{Grandis2024,
       author = {{Grandis}, S. and {Ghirardini}, V. and {Bocquet}, S. and {Garrel}, C. and {Mohr}, J.~J. and {Liu}, A. and {Kluge}, M. and {Kimmig}, L. and {Reiprich}, T.~H. and {Alarcon}, A. and {Amon}, A. and {Artis}, E. and {Bahar}, Y.~E. and {Balzer}, F. and {Bechtol}, K. and {Becker}, M.~R. and {Bernstein}, G. and {Bulbul}, E. and {Campos}, A. and {Carnero Rosell}, A. and {Carrasco Kind}, M. and {Cawthon}, R. and {Chang}, C. and {Chen}, R. and {Chiu}, I. and {Choi}, A. and {Clerc}, N. and {Comparat}, J. and {Cordero}, J. and {Davis}, C. and {Derose}, J. and {Diehl}, H.~T. and {Dodelson}, S. and {Doux}, C. and {Drlica-Wagner}, A. and {Eckert}, K. and {Elvin-Poole}, J. and {Everett}, S. and {Ferte}, A. and {Gatti}, M. and {Giannini}, G. and {Giles}, P. and {Gruen}, D. and {Gruendl}, R.~A. and {Harrison}, I. and {Hartley}, W.~G. and {Herner}, K. and {Huff}, E.~M. and {Kleinebreil}, F. and {Kuropatkin}, N. and {Leget}, P.~F. and {Maccrann}, N. and {Mccullough}, J. and {Merloni}, A. and {Myles}, J. and {Nandra}, K. and {Navarro-Alsina}, A. and {Okabe}, N. and {Pacaud}, F. and {Pandey}, S. and {Prat}, J. and {Predehl}, P. and {Ramos}, M. and {Raveri}, M. and {Rollins}, R.~P. and {Roodman}, A. and {Ross}, A.~J. and {Rykoff}, E.~S. and {Sanchez}, C. and {Sanders}, J. and {Schrabback}, T. and {Secco}, L.~F. and {Seppi}, R. and {Sevilla-Noarbe}, I. and {Sheldon}, E. and {Shin}, T. and {Troxel}, M. and {Tutusaus}, I. and {Varga}, T.~N. and {Wu}, H. and {Yanny}, B. and {Yin}, B. and {Zhang}, X. and {Zhang}, Y. and {Alves}, O. and {Bhargava}, S. and {Brooks}, D. and {Burke}, D.~L. and {Carretero}, J. and {Costanzi}, M. and {da Costa}, L.~N. and {Pereira}, M.~E.~S. and {De Vicente}, J. and {Desai}, S. and {Doel}, P. and {Ferrero}, I. and {Flaugher}, B. and {Friedel}, D. and {Frieman}, J. and {Garc{\'\i}a-Bellido}, J. and {Gutierrez}, G. and {Hinton}, S.~R. and {Hollowood}, D.~L. and {Honscheid}, K. and {James}, D.~J. and {Jeffrey}, N. and {Lahav}, O. and {Lee}, S. and {Marshall}, J.~L. and {Menanteau}, F. and {Ogando}, R.~L.~C. and {Pieres}, A. and {Plazas Malag{\'o}n}, A.~A. and {Romer}, A.~K. and {Sanchez}, E. and {Schubnell}, M. and {Smith}, M. and {Suchyta}, E. and {Swanson}, M.~E.~C. and {Tarle}, G. and {Weaverdyck}, N. and {Weller}, J.},
        title = "{The SRG/eROSITA All-Sky Survey: Dark Energy Survey year 3 weak gravitational lensing by eRASS1 selected galaxy clusters}",
      journal = {\aap},
         year = 2024,
        month = jul,
       volume = {687},
          eid = {A178},
        pages = {A178},
          doi = {10.1051/0004-6361/202348615},
archivePrefix = {arXiv},
       eprint = {2402.08455},
 primaryClass = {astro-ph.CO},
       adsurl = {https://ui.adsabs.harvard.edu/abs/2024A&A...687A.178G}
}

@ARTICLE{Hadzhiyska2025,
       author = {{Hadzhiyska}, B. and {Ferraro}, S. and {Ried Guachalla}, B. and {Schaan}, E. and {Aguilar}, J. and {Ahlen}, S. and {Battaglia}, N. and {Bond}, J.~R. and {Brooks}, D. and {Calabrese}, E. and {Choi}, S.~K. and {Claybaugh}, T. and {Coulton}, W.~R. and {Dawson}, K. and {Devlin}, M. and {Dey}, B. and {Doel}, P. and {Duivenvoorden}, A.~J. and {Dunkley}, J. and {Farren}, G.~S. and {Font-Ribera}, A. and {Forero-Romero}, J.~E. and {Gallardo}, P.~A. and {Gazta{\~n}aga}, E. and {Gontcho Gontcho}, S. and {Gralla}, M. and {Le Guillou}, L. and {Gutierrez}, G. and {Guy}, J. and {Hill}, J.~C. and {Hlo{\v{z}}ek}, R. and {Honscheid}, K. and {Juneau}, S. and {Kehoe}, R. and {Kisner}, T. and {Kremin}, A. and {Landriau}, M. and {Liu}, R.~H. and {Louis}, T. and {MacCrann}, N. and {de Macorra}, A. and {Madhavacheril}, M. and {Manera}, M. and {Meisner}, A. and {Miquel}, R. and {Moodley}, K. and {Moustakas}, J. and {Mroczkowski}, T. and {Naess}, S. and {Newman}, J. and {Niemack}, M.~D. and {Niz}, G. and {Page}, L. and {Palanque-Delabrouille}, N. and {Partridge}, B. and {Percival}, W.~J. and {Prada}, F. and {Qu}, F.~J. and {Rossi}, G. and {Sanchez}, E. and {Schlegel}, D. and {Schubnell}, M. and {Sherwin}, B. and {Sehgal}, N. and {Seo}, H. and {Sif{\'o}n}, C. and {Spergel}, D. and {Sprayberry}, D. and {Staggs}, S. and {Tarl{\'e}}, G. and {Vargas}, C. and {Vavagiakis}, E.~M. and {Weaver}, B.~A. and {Wollack}, E.~J. and {Zhou}, R. and {Zou}, H.},
        title = "{Evidence for large baryonic feedback at low and intermediate redshifts from kinematic Sunyaev-Zel'dovich observations with ACT and DESI photometric galaxies}",
      journal = {\prd},
         year = 2025,
        month = oct,
       volume = {112},
       number = {8},
          eid = {083509},
        pages = {083509},
          doi = {10.1103/kclp-x5j1},
archivePrefix = {arXiv},
       eprint = {2407.07152},
 primaryClass = {astro-ph.CO},
       adsurl = {https://ui.adsabs.harvard.edu/abs/2025PhRvD.112h3509H}
}

@article{iminuit,
  author={{Dembinski}, Hans and {Ongmongkolkul}, Piti and et al.},
  title={scikit-hep/iminuit},
  DOI={10.5281/zenodo.3949207},
  publisher={Zenodo},
  year={2020},
  month={Dec},
  url={https://doi.org/10.5281/zenodo.3949207}
}

@ARTICLE{Ishiyama2021,
       author = {{Ishiyama}, Tomoaki and {Prada}, Francisco and {Klypin}, Anatoly A. and {Sinha}, Manodeep and {Metcalf}, R. Benton and {Jullo}, Eric and {Altieri}, Bruno and {Cora}, Sof{\'\i}a A. and {Croton}, Darren and {de la Torre}, Sylvain and {Mill{\'a}n-Calero}, David E. and {Oogi}, Taira and {Ruedas}, Jos{\'e} and {Vega-Mart{\'\i}nez}, Cristian A.},
        title = "{The Uchuu simulations: Data Release 1 and dark matter halo concentrations}",
      journal = {\mnras},
         year = 2021,
        month = sep,
       volume = {506},
       number = {3},
        pages = {4210-4231},
          doi = {10.1093/mnras/stab1755},
archivePrefix = {arXiv},
       eprint = {2007.14720},
 primaryClass = {astro-ph.CO},
       adsurl = {https://ui.adsabs.harvard.edu/abs/2021MNRAS.506.4210I}
}

@book{Jeffreys1961,
  title={Theory of Probability},
  author={Jeffreys, H.},
  lccn={62000074},
  series={International series of monographs on physics},
  url={https://books.google.de/books?id=AavQAAAAMAAJ},
  year={1961},
  publisher={Clarendon Press}
}

@ARTICLE{Karamanis2022_pocomc,
       author = {{Karamanis}, Minas and {Nabergoj}, David and {Beutler}, Florian and {Peacock}, John and {Seljak}, Uro{\v{s}}},
        title = "{pocoMC: A Python package for accelerated Bayesian inference in astronomy and cosmology}",
      journal = {The Journal of Open Source Software},
         year = 2022,
        month = nov,
       volume = {7},
       number = {79},
          eid = {4634},
        pages = {4634},
          doi = {10.21105/joss.04634},
archivePrefix = {arXiv},
       eprint = {2207.05660},
 primaryClass = {astro-ph.IM},
       adsurl = {https://ui.adsabs.harvard.edu/abs/2022JOSS....7.4634K}
}

@ARTICLE{Karamanis2022_method,
       author = {{Karamanis}, Minas and {Beutler}, Florian and {Peacock}, John A. and {Nabergoj}, David and {Seljak}, Uro{\v{s}}},
        title = "{Accelerating astronomical and cosmological inference with preconditioned Monte Carlo}",
      journal = {\mnras},
         year = 2022,
        month = oct,
       volume = {516},
       number = {2},
        pages = {1644-1653},
          doi = {10.1093/mnras/stac2272},
archivePrefix = {arXiv},
       eprint = {2207.05652},
 primaryClass = {astro-ph.IM},
       adsurl = {https://ui.adsabs.harvard.edu/abs/2022MNRAS.516.1644K}
}

@article{Kass1995,
  title={Bayes factors},
  author={Kass, Robert E and Raftery, Adrian E},
  journal={Journal of the american statistical association},
  volume={90},
  number={430},
  pages={773--795},
  year={1995},
  publisher={Taylor \& Francis}
}

@ARTICLE{Kleinebreil2025,
       author = {{Kleinebreil}, F. and {Grandis}, S. and {Schrabback}, T. and {Ghirardini}, V. and {Chiu}, I. and {Liu}, A. and {Kluge}, M. and {Reiprich}, T.~H. and {Artis}, E. and {Bahar}, Y.~E. and {Balzer}, F. and {Bulbul}, E. and {Clerc}, N. and {Comparat}, J. and {Garrel}, C. and {Gruen}, D. and {Li}, X. and {Miyatake}, H. and {Miyazaki}, S. and {Ramos-Ceja}, M.~E. and {Sanders}, J. and {Seppi}, R. and {Okabe}, N. and {Zhang}, X.},
        title = "{The SRG/eROSITA All-Sky Survey: Weak lensing of eRASS1 galaxy clusters in KiDS-1000 and consistency checks with DES Y3 and HSC-Y3}",
      journal = {\aap},
         year = 2025,
        month = mar,
       volume = {695},
          eid = {A216},
        pages = {A216},
          doi = {10.1051/0004-6361/202449599},
archivePrefix = {arXiv},
       eprint = {2402.08456},
 primaryClass = {astro-ph.CO},
       adsurl = {https://ui.adsabs.harvard.edu/abs/2025A&A...695A.216K}
}

@ARTICLE{Kluge2024,
       author = {{Kluge}, M. and {Comparat}, J. and {Liu}, A. and {Balzer}, F. and {Bulbul}, E. and {Ider Chitham}, J. and {Ghirardini}, V. and {Garrel}, C. and {Bahar}, Y.~E. and {Artis}, E. and {Bender}, R. and {Clerc}, N. and {Dwelly}, T. and {Fabricius}, M.~H. and {Grandis}, S. and {Hern{\'a}ndez-Lang}, D. and {Hill}, G.~J. and {Joshi}, J. and {Lamer}, G. and {Merloni}, A. and {Nandra}, K. and {Pacaud}, F. and {Predehl}, P. and {Ramos-Ceja}, M.~E. and {Reiprich}, T.~H. and {Salvato}, M. and {Sanders}, J.~S. and {Schrabback}, T. and {Seppi}, R. and {Zelmer}, S. and {Zenteno}, A. and {Zhang}, X.},
        title = "{The SRG/eROSITA All-Sky Survey. Optical identification and properties of galaxy clusters and groups in the western galactic hemisphere}",
      journal = {\aap},
         year = 2024,
        month = aug,
       volume = {688},
          eid = {A210},
        pages = {A210},
          doi = {10.1051/0004-6361/202349031},
archivePrefix = {arXiv},
       eprint = {2402.08453},
 primaryClass = {astro-ph.CO},
       adsurl = {https://ui.adsabs.harvard.edu/abs/2024A&A...688A.210K}
}

@ARTICLE{Kovacs2023,
       author = {{Kov{\'a}cs}, Orsolya Eszter and {Zhu}, Zhenlin and {Werner}, Norbert and {Simionescu}, Aurora and {Bogd{\'a}n}, {\'A}kos},
        title = "{Outskirts of Abell 1795: Probing gas clumping in the intracluster medium}",
      journal = {\aap},
         year = 2023,
        month = oct,
       volume = {678},
          eid = {A91},
        pages = {A91},
          doi = {10.1051/0004-6361/202347201},
archivePrefix = {arXiv},
       eprint = {2306.10101},
 primaryClass = {astro-ph.GA},
       adsurl = {https://ui.adsabs.harvard.edu/abs/2023A&A...678A..91K}
}

@ARTICLE{Lau2015,
       author = {{Lau}, Erwin T. and {Nagai}, Daisuke and {Avestruz}, Camille and {Nelson}, Kaylea and {Vikhlinin}, Alexey},
        title = "{Mass Accretion and its Effects on the Self-similarity of Gas Profiles in the Outskirts of Galaxy Clusters}",
      journal = {\apj},
         year = 2015,
        month = jun,
       volume = {806},
       number = {1},
          eid = {68},
        pages = {68},
          doi = {10.1088/0004-637X/806/1/68},
archivePrefix = {arXiv},
       eprint = {1411.5361},
 primaryClass = {astro-ph.CO},
       adsurl = {https://ui.adsabs.harvard.edu/abs/2015ApJ...806...68L}
}

@ARTICLE{Li2025,
       author = {{Li}, Renjie and {Cui}, Weiguang and {Liu}, Ang and {Wang}, Huiyuan and {Srivastava}, Atulit and {Dave}, Romeel and {Pearce}, Frazer R.},
        title = "{THE THREE HUNDRED project: Gas properties outside of galaxy clusters with the WHIM contribution and detection}",
      journal = {\aap},
         year = 2025,
        month = sep,
       volume = {701},
          eid = {A37},
        pages = {A37},
          doi = {10.1051/0004-6361/202554390},
archivePrefix = {arXiv},
       eprint = {2503.05011},
 primaryClass = {astro-ph.CO},
       adsurl = {https://ui.adsabs.harvard.edu/abs/2025A&A...701A..37L}
}

@ARTICLE{Liu2022,
       author = {{Liu}, A. and {Bulbul}, E. and {Ghirardini}, V. and {Liu}, T. and {Klein}, M. and {Clerc}, N. and {{\"O}zsoy}, Y. and {Ramos-Ceja}, M.~E. and {Pacaud}, F. and {Comparat}, J. and {Okabe}, N. and {Bahar}, Y.~E. and {Biffi}, V. and {Brunner}, H. and {Br{\"u}ggen}, M. and {Buchner}, J. and {Ider Chitham}, J. and {Chiu}, I. and {Dolag}, K. and {Gatuzz}, E. and {Gonzalez}, J. and {Hoang}, D.~N. and {Lamer}, G. and {Merloni}, A. and {Nandra}, K. and {Oguri}, M. and {Ota}, N. and {Predehl}, P. and {Reiprich}, T.~H. and {Salvato}, M. and {Schrabback}, T. and {Sanders}, J.~S. and {Seppi}, R. and {Thibaud}, Q.},
        title = "{The eROSITA Final Equatorial-Depth Survey (eFEDS). Catalog of galaxy clusters and groups}",
      journal = {\aap},
         year = 2022,
        month = may,
       volume = {661},
          eid = {A2},
        pages = {A2},
          doi = {10.1051/0004-6361/202141120},
archivePrefix = {arXiv},
       eprint = {2106.14518},
 primaryClass = {astro-ph.CO},
       adsurl = {https://ui.adsabs.harvard.edu/abs/2022A&A...661A...2L}
}

@ARTICLE{Lodders09,
       author = {{Lodders}, K. and {Palme}, H. and {Gail}, H. -P.},
        title = "{Abundances of the Elements in the Solar System}",
      journal = {Landolt B{\"o}rnstein},
         year = 2009,
        month = jan,
       volume = {4B},
        pages = {712},
          doi = {10.1007/978-3-540-88055-4_34},
archivePrefix = {arXiv},
       eprint = {0901.1149},
 primaryClass = {astro-ph.EP},
       adsurl = {https://ui.adsabs.harvard.edu/abs/2009LanB...4B..712L}
}

@ARTICLE{Lovisari2015,
       author = {{Lovisari}, L. and {Reiprich}, T.~H. and {Schellenberger}, G.},
        title = "{Scaling properties of a complete X-ray selected galaxy group sample}",
      journal = {\aap},
         year = 2015,
        month = jan,
       volume = {573},
          eid = {A118},
        pages = {A118},
          doi = {10.1051/0004-6361/201423954},
archivePrefix = {arXiv},
       eprint = {1409.3845},
 primaryClass = {astro-ph.CO},
       adsurl = {https://ui.adsabs.harvard.edu/abs/2015A&A...573A.118L}
}

@ARTICLE{Lovisari2020,
       author = {{Lovisari}, Lorenzo and {Schellenberger}, Gerrit and {Sereno}, Mauro and {Ettori}, Stefano and {Pratt}, Gabriel W. and {Forman}, William R. and {Jones}, Christine and {Andrade-Santos}, Felipe and {Randall}, Scott and {Kraft}, Ralph},
        title = "{X-Ray Scaling Relations for a Representative Sample of Planck-selected Clusters Observed with XMM-Newton}",
      journal = {\apj},
         year = 2020,
        month = apr,
       volume = {892},
       number = {2},
          eid = {102},
        pages = {102},
          doi = {10.3847/1538-4357/ab7997},
archivePrefix = {arXiv},
       eprint = {2002.11740},
 primaryClass = {astro-ph.CO},
       adsurl = {https://ui.adsabs.harvard.edu/abs/2020ApJ...892..102L}
}

@ARTICLE{Lyskova2023,
       author = {{Lyskova}, N. and {Churazov}, E. and {Khabibullin}, I.~I. and {Burenin}, R. and {Starobinsky}, A.~A. and {Sunyaev}, R.},
        title = "{X-ray surface brightness and gas density profiles of galaxy clusters up to 3 {\texttimes} R$_{500c}$ with SRG/eROSITA}",
      journal = {\mnras},
         year = 2023,
        month = oct,
       volume = {525},
       number = {1},
        pages = {898-907},
          doi = {10.1093/mnras/stad2305},
archivePrefix = {arXiv},
       eprint = {2305.07080},
 primaryClass = {astro-ph.CO},
       adsurl = {https://ui.adsabs.harvard.edu/abs/2023MNRAS.525..898L}
}

@ARTICLE{Malavasi2020,
       author = {{Malavasi}, Nicola and {Aghanim}, Nabila and {Tanimura}, Hideki and {Bonjean}, Victor and {Douspis}, Marian},
        title = "{Like a spider in its web: a study of the large-scale structure around the Coma cluster}",
      journal = {\aap},
         year = 2020,
        month = feb,
       volume = {634},
          eid = {A30},
        pages = {A30},
          doi = {10.1051/0004-6361/201936629},
archivePrefix = {arXiv},
       eprint = {1910.11879},
 primaryClass = {astro-ph.CO},
       adsurl = {https://ui.adsabs.harvard.edu/abs/2020A&A...634A..30M}
}

@ARTICLE{Malavasi2023,
       author = {{Malavasi}, Nicola and {Sorce}, Jenny G. and {Dolag}, Klaus and {Aghanim}, Nabila},
        title = "{The cosmic web around the Coma cluster from constrained cosmological simulations. I. Filaments connected to Coma at z = 0}",
      journal = {\aap},
         year = 2023,
        month = jul,
       volume = {675},
          eid = {A76},
        pages = {A76},
          doi = {10.1051/0004-6361/202245777},
archivePrefix = {arXiv},
       eprint = {2306.03124},
 primaryClass = {astro-ph.CO},
       adsurl = {https://ui.adsabs.harvard.edu/abs/2023A&A...675A..76M}
}

@ARTICLE{Marinacci2018,
       author = {{Marinacci}, Federico and {Vogelsberger}, Mark and {Pakmor}, R{\"u}diger and {Torrey}, Paul and {Springel}, Volker and {Hernquist}, Lars and {Nelson}, Dylan and {Weinberger}, Rainer and {Pillepich}, Annalisa and {Naiman}, Jill and {Genel}, Shy},
        title = "{First results from the IllustrisTNG simulations: radio haloes and magnetic fields}",
      journal = {\mnras},
         year = 2018,
        month = nov,
       volume = {480},
       number = {4},
        pages = {5113-5139},
          doi = {10.1093/mnras/sty2206},
archivePrefix = {arXiv},
       eprint = {1707.03396},
 primaryClass = {astro-ph.CO},
       adsurl = {https://ui.adsabs.harvard.edu/abs/2018MNRAS.480.5113M}
}

@ARTICLE{Mansfield2017,
       author = {{Mansfield}, Philip and {Kravtsov}, Andrey V. and {Diemer}, Benedikt},
        title = "{Splashback Shells of Cold Dark Matter Halos}",
      journal = {\apj},
         year = 2017,
        month = may,
       volume = {841},
       number = {1},
          eid = {34},
        pages = {34},
          doi = {10.3847/1538-4357/aa7047},
archivePrefix = {arXiv},
       eprint = {1612.01531},
 primaryClass = {astro-ph.CO},
       adsurl = {https://ui.adsabs.harvard.edu/abs/2017ApJ...841...34M}
}

@ARTICLE{McCall2024,
       author = {{McCall}, Hannah and {Reiprich}, Thomas H. and {Veronica}, Angie and {Pacaud}, Florian and {Sanders}, Jeremy and {Edler}, Henrik W. and {Br{\"u}ggen}, Marcus and {Bulbul}, Esra and {de Gasperin}, Francesco and {Gatuzz}, Efrain and {Liu}, Ang and {Merloni}, Andrea and {Migkas}, Konstantinos and {Zhang}, Xiaoyuan},
        title = "{The SRG/eROSITA All-Sky Survey: View of the Virgo Cluster}",
      journal = {\aap},
         year = 2024,
        month = sep,
       volume = {689},
          eid = {A113},
        pages = {A113},
          doi = {10.1051/0004-6361/202449391},
archivePrefix = {arXiv},
       eprint = {2401.17296},
 primaryClass = {astro-ph.CO},
       adsurl = {https://ui.adsabs.harvard.edu/abs/2024A&A...689A.113M}
}

@ARTICLE{Merloni2024,
       author = {{Merloni}, A. and {Lamer}, G. and {Liu}, T. and {Ramos-Ceja}, M.~E. and {Brunner}, H. and {Bulbul}, E. and {Dennerl}, K. and {Doroshenko}, V. and {Freyberg}, M.~J. and {Friedrich}, S. and {Gatuzz}, E. and {Georgakakis}, A. and {Haberl}, F. and {Igo}, Z. and {Kreykenbohm}, I. and {Liu}, A. and {Maitra}, C. and {Malyali}, A. and {Mayer}, M.~G.~F. and {Nandra}, K. and {Predehl}, P. and {Robrade}, J. and {Salvato}, M. and {Sanders}, J.~S. and {Stewart}, I. and {Tub{\'\i}n-Arenas}, D. and {Weber}, P. and {Wilms}, J. and {Arcodia}, R. and {Artis}, E. and {Aschersleben}, J. and {Avakyan}, A. and {Aydar}, C. and {Bahar}, Y.~E. and {Balzer}, F. and {Becker}, W. and {Berger}, K. and {Boller}, T. and {Bornemann}, W. and {Br{\"u}ggen}, M. and {Brusa}, M. and {Buchner}, J. and {Burwitz}, V. and {Camilloni}, F. and {Clerc}, N. and {Comparat}, J. and {Coutinho}, D. and {Czesla}, S. and {Dannhauer}, S.~M. and {Dauner}, L. and {Dauser}, T. and {Dietl}, J. and {Dolag}, K. and {Dwelly}, T. and {Egg}, K. and {Ehl}, E. and {Freund}, S. and {Friedrich}, P. and {Gaida}, R. and {Garrel}, C. and {Ghirardini}, V. and {Gokus}, A. and {Gr{\"u}nwald}, G. and {Grandis}, S. and {Grotova}, I. and {Gruen}, D. and {Gueguen}, A. and {H{\"a}mmerich}, S. and {Hamaus}, N. and {Hasinger}, G. and {Haubner}, K. and {Homan}, D. and {Ider Chitham}, J. and {Joseph}, W.~M. and {Joyce}, A. and {K{\"o}nig}, O. and {Kaltenbrunner}, D.~M. and {Khokhriakova}, A. and {Kink}, W. and {Kirsch}, C. and {Kluge}, M. and {Knies}, J. and {Krippendorf}, S. and {Krumpe}, M. and {Kurpas}, J. and {Li}, P. and {Liu}, Z. and {Locatelli}, N. and {Lorenz}, M. and {M{\"u}ller}, S. and {Magaudda}, E. and {Mannes}, C. and {McCall}, H. and {Meidinger}, N. and {Michailidis}, M. and {Migkas}, K. and {Mu{\~n}oz-Giraldo}, D. and {Musiimenta}, B. and {Nguyen-Dang}, N.~T. and {Ni}, Q. and {Olechowska}, A. and {Ota}, N. and {Pacaud}, F. and {Pasini}, T. and {Perinati}, E. and {Pires}, A.~M. and {Pommranz}, C. and {Ponti}, G. and {Poppenhaeger}, K. and {P{\"u}hlhofer}, G. and {Rau}, A. and {Reh}, M. and {Reiprich}, T.~H. and {Roster}, W. and {Saeedi}, S. and {Santangelo}, A. and {Sasaki}, M. and {Schmitt}, J. and {Schneider}, P.~C. and {Schrabback}, T. and {Schuster}, N. and {Schwope}, A. and {Seppi}, R. and {Serim}, M.~M. and {Shreeram}, S. and {Sokolova-Lapa}, E. and {Starck}, H. and {Stelzer}, B. and {Stierhof}, J. and {Suleimanov}, V. and {Tenzer}, C. and {Traulsen}, I. and {Tr{\"u}mper}, J. and {Tsuge}, K. and {Urrutia}, T. and {Veronica}, A. and {Waddell}, S.~G.~H. and {Willer}, R. and {Wolf}, J. and {Yeung}, M.~C.~H. and {Zainab}, A. and {Zangrandi}, F. and {Zhang}, X. and {Zhang}, Y. and {Zheng}, X.},
        title = "{The SRG/eROSITA all-sky survey. First X-ray catalogues and data release of the western Galactic hemisphere}",
      journal = {\aap},
         year = 2024,
        month = feb,
       volume = {682},
          eid = {A34},
        pages = {A34},
          doi = {10.1051/0004-6361/202347165},
archivePrefix = {arXiv},
       eprint = {2401.17274},
 primaryClass = {astro-ph.HE},
       adsurl = {https://ui.adsabs.harvard.edu/abs/2024A&A...682A..34M}
}

@ARTICLE{Mirakhor2020,
       author = {{Mirakhor}, M.~S. and {Walker}, S.~A.},
        title = "{A complete view of the outskirts of the Coma cluster}",
      journal = {\mnras},
         year = 2020,
        month = sep,
       volume = {497},
       number = {3},
        pages = {3204-3220},
          doi = {10.1093/mnras/staa2203},
archivePrefix = {arXiv},
       eprint = {2007.12194},
 primaryClass = {astro-ph.CO},
       adsurl = {https://ui.adsabs.harvard.edu/abs/2020MNRAS.497.3204M}
}

@ARTICLE{Moser2022,
       author = {{Moser}, Emily and {Battaglia}, Nicholas and {Nagai}, Daisuke and {Lau}, Erwin and {Machado Poletti Valle}, Luis Fernando and {Villaescusa-Navarro}, Francisco and {Amodeo}, Stefania and {Angl{\'e}s-Alc{\'a}zar}, Daniel and {Bryan}, Greg L. and {Dave}, Romeel and {Hernquist}, Lars and {Vogelsberger}, Mark},
        title = "{The Circumgalactic Medium from the CAMELS Simulations: Forecasting Constraints on Feedback Processes from Future Sunyaev-Zeldovich Observations}",
      journal = {\apj},
         year = 2022,
        month = jul,
       volume = {933},
       number = {2},
          eid = {133},
        pages = {133},
          doi = {10.3847/1538-4357/ac70c6},
archivePrefix = {arXiv},
       eprint = {2201.02708},
 primaryClass = {astro-ph.CO},
       adsurl = {https://ui.adsabs.harvard.edu/abs/2022ApJ...933..133M}
}

@ARTICLE{Nagai2007,
       author = {{Nagai}, Daisuke and {Kravtsov}, Andrey V. and {Vikhlinin}, Alexey},
        title = "{Effects of Galaxy Formation on Thermodynamics of the Intracluster Medium}",
      journal = {\apj},
         year = 2007,
        month = oct,
       volume = {668},
       number = {1},
        pages = {1-14},
          doi = {10.1086/521328},
archivePrefix = {arXiv},
       eprint = {astro-ph/0703661},
 primaryClass = {astro-ph},
       adsurl = {https://ui.adsabs.harvard.edu/abs/2007ApJ...668....1N}
}

@ARTICLE{Nagai2011,
       author = {{Nagai}, Daisuke and {Lau}, Erwin T.},
        title = "{Gas Clumping in the Outskirts of {\ensuremath{\Lambda}}CDM Clusters}",
      journal = {\apjl},
         year = 2011,
        month = apr,
       volume = {731},
       number = {1},
          eid = {L10},
        pages = {L10},
          doi = {10.1088/2041-8205/731/1/L10},
archivePrefix = {arXiv},
       eprint = {1103.0280},
 primaryClass = {astro-ph.CO},
       adsurl = {https://ui.adsabs.harvard.edu/abs/2011ApJ...731L..10N}
}

@ARTICLE{Naiman2018,
       author = {{Naiman}, Jill P. and {Pillepich}, Annalisa and {Springel}, Volker and {Ramirez-Ruiz}, Enrico and {Torrey}, Paul and {Vogelsberger}, Mark and {Pakmor}, R{\"u}diger and {Nelson}, Dylan and {Marinacci}, Federico and {Hernquist}, Lars and {Weinberger}, Rainer and {Genel}, Shy},
        title = "{First results from the IllustrisTNG simulations: a tale of two elements - chemical evolution of magnesium and europium}",
      journal = {\mnras},
         year = 2018,
        month = jun,
       volume = {477},
       number = {1},
        pages = {1206-1224},
          doi = {10.1093/mnras/sty618},
archivePrefix = {arXiv},
       eprint = {1707.03401},
 primaryClass = {astro-ph.GA},
       adsurl = {https://ui.adsabs.harvard.edu/abs/2018MNRAS.477.1206N}
}

@ARTICLE{Nandra2013,
       author = {{Nandra}, Kirpal and {Barret}, Didier and {Barcons}, Xavier and {Fabian}, Andy and {den Herder}, Jan-Willem and {Piro}, Luigi and {Watson}, Mike and {Adami}, Christophe and {Aird}, James and {Afonso}, Jose Manuel and {Alexander}, Dave and {Argiroffi}, Costanza and {Amati}, Lorenzo and {Arnaud}, Monique and {Atteia}, Jean-Luc and {Audard}, Marc and {Badenes}, Carles and {Ballet}, Jean and {Ballo}, Lucia and {Bamba}, Aya and {Bhardwaj}, Anil and {Stefano Battistelli}, Elia and {Becker}, Werner and {De Becker}, Micha{\"e}l and {Behar}, Ehud and {Bianchi}, Stefano and {Biffi}, Veronica and {B{\^\i}rzan}, Laura and {Bocchino}, Fabrizio and {Bogdanov}, Slavko and {Boirin}, Laurence and {Boller}, Thomas and {Borgani}, Stefano and {Borm}, Katharina and {Bouch{\'e}}, Nicolas and {Bourdin}, Herv{\'e} and {Bower}, Richard and {Braito}, Valentina and {Branchini}, Enzo and {Branduardi-Raymont}, Graziella and {Bregman}, Joel and {Brenneman}, Laura and {Brightman}, Murray and {Br{\"u}ggen}, Marcus and {Buchner}, Johannes and {Bulbul}, Esra and {Brusa}, Marcella and {Bursa}, Michal and {Caccianiga}, Alessandro and {Cackett}, Ed and {Campana}, Sergio and {Cappelluti}, Nico and {Cappi}, Massimo and {Carrera}, Francisco and {Ceballos}, Maite and {Christensen}, Finn and {Chu}, You-Hua and {Churazov}, Eugene and {Clerc}, Nicolas and {Corbel}, Stephane and {Corral}, Amalia and {Comastri}, Andrea and {Costantini}, Elisa and {Croston}, Judith and {Dadina}, Mauro and {D'Ai}, Antonino and {Decourchelle}, Anne and {Della Ceca}, Roberto and {Dennerl}, Konrad and {Dolag}, Klaus and {Done}, Chris and {Dovciak}, Michal and {Drake}, Jeremy and {Eckert}, Dominique and {Edge}, Alastair and {Ettori}, Stefano and {Ezoe}, Yuichiro and {Feigelson}, Eric and {Fender}, Rob and {Feruglio}, Chiara and {Finoguenov}, Alexis and {Fiore}, Fabrizio and {Galeazzi}, Massimiliano and {Gallagher}, Sarah and {Gandhi}, Poshak and {Gaspari}, Massimo and {Gastaldello}, Fabio and {Georgakakis}, Antonis and {Georgantopoulos}, Ioannis and {Gilfanov}, Marat and {Gitti}, Myriam and {Gladstone}, Randy and {Goosmann}, Rene and {Gosset}, Eric and {Grosso}, Nicolas and {Guedel}, Manuel and {Guerrero}, Martin and {Haberl}, Frank and {Hardcastle}, Martin and {Heinz}, Sebastian and {Alonso Herrero}, Almudena and {Herv{\'e}}, Anthony and {Holmstrom}, Mats and {Iwasawa}, Kazushi and {Jonker}, Peter and {Kaastra}, Jelle and {Kara}, Erin and {Karas}, Vladimir and {Kastner}, Joel and {King}, Andrew and {Kosenko}, Daria and {Koutroumpa}, Dimita and {Kraft}, Ralph and {Kreykenbohm}, Ingo and {Lallement}, Rosine and {Lanzuisi}, Giorgio and {Lee}, J. and {Lemoine-Goumard}, Marianne and {Lobban}, Andrew and {Lodato}, Giuseppe and {Lovisari}, Lorenzo and {Lotti}, Simone and {McCharthy}, Ian and {McNamara}, Brian and {Maggio}, Antonio and {Maiolino}, Roberto and {De Marco}, Barbara and {de Martino}, Domitilla and {Mateos}, Silvia and {Matt}, Giorgio and {Maughan}, Ben and {Mazzotta}, Pasquale and {Mendez}, Mariano and {Merloni}, Andrea and {Micela}, Giuseppina and {Miceli}, Marco and {Mignani}, Robert and {Miller}, Jon and {Miniutti}, Giovanni and {Molendi}, Silvano and {Montez}, Rodolfo and {Moretti}, Alberto and {Motch}, Christian and {Naz{\'e}}, Ya{\"e}l and {Nevalainen}, Jukka and {Nicastro}, Fabrizio and {Nulsen}, Paul and {Ohashi}, Takaya and {O'Brien}, Paul and {Osborne}, Julian and {Oskinova}, Lida and {Pacaud}, Florian and {Paerels}, Frederik and {Page}, Mat and {Papadakis}, Iossif and {Pareschi}, Giovanni and {Petre}, Robert and {Petrucci}, Pierre-Olivier and {Piconcelli}, Enrico and {Pillitteri}, Ignazio and {Pinto}, C. and {de Plaa}, Jelle and {Pointecouteau}, Etienne and {Ponman}, Trevor and {Ponti}, Gabriele and {Porquet}, Delphine and {Pounds}, Ken and {Pratt}, Gabriel and {Predehl}, Peter and {Proga}, Daniel and {Psaltis}, Dimitrios and {Rafferty}, David and {Ramos-Ceja}, Miriam and {Ranalli}, Piero and {Rasia}, Elena and {Rau}, Arne and {Rauw}, Gregor and {Rea}, Nanda and {Read}, Andy and {Reeves}, James and {Reiprich}, Thomas and {Renaud}, Matthieu and {Reynolds}, Chris and {Risaliti}, Guido and {Rodriguez}, Jerome and {Rodriguez Hidalgo}, Paola and {Roncarelli}, Mauro and {Rosario}, David and {Rossetti}, Mariachiara and {Rozanska}, Agata and {Rovilos}, Emmanouil and {Salvaterra}, Ruben and {Salvato}, Mara and {Di Salvo}, Tiziana and {Sanders}, Jeremy and {Sanz-Forcada}, Jorge and {Schawinski}, Kevin and {Schaye}, Joop and {Schwope}, Axel and {Sciortino}, Salvatore},
        title = "{The Hot and Energetic Universe: A White Paper presenting the science theme motivating the Athena+ mission}",
      journal = {arXiv e-prints},
         year = 2013,
        month = jun,
          eid = {arXiv:1306.2307},
        pages = {arXiv:1306.2307},
          doi = {10.48550/arXiv.1306.2307},
archivePrefix = {arXiv},
       eprint = {1306.2307},
 primaryClass = {astro-ph.HE},
       adsurl = {https://ui.adsabs.harvard.edu/abs/2013arXiv1306.2307N}
}

@ARTICLE{Navarro1997,
       author = {{Navarro}, Julio F. and {Frenk}, Carlos S. and {White}, Simon D.~M.},
        title = "{A Universal Density Profile from Hierarchical Clustering}",
      journal = {\apj},
         year = 1997,
        month = dec,
       volume = {490},
       number = {2},
        pages = {493-508},
          doi = {10.1086/304888},
archivePrefix = {arXiv},
       eprint = {astro-ph/9611107},
 primaryClass = {astro-ph},
       adsurl = {https://ui.adsabs.harvard.edu/abs/1997ApJ...490..493N}
}

@ARTICLE{Nelson2018,
       author = {{Nelson}, Dylan and {Pillepich}, Annalisa and {Springel}, Volker and {Weinberger}, Rainer and {Hernquist}, Lars and {Pakmor}, R{\"u}diger and {Genel}, Shy and {Torrey}, Paul and {Vogelsberger}, Mark and {Kauffmann}, Guinevere and {Marinacci}, Federico and {Naiman}, Jill},
        title = "{First results from the IllustrisTNG simulations: the galaxy colour bimodality}",
      journal = {\mnras},
         year = 2018,
        month = mar,
       volume = {475},
       number = {1},
        pages = {624-647},
          doi = {10.1093/mnras/stx3040},
archivePrefix = {arXiv},
       eprint = {1707.03395},
 primaryClass = {astro-ph.GA},
       adsurl = {https://ui.adsabs.harvard.edu/abs/2018MNRAS.475..624N}
}

@ARTICLE{Okabe2025,
       author = {{Okabe}, Nobuhiro and {Reiprich}, Thomas H. and {Grandis}, Sebastian and {Chiu}, I.-Non and {Oguri}, Masamune and {Umetsu}, Keiichi and {Bulbul}, Esra and {Bahar}, Emre and {Balzer}, Fabian and {Clerc}, Nicolas and {Comparat}, Johan and {Ghirardini}, Vittorio and {Kleinebreil}, Florian and {Kluge}, Matthias and {Liu}, Ang and {Monteiro-Oliveira}, Rog{\'e}rio and {Pacaud}, Florian and {Ceja}, Miriam Ramos and {Sanders}, Jeremy and {Schrabback}, Tim and {Seppi}, Riccardo and {Sommer}, Martin and {Zhang}, Xiaoyuan},
        title = "{The SRG/eROSITA all-sky survey: Subaru/HSC-SSP weak-lensing mass measurements for eRASS1 galaxy clusters}",
      journal = {\aap},
         year = 2025,
        month = aug,
       volume = {700},
          eid = {A46},
        pages = {A46},
          doi = {10.1051/0004-6361/202553708},
archivePrefix = {arXiv},
       eprint = {2503.09952},
 primaryClass = {astro-ph.CO},
       adsurl = {https://ui.adsabs.harvard.edu/abs/2025A&A...700A..46O}
}

@ARTICLE{ONeil2021,
       author = {{O'Neil}, Stephanie and {Barnes}, David J. and {Vogelsberger}, Mark and {Diemer}, Benedikt},
        title = "{The splashback boundary of haloes in hydrodynamic simulations}",
      journal = {\mnras},
         year = 2021,
        month = jul,
       volume = {504},
       number = {3},
        pages = {4649-4666},
          doi = {10.1093/mnras/stab1221},
archivePrefix = {arXiv},
       eprint = {2012.00025},
 primaryClass = {astro-ph.GA},
       adsurl = {https://ui.adsabs.harvard.edu/abs/2021MNRAS.504.4649O}
}

@ARTICLE{Ostriker1988,
       author = {{Ostriker}, Jeremiah P. and {McKee}, Christopher F.},
        title = "{Astrophysical blastwaves}",
      journal = {Reviews of Modern Physics},
         year = 1988,
        month = jan,
       volume = {60},
       number = {1},
        pages = {1-68},
          doi = {10.1103/RevModPhys.60.1},
       adsurl = {https://ui.adsabs.harvard.edu/abs/1988RvMP...60....1O}
}

@ARTICLE{Pillepich2010,
       author = {{Pillepich}, Annalisa and {Porciani}, Cristiano and {Hahn}, Oliver},
        title = "{Halo mass function and scale-dependent bias from N-body simulations with non-Gaussian initial conditions}",
      journal = {\mnras},
         year = 2010,
        month = feb,
       volume = {402},
       number = {1},
        pages = {191-206},
          doi = {10.1111/j.1365-2966.2009.15914.x},
archivePrefix = {arXiv},
       eprint = {0811.4176},
 primaryClass = {astro-ph},
       adsurl = {https://ui.adsabs.harvard.edu/abs/2010MNRAS.402..191P}
}

@ARTICLE{Pillepich2018,
       author = {{Pillepich}, Annalisa and {Nelson}, Dylan and {Hernquist}, Lars and {Springel}, Volker and {Pakmor}, R{\"u}diger and {Torrey}, Paul and {Weinberger}, Rainer and {Genel}, Shy and {Naiman}, Jill P. and {Marinacci}, Federico and {Vogelsberger}, Mark},
        title = "{First results from the IllustrisTNG simulations: the stellar mass content of groups and clusters of galaxies}",
      journal = {\mnras},
         year = 2018,
        month = mar,
       volume = {475},
       number = {1},
        pages = {648-675},
          doi = {10.1093/mnras/stx3112},
archivePrefix = {arXiv},
       eprint = {1707.03406},
 primaryClass = {astro-ph.GA},
       adsurl = {https://ui.adsabs.harvard.edu/abs/2018MNRAS.475..648P}
}

@ARTICLE{Planck2013,
       author = {{Planck Collaboration} and {Ade}, P.~A.~R. and {Aghanim}, N. and {Arnaud}, M. and {Ashdown}, M. and {Atrio-Barandela}, F. and {Aumont}, J. and {Baccigalupi}, C. and {Balbi}, A. and {Banday}, A.~J. and {Barreiro}, R.~B. and {Bartlett}, J.~G. and {Battaner}, E. and {Benabed}, K. and {Beno{\^\i}t}, A. and {Bernard}, J. -P. and {Bersanelli}, M. and {Bhatia}, R. and {Bikmaev}, I. and {Bobin}, J. and {B{\"o}hringer}, H. and {Bonaldi}, A. and {Bond}, J.~R. and {Borgani}, S. and {Borrill}, J. and {Bouchet}, F.~R. and {Bourdin}, H. and {Brown}, M.~L. and {Burenin}, R. and {Burigana}, C. and {Cabella}, P. and {Cardoso}, J. -F. and {Carvalho}, P. and {Castex}, G. and {Catalano}, A. and {Cay{\'o}n}, L. and {Chamballu}, A. and {Chiang}, L. -Y. and {Chon}, G. and {Christensen}, P.~R. and {Churazov}, E. and {Clements}, D.~L. and {Colafrancesco}, S. and {Colombi}, S. and {Colombo}, L.~P.~L. and {Comis}, B. and {Coulais}, A. and {Crill}, B.~P. and {Cuttaia}, F. and {Da Silva}, A. and {Dahle}, H. and {Danese}, L. and {Davis}, R.~J. and {de Bernardis}, P. and {de Gasperis}, G. and {de Zotti}, G. and {Delabrouille}, J. and {D{\'e}mocl{\`e}s}, J. and {D{\'e}sert}, F. -X. and {Diego}, J.~M. and {Dolag}, K. and {Dole}, H. and {Donzelli}, S. and {Dor{\'e}}, O. and {D{\"o}rl}, U. and {Douspis}, M. and {Dupac}, X. and {Efstathiou}, G. and {En{\ss}lin}, T.~A. and {Eriksen}, H.~K. and {Finelli}, F. and {Flores-Cacho}, I. and {Forni}, O. and {Fosalba}, P. and {Frailis}, M. and {Franceschi}, E. and {Frommert}, M. and {Galeotta}, S. and {Ganga}, K. and {G{\'e}nova-Santos}, R.~T. and {Giard}, M. and {Giraud-H{\'e}raud}, Y. and {Gonz{\'a}lez-Nuevo}, J. and {G{\'o}rski}, K.~M. and {Gregorio}, A. and {Gruppuso}, A. and {Hansen}, F.~K. and {Harrison}, D. and {Hempel}, A. and {Henrot-Versill{\'e}}, S. and {Hern{\'a}ndez-Monteagudo}, C. and {Herranz}, D. and {Hildebrandt}, S.~R. and {Hivon}, E. and {Hobson}, M. and {Holmes}, W.~A. and {Hurier}, G. and {Jaffe}, T.~R. and {Jaffe}, A.~H. and {Jagemann}, T. and {Jones}, W.~C. and {Juvela}, M. and {Keih{\"a}nen}, E. and {Khamitov}, I. and {Kisner}, T.~S. and {Kneissl}, R. and {Knoche}, J. and {Knox}, L. and {Kunz}, M. and {Kurki-Suonio}, H. and {Lagache}, G. and {L{\"a}hteenm{\"a}ki}, A. and {Lamarre}, J. -M. and {Lasenby}, A. and {Lawrence}, C.~R. and {Le Jeune}, M. and {Leonardi}, R. and {Liddle}, A. and {Lilje}, P.~B. and {L{\'o}pez-Caniego}, M. and {Luzzi}, G. and {Mac{\'\i}as-P{\'e}rez}, J.~F. and {Maino}, D. and {Mandolesi}, N. and {Maris}, M. and {Marleau}, F. and {Marshall}, D.~J. and {Mart{\'\i}nez-Gonz{\'a}lez}, E. and {Masi}, S. and {Massardi}, M. and {Matarrese}, S. and {Mazzotta}, P. and {Mei}, S. and {Melchiorri}, A. and {Melin}, J. -B. and {Mendes}, L. and {Mennella}, A. and {Mitra}, S. and {Miville-Desch{\^e}nes}, M. -A. and {Moneti}, A. and {Montier}, L. and {Morgante}, G. and {Mortlock}, D. and {Munshi}, D. and {Murphy}, J.~A. and {Naselsky}, P. and {Nati}, F. and {Natoli}, P. and {N{\o}rgaard-Nielsen}, H.~U. and {Noviello}, F. and {Novikov}, D. and {Novikov}, I. and {Osborne}, S. and {Pajot}, F. and {Paoletti}, D. and {Pasian}, F. and {Patanchon}, G. and {Perdereau}, O. and {Perotto}, L. and {Perrotta}, F. and {Piacentini}, F. and {Piat}, M. and {Pierpaoli}, E. and {Piffaretti}, R. and {Plaszczynski}, S. and {Pointecouteau}, E. and {Polenta}, G. and {Ponthieu}, N. and {Popa}, L. and {Poutanen}, T. and {Pratt}, G.~W. and {Prunet}, S. and {Puget}, J. -L. and {Rachen}, J.~P. and {Reach}, W.~T. and {Rebolo}, R. and {Reinecke}, M. and {Remazeilles}, M. and {Renault}, C. and {Ricciardi}, S. and {Riller}, T. and {Ristorcelli}, I. and {Rocha}, G. and {Roman}, M. and {Rosset}, C. and {Rossetti}, M. and {Rubi{\~n}o-Mart{\'\i}n}, J.~A. and {Rusholme}, B. and {Sandri}, M. and {Savini}, G. and {Scott}, D. and {Smoot}, G.~F. and {Starck}, J. -L. and {Sudiwala}, R. and {Sunyaev}, R. and {Sutton}, D. and {Suur-Uski}, A. -S. and {Sygnet}, J. -F. and {Tauber}, J.~A. and {Terenzi}, L.},
        title = "{Planck intermediate results. V. Pressure profiles of galaxy clusters from the Sunyaev-Zeldovich effect}",
      journal = {\aap},
         year = 2013,
        month = feb,
       volume = {550},
          eid = {A131},
        pages = {A131},
          doi = {10.1051/0004-6361/201220040},
archivePrefix = {arXiv},
       eprint = {1207.4061},
 primaryClass = {astro-ph.CO},
       adsurl = {https://ui.adsabs.harvard.edu/abs/2013A&A...550A.131P}
}

@ARTICLE{Planck2016,
       author = {{Planck Collaboration} and {Ade}, P.~A.~R. and {Aghanim}, N. and {Arnaud}, M. and {Ashdown}, M. and {Aumont}, J. and {Baccigalupi}, C. and {Banday}, A.~J. and {Barreiro}, R.~B. and {Bartlett}, J.~G. and {Bartolo}, N. and {Battaner}, E. and {Battye}, R. and {Benabed}, K. and {Beno{\^\i}t}, A. and {Benoit-L{\'e}vy}, A. and {Bernard}, J. -P. and {Bersanelli}, M. and {Bielewicz}, P. and {Bock}, J.~J. and {Bonaldi}, A. and {Bonavera}, L. and {Bond}, J.~R. and {Borrill}, J. and {Bouchet}, F.~R. and {Boulanger}, F. and {Bucher}, M. and {Burigana}, C. and {Butler}, R.~C. and {Calabrese}, E. and {Cardoso}, J. -F. and {Catalano}, A. and {Challinor}, A. and {Chamballu}, A. and {Chary}, R. -R. and {Chiang}, H.~C. and {Chluba}, J. and {Christensen}, P.~R. and {Church}, S. and {Clements}, D.~L. and {Colombi}, S. and {Colombo}, L.~P.~L. and {Combet}, C. and {Coulais}, A. and {Crill}, B.~P. and {Curto}, A. and {Cuttaia}, F. and {Danese}, L. and {Davies}, R.~D. and {Davis}, R.~J. and {de Bernardis}, P. and {de Rosa}, A. and {de Zotti}, G. and {Delabrouille}, J. and {D{\'e}sert}, F. -X. and {Di Valentino}, E. and {Dickinson}, C. and {Diego}, J.~M. and {Dolag}, K. and {Dole}, H. and {Donzelli}, S. and {Dor{\'e}}, O. and {Douspis}, M. and {Ducout}, A. and {Dunkley}, J. and {Dupac}, X. and {Efstathiou}, G. and {Elsner}, F. and {En{\ss}lin}, T.~A. and {Eriksen}, H.~K. and {Farhang}, M. and {Fergusson}, J. and {Finelli}, F. and {Forni}, O. and {Frailis}, M. and {Fraisse}, A.~A. and {Franceschi}, E. and {Frejsel}, A. and {Galeotta}, S. and {Galli}, S. and {Ganga}, K. and {Gauthier}, C. and {Gerbino}, M. and {Ghosh}, T. and {Giard}, M. and {Giraud-H{\'e}raud}, Y. and {Giusarma}, E. and {Gjerl{\o}w}, E. and {Gonz{\'a}lez-Nuevo}, J. and {G{\'o}rski}, K.~M. and {Gratton}, S. and {Gregorio}, A. and {Gruppuso}, A. and {Gudmundsson}, J.~E. and {Hamann}, J. and {Hansen}, F.~K. and {Hanson}, D. and {Harrison}, D.~L. and {Helou}, G. and {Henrot-Versill{\'e}}, S. and {Hern{\'a}ndez-Monteagudo}, C. and {Herranz}, D. and {Hildebrandt}, S.~R. and {Hivon}, E. and {Hobson}, M. and {Holmes}, W.~A. and {Hornstrup}, A. and {Hovest}, W. and {Huang}, Z. and {Huffenberger}, K.~M. and {Hurier}, G. and {Jaffe}, A.~H. and {Jaffe}, T.~R. and {Jones}, W.~C. and {Juvela}, M. and {Keih{\"a}nen}, E. and {Keskitalo}, R. and {Kisner}, T.~S. and {Kneissl}, R. and {Knoche}, J. and {Knox}, L. and {Kunz}, M. and {Kurki-Suonio}, H. and {Lagache}, G. and {L{\"a}hteenm{\"a}ki}, A. and {Lamarre}, J. -M. and {Lasenby}, A. and {Lattanzi}, M. and {Lawrence}, C.~R. and {Leahy}, J.~P. and {Leonardi}, R. and {Lesgourgues}, J. and {Levrier}, F. and {Lewis}, A. and {Liguori}, M. and {Lilje}, P.~B. and {Linden-V{\o}rnle}, M. and {L{\'o}pez-Caniego}, M. and {Lubin}, P.~M. and {Mac{\'\i}as-P{\'e}rez}, J.~F. and {Maggio}, G. and {Maino}, D. and {Mandolesi}, N. and {Mangilli}, A. and {Marchini}, A. and {Maris}, M. and {Martin}, P.~G. and {Martinelli}, M. and {Mart{\'\i}nez-Gonz{\'a}lez}, E. and {Masi}, S. and {Matarrese}, S. and {McGehee}, P. and {Meinhold}, P.~R. and {Melchiorri}, A. and {Melin}, J. -B. and {Mendes}, L. and {Mennella}, A. and {Migliaccio}, M. and {Millea}, M. and {Mitra}, S. and {Miville-Desch{\^e}nes}, M. -A. and {Moneti}, A. and {Montier}, L. and {Morgante}, G. and {Mortlock}, D. and {Moss}, A. and {Munshi}, D. and {Murphy}, J.~A. and {Naselsky}, P. and {Nati}, F. and {Natoli}, P. and {Netterfield}, C.~B. and {N{\o}rgaard-Nielsen}, H.~U. and {Noviello}, F. and {Novikov}, D. and {Novikov}, I. and {Oxborrow}, C.~A. and {Paci}, F. and {Pagano}, L. and {Pajot}, F. and {Paladini}, R. and {Paoletti}, D. and {Partridge}, B. and {Pasian}, F. and {Patanchon}, G. and {Pearson}, T.~J. and {Perdereau}, O. and {Perotto}, L. and {Perrotta}, F. and {Pettorino}, V. and {Piacentini}, F. and {Piat}, M. and {Pierpaoli}, E. and {Pietrobon}, D. and {Plaszczynski}, S. and {Pointecouteau}, E. and {Polenta}, G. and {Popa}, L. and {Pratt}, G.~W. and {Pr{\'e}zeau}, G.},
        title = "{Planck 2015 results. XIII. Cosmological parameters}",
      journal = {\aap},
         year = 2016,
        month = sep,
       volume = {594},
          eid = {A13},
        pages = {A13},
          doi = {10.1051/0004-6361/201525830},
archivePrefix = {arXiv},
       eprint = {1502.01589},
 primaryClass = {astro-ph.CO},
       adsurl = {https://ui.adsabs.harvard.edu/abs/2016A&A...594A..13P}
}

@ARTICLE{Planck2016-psz2,
       author = {{Planck Collaboration} and {Ade}, P.~A.~R. and {Aghanim}, N. and {Arnaud}, M. and {Ashdown}, M. and {Aumont}, J. and {Baccigalupi}, C. and {Banday}, A.~J. and {Barreiro}, R.~B. and {Barrena}, R. and {Bartlett}, J.~G. and {Bartolo}, N. and {Battaner}, E. and {Battye}, R. and {Benabed}, K. and {Beno{\^\i}t}, A. and {Benoit-L{\'e}vy}, A. and {Bernard}, J. -P. and {Bersanelli}, M. and {Bielewicz}, P. and {Bikmaev}, I. and {B{\"o}hringer}, H. and {Bonaldi}, A. and {Bonavera}, L. and {Bond}, J.~R. and {Borrill}, J. and {Bouchet}, F.~R. and {Bucher}, M. and {Burenin}, R. and {Burigana}, C. and {Butler}, R.~C. and {Calabrese}, E. and {Cardoso}, J. -F. and {Carvalho}, P. and {Catalano}, A. and {Challinor}, A. and {Chamballu}, A. and {Chary}, R. -R. and {Chiang}, H.~C. and {Chon}, G. and {Christensen}, P.~R. and {Clements}, D.~L. and {Colombi}, S. and {Colombo}, L.~P.~L. and {Combet}, C. and {Comis}, B. and {Couchot}, F. and {Coulais}, A. and {Crill}, B.~P. and {Curto}, A. and {Cuttaia}, F. and {Dahle}, H. and {Danese}, L. and {Davies}, R.~D. and {Davis}, R.~J. and {de Bernardis}, P. and {de Rosa}, A. and {de Zotti}, G. and {Delabrouille}, J. and {D{\'e}sert}, F. -X. and {Dickinson}, C. and {Diego}, J.~M. and {Dolag}, K. and {Dole}, H. and {Donzelli}, S. and {Dor{\'e}}, O. and {Douspis}, M. and {Ducout}, A. and {Dupac}, X. and {Efstathiou}, G. and {Eisenhardt}, P.~R.~M. and {Elsner}, F. and {En{\ss}lin}, T.~A. and {Eriksen}, H.~K. and {Falgarone}, E. and {Fergusson}, J. and {Feroz}, F. and {Ferragamo}, A. and {Finelli}, F. and {Forni}, O. and {Frailis}, M. and {Fraisse}, A.~A. and {Franceschi}, E. and {Frejsel}, A. and {Galeotta}, S. and {Galli}, S. and {Ganga}, K. and {G{\'e}nova-Santos}, R.~T. and {Giard}, M. and {Giraud-H{\'e}raud}, Y. and {Gjerl{\o}w}, E. and {Gonz{\'a}lez-Nuevo}, J. and {G{\'o}rski}, K.~M. and {Grainge}, K.~J.~B. and {Gratton}, S. and {Gregorio}, A. and {Gruppuso}, A. and {Gudmundsson}, J.~E. and {Hansen}, F.~K. and {Hanson}, D. and {Harrison}, D.~L. and {Hempel}, A. and {Henrot-Versill{\'e}}, S. and {Hern{\'a}ndez-Monteagudo}, C. and {Herranz}, D. and {Hildebrandt}, S.~R. and {Hivon}, E. and {Hobson}, M. and {Holmes}, W.~A. and {Hornstrup}, A. and {Hovest}, W. and {Huffenberger}, K.~M. and {Hurier}, G. and {Jaffe}, A.~H. and {Jaffe}, T.~R. and {Jin}, T. and {Jones}, W.~C. and {Juvela}, M. and {Keih{\"a}nen}, E. and {Keskitalo}, R. and {Khamitov}, I. and {Kisner}, T.~S. and {Kneissl}, R. and {Knoche}, J. and {Kunz}, M. and {Kurki-Suonio}, H. and {Lagache}, G. and {Lamarre}, J. -M. and {Lasenby}, A. and {Lattanzi}, M. and {Lawrence}, C.~R. and {Leonardi}, R. and {Lesgourgues}, J. and {Levrier}, F. and {Liguori}, M. and {Lilje}, P.~B. and {Linden-V{\o}rnle}, M. and {L{\'o}pez-Caniego}, M. and {Lubin}, P.~M. and {Mac{\'\i}as-P{\'e}rez}, J.~F. and {Maggio}, G. and {Maino}, D. and {Mak}, D.~S.~Y. and {Mandolesi}, N. and {Mangilli}, A. and {Martin}, P.~G. and {Mart{\'\i}nez-Gonz{\'a}lez}, E. and {Masi}, S. and {Matarrese}, S. and {Mazzotta}, P. and {McGehee}, P. and {Mei}, S. and {Melchiorri}, A. and {Melin}, J. -B. and {Mendes}, L. and {Mennella}, A. and {Migliaccio}, M. and {Mitra}, S. and {Miville-Desch{\^e}nes}, M. -A. and {Moneti}, A. and {Montier}, L. and {Morgante}, G. and {Mortlock}, D. and {Moss}, A. and {Munshi}, D. and {Murphy}, J.~A. and {Naselsky}, P. and {Nastasi}, A. and {Nati}, F. and {Natoli}, P. and {Netterfield}, C.~B. and {N{\o}rgaard-Nielsen}, H.~U. and {Noviello}, F. and {Novikov}, D. and {Novikov}, I. and {Olamaie}, M. and {Oxborrow}, C.~A. and {Paci}, F. and {Pagano}, L. and {Pajot}, F. and {Paoletti}, D. and {Pasian}, F. and {Patanchon}, G. and {Pearson}, T.~J. and {Perdereau}, O. and {Perotto}, L. and {Perrott}, Y.~C. and {Perrotta}, F. and {Pettorino}, V. and {Piacentini}, F. and {Piat}, M. and {Pierpaoli}, E. and {Pietrobon}, D. and {Plaszczynski}, S. and {Pointecouteau}, E. and {Polenta}, G. and {Pratt}, G.~W. and {Pr{\'e}zeau}, G. and {Prunet}, S. and {Puget}, J. -L.},
        title = "{Planck 2015 results. XXVII. The second Planck catalogue of Sunyaev-Zeldovich sources}",
      journal = {\aap},
         year = 2016,
        month = sep,
       volume = {594},
          eid = {A27},
        pages = {A27},
          doi = {10.1051/0004-6361/201525823},
archivePrefix = {arXiv},
       eprint = {1502.01598},
 primaryClass = {astro-ph.CO},
       adsurl = {https://ui.adsabs.harvard.edu/abs/2016A&A...594A..27P}
}

@ARTICLE{Planck2020,
       author = {{Planck Collaboration} and {Aghanim}, N. and {Akrami}, Y. and {Ashdown}, M. and {Aumont}, J. and {Baccigalupi}, C. and {Ballardini}, M. and {Banday}, A.~J. and {Barreiro}, R.~B. and {Bartolo}, N. and {Basak}, S. and {Battye}, R. and {Benabed}, K. and {Bernard}, J. -P. and {Bersanelli}, M. and {Bielewicz}, P. and {Bock}, J.~J. and {Bond}, J.~R. and {Borrill}, J. and {Bouchet}, F.~R. and {Boulanger}, F. and {Bucher}, M. and {Burigana}, C. and {Butler}, R.~C. and {Calabrese}, E. and {Cardoso}, J. -F. and {Carron}, J. and {Challinor}, A. and {Chiang}, H.~C. and {Chluba}, J. and {Colombo}, L.~P.~L. and {Combet}, C. and {Contreras}, D. and {Crill}, B.~P. and {Cuttaia}, F. and {de Bernardis}, P. and {de Zotti}, G. and {Delabrouille}, J. and {Delouis}, J. -M. and {Di Valentino}, E. and {Diego}, J.~M. and {Dor{\'e}}, O. and {Douspis}, M. and {Ducout}, A. and {Dupac}, X. and {Dusini}, S. and {Efstathiou}, G. and {Elsner}, F. and {En{\ss}lin}, T.~A. and {Eriksen}, H.~K. and {Fantaye}, Y. and {Farhang}, M. and {Fergusson}, J. and {Fernandez-Cobos}, R. and {Finelli}, F. and {Forastieri}, F. and {Frailis}, M. and {Fraisse}, A.~A. and {Franceschi}, E. and {Frolov}, A. and {Galeotta}, S. and {Galli}, S. and {Ganga}, K. and {G{\'e}nova-Santos}, R.~T. and {Gerbino}, M. and {Ghosh}, T. and {Gonz{\'a}lez-Nuevo}, J. and {G{\'o}rski}, K.~M. and {Gratton}, S. and {Gruppuso}, A. and {Gudmundsson}, J.~E. and {Hamann}, J. and {Handley}, W. and {Hansen}, F.~K. and {Herranz}, D. and {Hildebrandt}, S.~R. and {Hivon}, E. and {Huang}, Z. and {Jaffe}, A.~H. and {Jones}, W.~C. and {Karakci}, A. and {Keih{\"a}nen}, E. and {Keskitalo}, R. and {Kiiveri}, K. and {Kim}, J. and {Kisner}, T.~S. and {Knox}, L. and {Krachmalnicoff}, N. and {Kunz}, M. and {Kurki-Suonio}, H. and {Lagache}, G. and {Lamarre}, J. -M. and {Lasenby}, A. and {Lattanzi}, M. and {Lawrence}, C.~R. and {Le Jeune}, M. and {Lemos}, P. and {Lesgourgues}, J. and {Levrier}, F. and {Lewis}, A. and {Liguori}, M. and {Lilje}, P.~B. and {Lilley}, M. and {Lindholm}, V. and {L{\'o}pez-Caniego}, M. and {Lubin}, P.~M. and {Ma}, Y. -Z. and {Mac{\'\i}as-P{\'e}rez}, J.~F. and {Maggio}, G. and {Maino}, D. and {Mandolesi}, N. and {Mangilli}, A. and {Marcos-Caballero}, A. and {Maris}, M. and {Martin}, P.~G. and {Martinelli}, M. and {Mart{\'\i}nez-Gonz{\'a}lez}, E. and {Matarrese}, S. and {Mauri}, N. and {McEwen}, J.~D. and {Meinhold}, P.~R. and {Melchiorri}, A. and {Mennella}, A. and {Migliaccio}, M. and {Millea}, M. and {Mitra}, S. and {Miville-Desch{\^e}nes}, M. -A. and {Molinari}, D. and {Montier}, L. and {Morgante}, G. and {Moss}, A. and {Natoli}, P. and {N{\o}rgaard-Nielsen}, H.~U. and {Pagano}, L. and {Paoletti}, D. and {Partridge}, B. and {Patanchon}, G. and {Peiris}, H.~V. and {Perrotta}, F. and {Pettorino}, V. and {Piacentini}, F. and {Polastri}, L. and {Polenta}, G. and {Puget}, J. -L. and {Rachen}, J.~P. and {Reinecke}, M. and {Remazeilles}, M. and {Renzi}, A. and {Rocha}, G. and {Rosset}, C. and {Roudier}, G. and {Rubi{\~n}o-Mart{\'\i}n}, J.~A. and {Ruiz-Granados}, B. and {Salvati}, L. and {Sandri}, M. and {Savelainen}, M. and {Scott}, D. and {Shellard}, E.~P.~S. and {Sirignano}, C. and {Sirri}, G. and {Spencer}, L.~D. and {Sunyaev}, R. and {Suur-Uski}, A. -S. and {Tauber}, J.~A. and {Tavagnacco}, D. and {Tenti}, M. and {Toffolatti}, L. and {Tomasi}, M. and {Trombetti}, T. and {Valenziano}, L. and {Valiviita}, J. and {Van Tent}, B. and {Vibert}, L. and {Vielva}, P. and {Villa}, F. and {Vittorio}, N. and {Wandelt}, B.~D. and {Wehus}, I.~K. and {White}, M. and {White}, S.~D.~M. and {Zacchei}, A. and {Zonca}, A.},
        title = "{Planck 2018 results. VI. Cosmological parameters}",
      journal = {\aap},
         year = 2020,
        month = sep,
       volume = {641},
          eid = {A6},
        pages = {A6},
          doi = {10.1051/0004-6361/201833910},
archivePrefix = {arXiv},
       eprint = {1807.06209},
 primaryClass = {astro-ph.CO},
       adsurl = {https://ui.adsabs.harvard.edu/abs/2020A&A...641A...6P}
}

@ARTICLE{Pop2022,
       author = {{Pop}, Ana-Roxana and {Hernquist}, Lars and {Nagai}, Daisuke and {Kannan}, Rahul and {Weinberger}, Rainer and {Springel}, Volker and {Vogelsberger}, Mark and {Nelson}, Dylan and {Pakmor}, R{\"u}diger and {Pillepich}, Annalisa and {Torrey}, Paul},
        title = "{Sunyaev-Zel'dovich effect and X-ray scaling relations of galaxies, groups and clusters in the IllustrisTNG simulations}",
      journal = {arXiv e-prints},
         year = 2022,
        month = may,
          eid = {arXiv:2205.11528},
        pages = {arXiv:2205.11528},
          doi = {10.48550/arXiv.2205.11528},
archivePrefix = {arXiv},
       eprint = {2205.11528},
 primaryClass = {astro-ph.GA},
       adsurl = {https://ui.adsabs.harvard.edu/abs/2022arXiv220511528P}
}

@ARTICLE{Reiprich2013,
       author = {{Reiprich}, Thomas H. and {Basu}, Kaustuv and {Ettori}, Stefano and {Israel}, Holger and {Lovisari}, Lorenzo and {Molendi}, Silvano and {Pointecouteau}, Etienne and {Roncarelli}, Mauro},
        title = "{Outskirts of Galaxy Clusters}",
      journal = {\ssr},
         year = 2013,
        month = aug,
       volume = {177},
       number = {1-4},
        pages = {195-245},
          doi = {10.1007/s11214-013-9983-8},
archivePrefix = {arXiv},
       eprint = {1303.3286},
 primaryClass = {astro-ph.CO},
       adsurl = {https://ui.adsabs.harvard.edu/abs/2013SSRv..177..195R}
}

\begin{appendix}

\section{Profiles in four individual sectors}\label{app:four_sector}

In the inset of Fig.~\ref{fig:profile_main}, there is a plausible surface brightness jump at $\sim2r_\mathrm{200m}$. In this section, we explore the origin of this feature.

\begin{figure}
    \centering
    \includegraphics[width=\linewidth]{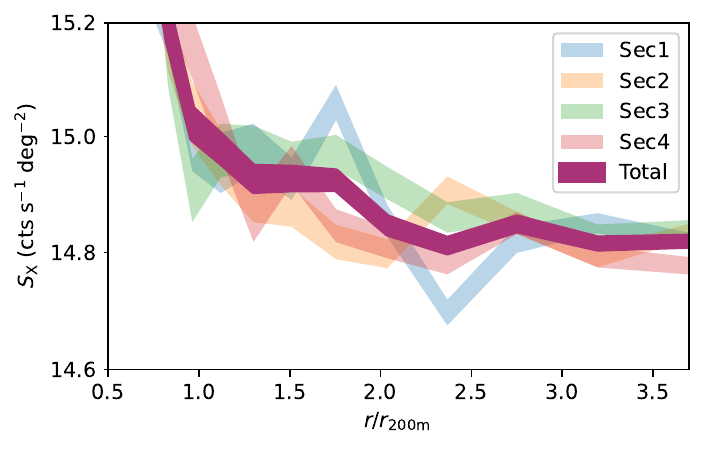}
    \caption{Comparison between the full stacked profile and those in four independent sectors. }
    \label{fig:prof_4sector}
\end{figure}

Following \citetalias{Lyskova2023}, we divided each cluster into four $90\degr$ sectors and stacked profiles of each. The comparison between the four sector profiles and the full profile is shown in Fig.~\ref{fig:prof_4sector}. The profile in sector 1 is the most scattered profile among the four. It shows features of a spike in the bin between 1.6 and 1.9 $r_\mathrm{200m}$ and a dip in the bin between 2.2 and 2.6 $r_\mathrm{200m}$. The two features result in a plausible jump in the full profile. Because the jump is not a universal feature across all four sectors, we do not overinterpret it further.

\section{Noise level from the uncorrelated components}\label{app:null_test}
The stacking of randomly distributed objects in the sky filters out signals from uncorrelated components, such as the instrumental background, fluctuations in Galactic halo emission, and foreground absorption. To validate it and quantify the contribution of the uncorrelated components in the error budget, we performed stacking analysis on a sky map free of sources.

\begin{figure*}
    \centering
    
    \includegraphics[width=0.45\textwidth]{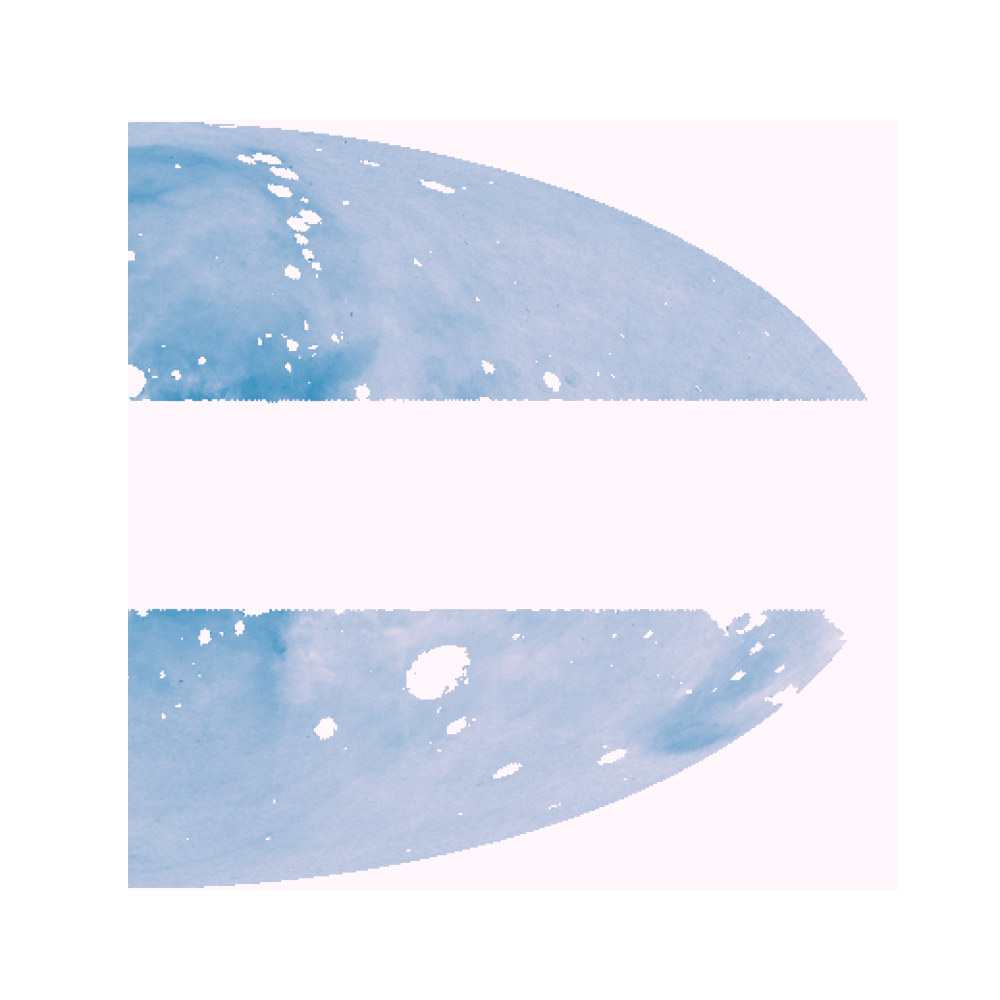}
    \includegraphics[width=0.4\textwidth]{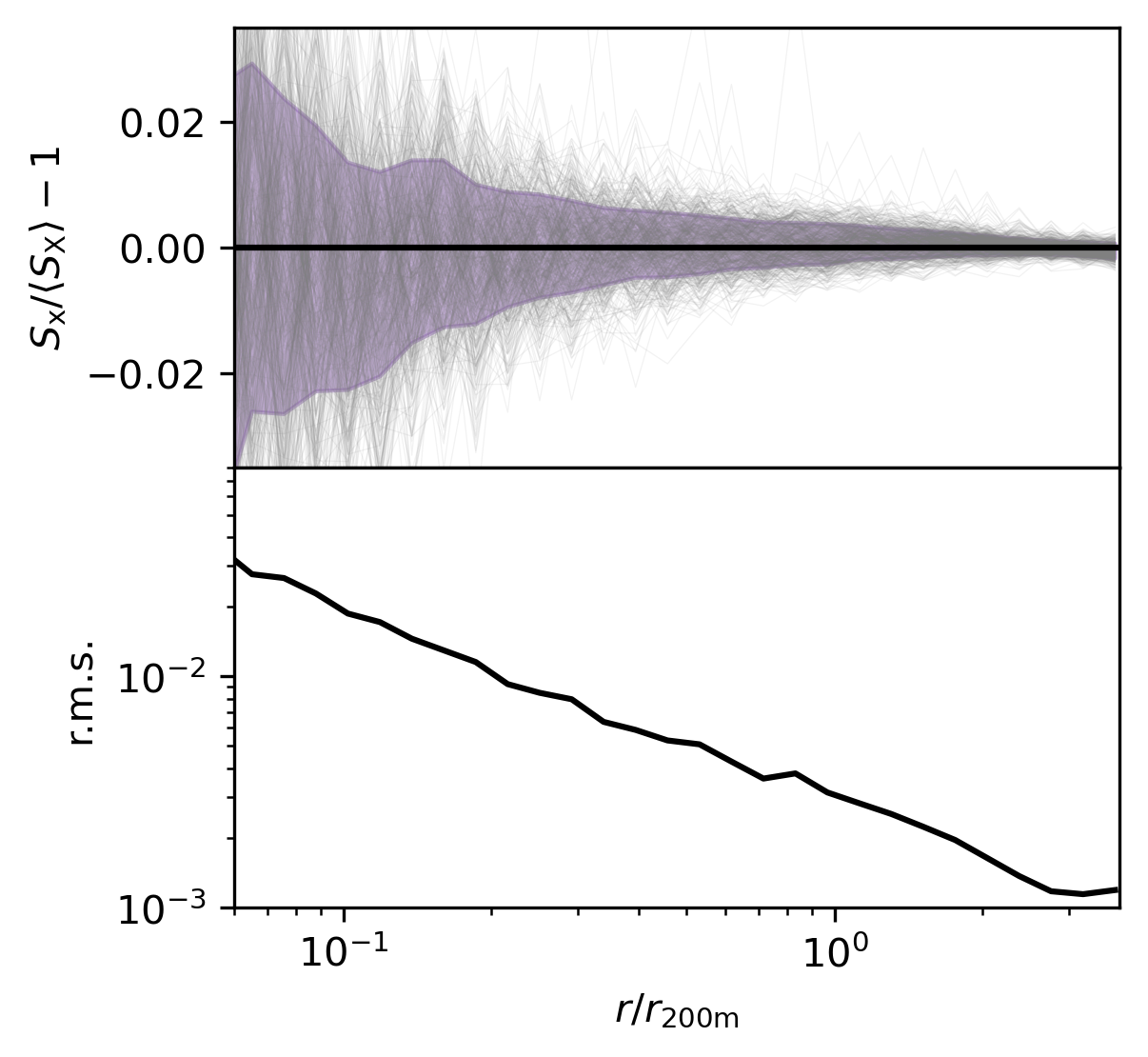}

    \caption{\emph{Left}: Source removed eRASS:4 sky map for studying the uncorrelated background noise. 
    \emph{Top right}: Five-hundred control sample relative surface brightness profiles stacked with cluster positions randomly distributed on the sky. The shade region denotes the scatter of the 500 control sample profiles. 
    \emph{Bottom right}: The scatter of the 500 control sample profile as a function of the radius. As the radius increases, the scatter decreases. Beyond $r_\mathrm{200m}$, the scatter is smaller than $0.3\%$} 
    \label{fig:null_test}
\end{figure*}

We first created an eRASS:4 HEALPix map in the 0.2-2.3~keV band with all sources removed (see Sect.~\ref{sect:mask_source} for the source catalogs). An example of the source masked sky rate map is shown in the left panel of Fig.~\ref{fig:null_test}. Then, we randomly distributed the positions of the 680 galaxy clusters on the half-sky map. With the randomized cluster positions, we obtained a stacked surface brightness profile with the weights provided in Eq.~\ref{eq:weight_average}. We repeated the randomization process 500 times to obtain a set of control sample profiles. 

We aimed to study relative surface brightness fluctuations relative to the profile average. Therefore, for each profile, we divided individual bin values by the profile mean.
On the right panel of Fig.~\ref{fig:null_test}, we plot the relative profiles of the 500 control samples, where the 68\% scatter is shown as the shaded region. On the bottom-right panel, we plot the relative scatter as a function of the radius. It shows that the 680 objects and the depth of the eRASS:4 sky result in a noise level below 0.3\% from uncorrelated components. This 0.3\% fluctuation can be treated as the systematic uncertainties of the stacked emission beyond $r_\mathrm{200m}$.

\section{Impact of using a constant cooling function}\label{app:cooling}

In this section, we investigate the impact and possible systematic uncertainties of adopting a constant $\Lambda_\mathrm{cf} A_\mathrm{eff}$ for fitting the stacked surface brightness profile. In reality, each object has its own temperature and metallicity profiles as well as a sky position-dependent foreground $n_\mathrm{H}$ column density that attenuates the observed count rate. 

\begin{figure}
    \centering
    \includegraphics[width=0.5\textwidth]{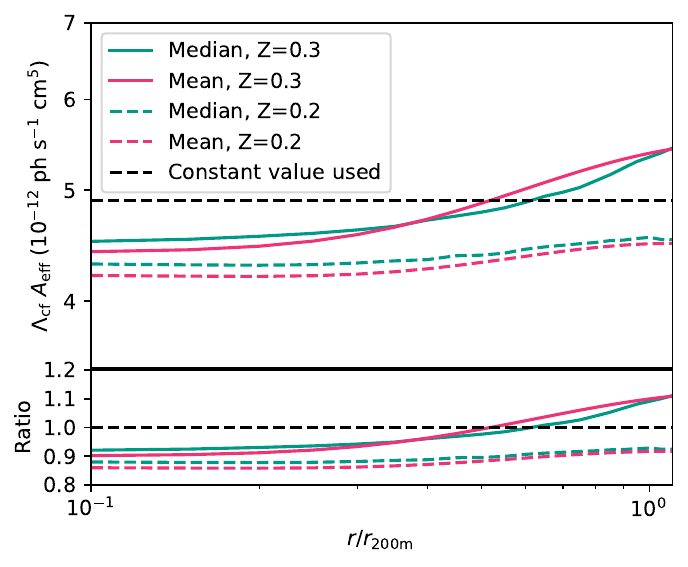}
    \caption{\emph{Top}: Sample averaged (teal) and median (red) radial cooling functions that take individual cluster temperature profiles and foreground absorption strength into account. Solid and dashed lines denote metallicity assumptions of 0.3 and 0.2 $Z_\sun$, respectively. The black-dashed line represents the constant cooling function used for model fitting in this work. 
    \emph{Bottom:} Ratios between radial cooling functions and the constant value used for model fitting. 
   }
    \label{fig:cf_profile}
\end{figure}

For each cluster, we predicted the temperature profile by adopting eqs. 10 and 11 and the best-fit parameters therein from \citet{Ghirardini2019}. The temperature-mass scaling relation is based on the self-similar theory $T\propto M^{2/3}$, and the profile shape is originally from \citet{Vikhlinin2006}. Then we calculated the radial $\Lambda_\mathrm{cf}A_\mathrm{eff}$ by plugging in the temperature profile, redshift, foreground $n_\mathrm{H,tot}$ value\footnote{Total effective hydrogen column density by taking molecular into account \citep{Willingale2013}.}, and gas metallicity. We calculated profiles with both the 0.2 and 0.3 $Z_\sun$ metallicity assumptions. The comparison between the constant value $4.9\times10^{-12}$~ph~s$^{-1}$~cm$^{5}$ used in this work and the sample mean/median is shown in Fig. \ref{fig:cf_profile}. 

With the $Z=0.3Z_\sun$ metallicity assumption, the sample average and median $\Lambda_\mathrm{cf}A_\mathrm{eff}$ profiles are $10\%$ lower than the constant value we used in the radial range $r\lesssim0.5\times r_\mathrm{200m}$ and are $10\%$ higher at large radii. The mild increase of the $\Lambda_\mathrm{cf}A_\mathrm{eff}$ with radius is due to the decrease of the gas temperature in the outskirts. With the $Z=0.2Z_\sun$ metallicity assumption, the $\Lambda_\mathrm{cf}A_\mathrm{eff}$ is $\sim10\%-15\%$ lower than the value used in this work at all radii. The lower radial dependence is due to the lower abundance of line emissions from the Fe-L complex, which reaches its highest emissivity at $kT\sim1$~keV.

\section{Fitting without two-halo term}

\begin{figure}
    \centering
    \includegraphics[width=0.45\textwidth]{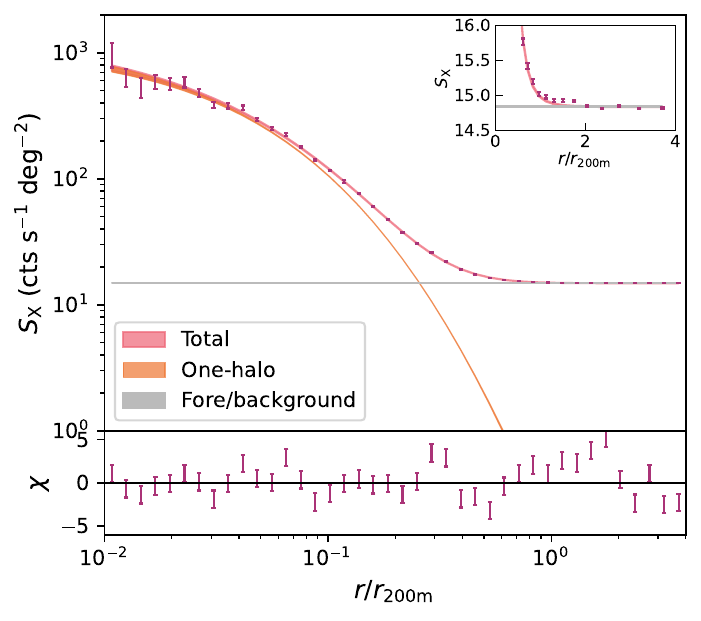} 
    \caption{Same as Fig.~\ref{fig:obs_prof_fitting} but without the two-halo term. Significant residuals at $r>r_\mathrm{200m}$ exist.
    }
    \label{fig:profile_fit_no2h}
\end{figure}

Fig.~\ref{fig:profile_fit_no2h} shows the best-fit components and residuals without the two-halo term. There are significant residuals at $r>r_\mathrm{200m}$ in both the inset and the residual plot.

\end{appendix}

\end{document}